\documentclass[aps,prc,showpacs,showkeys,twocolumn,letter]{revtex4}

\usepackage{graphicx,epsfig,amsmath,amssymb}
\usepackage[bookmarksnumbered,bookmarksopen]{hyperref}
\usepackage[capitalize]{cleveref}
\usepackage[utf8]{inputenc}

\AtBeginDocument{\usepackage{booktabs}}

\usepackage[usenames]{color}
\newcommand{\erf}{\operatorname{erf}}

\begin{document}

\title{Particle production and equilibrium properties within a new hadron transport approach for heavy-ion collisions}

\author{J.~Weil$^1$, V.~Steinberg$^1$, J. Staudenmaier$^{1,3}$, L.G.~Pang$^1$, D.~Oliinychenko$^{1,2}$, J.~Mohs$^{1,3}$, M.~Kretz$^{1,4}$, T.~Kehrenberg$^{1,3}$, A.~Goldschmidt$^{1,5}$, B.~Bäuchle$^1$, J.~Auvinen$^{1,6}$, M.~Attems$^{1,7}$ and H.~Petersen$^{1,3,4}$}

\affiliation{$^1$Frankfurt Institute for Advanced Studies, Ruth-Moufang-Strasse 1, 60438 Frankfurt am Main, Germany}
\affiliation{$^2$Bogolyubov Institute for Theoretical Physics, 14-b, Metrolohichna str., 03680 Kiev, Ukraine}
\affiliation{$^3$Institute for Theoretical Physics, Goethe University, Max-von-Laue-Strasse 1, 60438 Frankfurt am Main, Germany}
\affiliation{$^4$GSI Helmholtzzentrum für Schwerionenforschung, Planckstr. 1, 64291 Darmstadt, Germany}
\affiliation{$^5$Department of Physics, The Ohio State University, Columbus, OH 43210, USA}
\affiliation{$^6$Department of Physics, Duke University, Durham, North Carolina 27708-0305, United States}
\affiliation{$^7$ Departament de F\'\i sica Qu\`antica i Astrof\'\i sica \&
Institut de Ci\`encies del Cosmos (ICC), Universitat de Barcelona,
Mart\'{\i}  i Franqu\`es 1, 08028 Barcelona, Spain}
\date{\today}

\begin{abstract}
The microscopic description of heavy-ion reactions at low beam energies is achieved within hadronic transport approaches. In this article a new approach SMASH (Simulating Many Accelerated Strongly-interacting Hadrons) is introduced  and applied to study the production of non-strange particles in heavy-ion reactions at $E_{\rm kin}=0.4-2A$ GeV. First, the model is described including details about the collision criterion, the initial conditions and the resonance formation and decays. To validate the approach, equilibrium properties such as detailed balance are presented and the results are compared to experimental data for elementary cross sections.
Finally results for pion and proton production in C+C and Au+Au collisions is confronted with HADES and FOPI data. Predictions for particle production in $\pi+A$ collisions are made.
\end{abstract}



\maketitle


\section{Introduction}
Heavy-ion collisions offer the opportunity to study hot and dense strongly interacting matter under extreme conditions. High energy programs at the Large Hadron Collider (LHC) and the Relativistic Heavy Ion Collider (RHIC) are delivering a lot of detailed experimental data \cite{Adams:2005dq,Adcox:2004mh,Muller:2012zq} relevant for the high temperature and low net baryo-chemical potential part of the phase diagram which corresponds to the situation shortly after the Big Bang. Scanning the beam energies to lower values as currently done at the CERN-Super Proton Synchrotron (SPS) \cite{Gazdzicki:2011fx} and the RHIC beam energy scan \cite{Adamczyk:2014ipa,Adamczyk:2014fia,Adare:2015bua} program or in the future at FAIR and NICA provides access to regions in the phase diagram where a first order transition to the quark- gluon plasma is expected to take place. One of the goals of these programs is to search for a critical endpoint in the QCD phase diagram \cite{Friman:2011zz}.

Since there is no first principle solution of the many-body problem in quantum chromodynamics  including a non-equilibrium evolution through a phase transition up to date, effective theoretical approaches are necessary to describe the full dynamical evolution of heavy-ion reactions from the early to the late stages. By comparison of the output of these calculations with experimental data on particle distributions and their correlations in the final state, it is possible to draw conclusions about the properties of the hot and dense strongly interacting matter that was created for a very short time and in a very small volume.

Following the realization that the quark-gluon plasma behaves like an almost perfect fluid in contrast to the ideal gas expectation, within recent years the community has converged towards a standard model for the description of the evolution of heavy-ion reactions at high beam energies. The early stage of the collision is described by a non-equilibrium evolution likely based on fluctuating color fields/strings until approximate local equilibrium is reached \cite{Lappi:2015vta, Gelis:2013rba}. The hot and dense stage of the evolution is governed by relativistic dissipative hydrodynamics \cite{Song:2007ux,Luzum:2008cw,Gale:2013da,vanderSchee:2013pia} incorporating the QCD equation of state provided by lattice calculations \cite{Huovinen:2009yb,Borsanyi:2015waa,Moreland:2015dvc,Pratt:2015zsa}. The later dilute stages are described by a hadron transport approach \cite{Petersen:2014yqa}.
Even though most of the dynamical features are captured within the hydrodynamic calculation, the hadronic rescattering stage becomes necessary as soon as one wants to address identified particle spectra or correlation and fluctuation observables that are affected by resonance decays and baryon annihilation \cite{Werner:2010aa,Steinheimer:2012rd}.

The other limit where the description of the dynamical evolution of heavy-ion reactions is to some degree under control is at very low beam energies that are dominated by hadronic reactions and not yet affected by quark-gluon plasma formation. The region of intermediate beam energies that is of great interest with respect to the discovery of features in the QCD phase diagram poses a challenge to the current dynamical approaches. There are attempts to adapt the above described hybrid approaches and extend them to finite baryo-chemical potential \cite{Karpenko:2015xea}. The other option is to start from a vacuum hadronic transport approach that is extensible by including effects of the hot and dense medium such as many-body interactions. This second approach is the motivation for the development of a new hadronic transport approach, SMASH.

Hadronic transport approaches have been developed for 20-30 years and some models are still under active development \cite{Bass:1998ca, Nara:1999dz, Bratkovskaya:2011wp, Buss:2011mx}. The new experimental data that is available to constrain the resonance properties at low beam energies \cite{Agakishiev:2014wqa} and profiting from the experience of the existing transport approaches is the reason for developing a modern flexible open source code that can be adapted as a standard reference for a purely hadronic system with vacuum properties. To summarize, we have gained a lot of new experimental and theoretical insights over the past two decades that make the development of a new transport approach a timely endeavor. This new transport approach will also be highly relevant to provide a better understanding of the late stage evolution of hadronic rescattering at RHIC and LHC energies.

In this paper the newly developed approach is described in detail. In \cref{sec:model} the ingredients of the approach are explained including the general setup, the collision criterion, the initial conditions and treatment of potentials, Pauli blocking and resonance formation and decay. In \cref{sec:validation} basic checks of detailed balance and comparisons with elementary cross sections are shown. In \cref{sec:results} we present calculations of observables in comparison with experimental data from HADES and FOPI at $E_\text{kin} = 0.4-2A\,\text{GeV}$ and predictions for $\pi-A$ collisions.

\section{Model Description}
\label{sec:model}
\subsection{General Setup}
The main advantage of a microscopic transport approach is that the full phase-space information of all particles is available at all times. SMASH constitutes a solution of the non-equilibrium dynamics of hadrons in the regime where the inelastic interactions are treated by resonance excitations and decays with vacuum properties. The underlying equation is the relativistic Boltzmann equation
\begin{equation}
\label{eq:boltzmann_equation}
  p^\mu \partial_\mu f_i(x,p) + m_i F^{\alpha} \partial^{p}_{\alpha} f_i(x, p) = C^i_{\rm coll}
\end{equation}
where $C^i_{\rm coll}$ is the collision term, $F^{\alpha}$ is the force experienced by individual particles and $m_i$ is the particle mass.
For high beam energy collisions, $F^{\alpha} = 0$,  while for low beam energy collisions, $F^{\alpha} = -\partial^{\alpha} U(x)$ where $U(x)$ is the mean-field potential. 
The relativistic Boltzmann equation is an integro-differential equation in 6+1 dimensions.
$f_i(x,p)$ is the single particle distribution for each species~$i$ that is represented by test particles. Along the lines of quantum molecular dynamics each particle is in principle represented by a Gaussian wave packet. In practice, all particles are treated as point particles and the finite spatial extent is only invoked to calculate thermodynamic properties like the particle density. In our case, per default each real particle is represented by one test particle, but more test particles can be created if necessary. 

\subsubsection{Collision Criterion}
One of the major challenges for solving the Boltzmann equation in a relativistic situation is to define an appropriate collision criterion. The Kodama criterion \cite{Kodama:1983yk} is a fully covariant collision criterion, but since it involves boosts of several four vectors it is rather inefficient. In the current approach we have chosen to use the geometrical criterion employed in the UrQMD (Ultra-relativistic Quantum Molecular Dynamics) approach \cite{Bass:1998ca}, that is defined as follows:
\begin{equation} \label{eq:collision_criterion}
d_{\rm trans} < d_{\rm int} = \sqrt{\frac{\sigma_{\rm tot}}{\pi}}
\end{equation}
with
\begin{equation}
d_{\rm trans}^2 = (\vec{r_a}-\vec{r_b})^2-\frac{((\vec{r_a}-\vec{r_b})\cdot(\vec{p_a}-\vec{p_b}))^2}{(\vec{p_a}-\vec{p_b})^2}
\end{equation}
where $\vec{r}$ and $\vec{p}$ are the coordinates and momenta of the two particles $a$ and $b$ in the center of mass frame of the binary collision. The time of the collision is determined as the time of the closest approach in the computational frame:
\begin{equation} \label{eq:collision_time}
t_{\rm coll}=-\frac{(\vec{r_a}-\vec{r_b})\cdot (\vec{p_a}/E_a-\vec{p_b}/E_b)}{(\vec{p_a}/E_a-\vec{p_b}/E_b)^2}
\end{equation}
where now all coordinate and momentum vectors have to be taken in the computational frame. The computational frame is usually chosen to be the equal velocity frame of the two nuclei which is the same as the center of mass frame in case of symmetric systems. The computational system is the one that carries the clock that is relevant for ordering of the collisions, therefore it is crucial to transform the collision times to the same frame to decide which collision happens first.

This geometrical criterion effectively encodes an instantaneous interaction over a finite distance  and gives rise to causality violations \cite{Cheng:2001dz}. We have compared the UrQMD criterion to the covariant Kodama criterion and found no significant differences. Since the above explained criterion is numerically more efficient, we stick to this definition in the following.

A different option to include all relevant scatterings at high density is to implement the solution of the Boltzmann equation by stochastic rates \cite{Danielewicz:1991dh,Cassing:2001ds,Xu:2004mz}. This approach has the advantage that multi-particle scatterings can be taken into account in a straightforward way. On the hadronic level there are of course a lot of different possibilities that one would need to take into account in such an approach, therefore this is left for future work. Also, the stochastic rates approach is relying on having a large number of test particles in each cell, therefore it is not clear how to model event-by-event fluctuations properly.

\subsubsection{Test particles}
Another method to circumvent the locality issues is the test particle method: all cross sections are scaled by a factor $N_{\rm test}^{-1}$, while the number of initially sampled particles is increased by the same factor $N_{\rm test}$.
\begin{align}
\label{eq:testparticles_method1}
\sigma & \mapsto \sigma N_{\rm test}^{-1} \\
\label{eq:testparticles_method2}
N & \mapsto N N_{\rm test}
\end{align}
$N_{\rm test}$ is referred to as ''test particle number''. After substitution (\cref{eq:testparticles_method1,eq:testparticles_method2}) the scattering rate (number of collisions per unit time per particle) remains unchanged, but the cross sections become smaller and collisions are ''more local''. Locality is restored in the limit $N_{\rm test} \to \infty$. As shown in \cite{Cheng:2001dz}, experimental observables such as particle spectra and flow obtained using transport models depend on $N_{\rm test}$ and saturate when $N_{\rm test}$ is sufficiently large (in case of \cite{Cheng:2001dz} $N_{\rm test} = 16$ was large enough for saturation). Another important application of test particles is to provide statistics for density or phase-space density estimates, which are required for evaluating potentials accurately.

\subsubsection{Time Steps/Propagation}
To solve the Boltzmann equation numerically, time and space need to be divided into cells. The granularity of the time steps are crucial, since the time steps need to be small enough to catch all collisions (when assuming a maximum of one collision per particle in a timestep) and as large as possible to ensure a fast evaluation of the evolution. Therefore, SMASH has two different options for the propagation. Either fixed time steps are chosen or the time steps are dynamically determined from the collision times. The first setup has the advantage that calculations with nuclear potentials are feasible while the second one adapts nicely to high and low density regions and is more efficient. 

In an algorithm with fixed time step size the actions are only determined for a short time $dt$ in advance. In addition to the time step size $dt$ a start $t_\text{start}$ and an end time $t_\text{end}$ have to be chosen. The search for collisions needs to extrapolate the movement of the particles. The assumption in SMASH  is that the time step size $\Delta t$ is small enough that the effect of potentials on the trajectory of the particles can be neglected during this time interval. Therefore,  the movement is extrapolated without taking potentials into account.

Any interaction of two particles is called an action.  If the criterion \cref{eq:collision_criterion} is satisfied, then the collision is added to the list of collisions and decays with a time stamp $t_{\rm coll}$. After all actions are found, they are sorted according to their associated time. Iterating over the sorted list, all actions are first tested whether they are still valid. We rely on the assumption that each particle only interacts once during one time step. Valid actions are performed which involves replacing the incoming particles with the outgoing particles. The actions are performed before the propagation, which means that the global time of the particles is still the time of the beginning of the time step. When all valid actions have been performed, all particles are propagated taking potentials into account (if they are present).

When propagating without potentials or for the later dilute stages of the collision, it is useful to abandon any fixed time steps and rather switch to an algorithm that takes the actions themselves to determine the next propagation step. The general idea is that the program keeps a list of actions which is constantly updated. Actions are removed from the list as they are performed and added as they are newly discovered. At the beginning of the simulation, the end time of the simulation $t_\text{end}$ is specified.  This is used to find all possible actions for all particles until $t_\text{end}$ as described before. Next, there is a loop over all actions that starts with the first action according to the time of execution of the actions and checks if this action is still valid. The check consists of verifying that the incoming particles were not part of another action since this action was found. If they were, the action is discarded.

If the action is valid, all particles are propagated to the point in time where the action is supposed to happen. Then the action is performed as described before. As a result, all actions that involved the incoming particles are implicitly rendered invalid, since the state of these particles has changed. In the last step, all possible actions of the outgoing particles are added to the list of actions. This algorithm realizes all actions that are supposed to happen, assuming the
time ordering of the actions is correct (which depends on the collision criterion).

\begin{figure}
  \centering
  \includegraphics[width=0.48\textwidth]{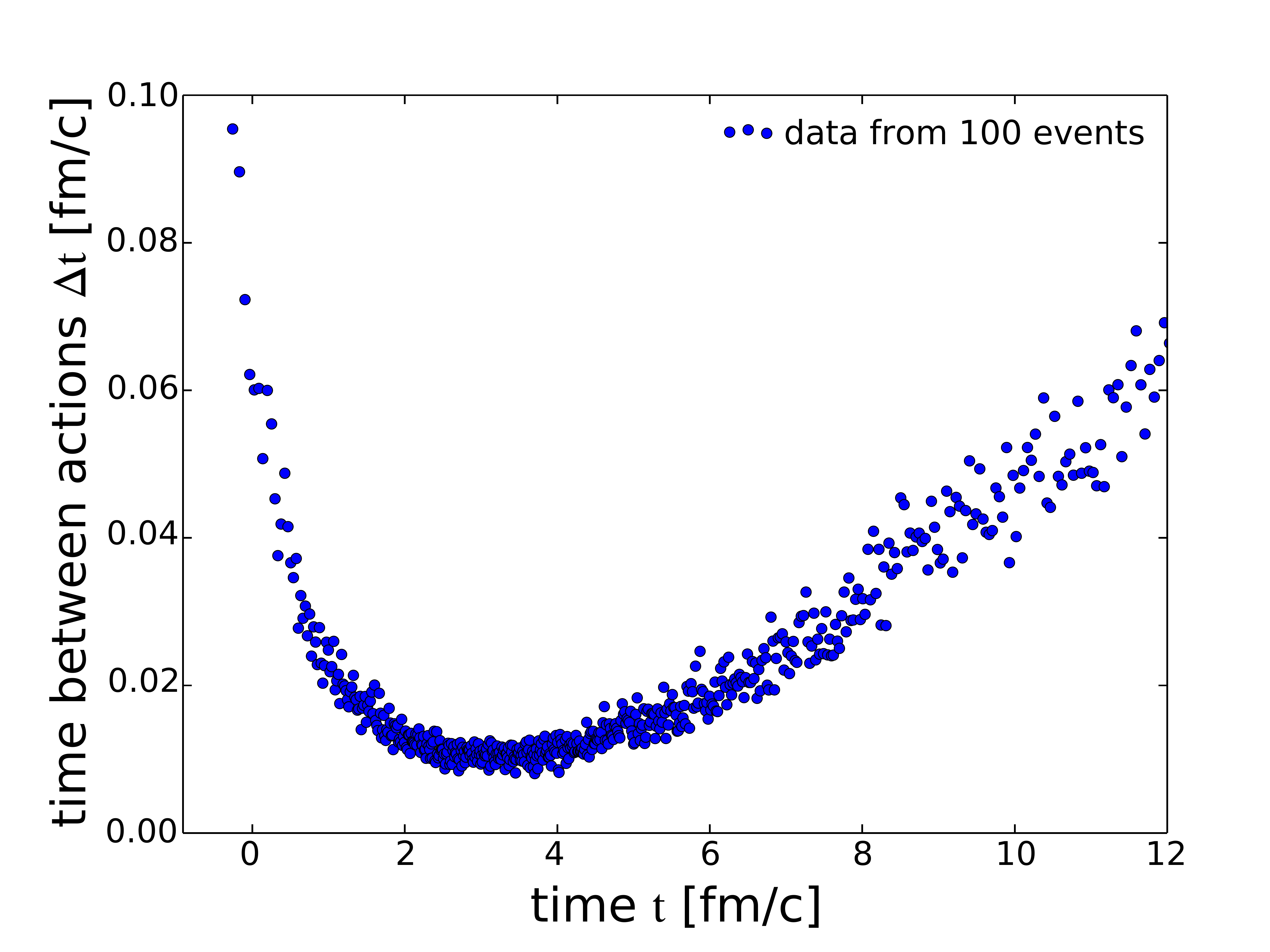}
  \caption{The time between consecutive actions for a central Cu-Cu collision at $\sqrt{s}_{NN}$
  = 3\,GeV, averaged over 100~events within the algorithm without fixed time steps. This plot shows the averaged time until the first, second, third,... $n^{\rm th}$ interaction, therefore, the results are still scattered.}
  \label{fig:time_between_actions}
\end{figure}

\cref{fig:time_between_actions} shows the time difference between
consecutive interactions $\Delta t$ for a Cu-Cu collision, averaged over 100~events. Each data
point corresponds to one action. The average of this time between actions becomes as
small as 0.01\,fm/c but the variance of $\Delta t$ is high and outliers can reach
0.0001\,fm/c.

\subsubsection{Mean-field potentials}
To create a more realistic simulation at low beam energies, a minimal version of mean-field potentials between nucleons is included. The equations of motions have to be adjusted according to the modified one-particle Hamiltonian $H_i$
\begin{equation}
H_i = \sqrt{\vec p_i^{\,2} + m_{\rm eff}^2} + U(\vec r_i)\,,
\end{equation}
where $m_{\rm eff}$ is the mass for stable hadrons and the effective mass for resonances in accordance with their mass distribution (e.g. Breit-Wigner). At this point, the potential depends only on the coordinates, but not on the momentum of the particles. The corresponding equations of motion are then
\begin{align}
  \frac{d\vec r_i}{dt} &=  \frac{\partial H_i}{\partial \vec p_i} = \frac{\vec p_i}{\sqrt{\vec p_i^{\,2} + m_{\rm eff}^2}} \; , \\
  \frac{d\vec p_i}{dt} &= -\frac{\partial H_i}{\partial \vec r_i} = -\frac{\partial U}{\partial \vec r_i} \; .
\end{align}

This formulation leads to the fact that momentum conservation is fulfilled only on average. Event by event momentum conservation requires that $d\vec p_i/dt =-\partial H_\text{tot}/\partial \vec r_i$, where $H_\text{tot} = \sum_i H_i$. The potential is calculated as a function of the local density
\begin{equation}
  \label{eq:potential}
  U = a (\rho/\rho_0) + b (\rho/\rho_0)^{\tau} \pm 2 S_\text{pot} \frac{\rho_{I3}}{\rho_0}
\end{equation}
Here $\rho$ is the Eckart rest frame baryon density and $\rho_{I3}$ is the Eckart rest frame baryon isospin density of the relative isospin projection $I_3/I$. $\rho_0=0.168 1/\text{fm}^3$ is the nuclear ground state density. Parameters for the Skyrme potential are by default set to $a=-209.2$ MeV, $b=156.4$ MeV and $\tau = 1.35$, while $S_\text{pot} = 18$ MeV is the default value for the symmetry potential. These parameters were agreed on for a recent transport code comparison \cite{Xu:2016lue} and correspond to a rather soft potential with an incompressibility of $K=240$ MeV. For the equations of motion one does not need the potential itself, but its gradient, $\partial U/\partial \vec r$. In the symmetry term the positive sign is applied for the potential acting on neutrons and the minus sign is applied for the potential acting on protons. Currently, the potential acts only on baryons. The potentials are always calculated after the actions are performed, right when the propagation happens.

We note that electromagnetic potentials (Coulomb and Lorentz force) are currently being neglected in the model, since they are typically much weaker than the hadronic mean fields (even if they are more long-ranged). The Coulomb potential can only play a role for collisions of large nuclei at very low energies and is completely negligible at higher energies (FAIR/RHIC/LHC).

\subsubsection{Nearest neighbor search}
\label{sec:grid}

To determine if two particles will scatter, their distance needs to be compared to the total cross section. In principle, every particle has to be paired with every other particle in the system and the complexity of the search will scale with $N^2$ where $N$ is the number of particles, which is computationally intensive. In order to reduce the combinatorics of this search, the space can be divided into cells whose sizes are chosen such that, accounting for the time step size $\Delta t$ and the maximal possible cross section $\sigma_\mathrm{tot}^\mathrm{max}$, all collisions will happen within one cell or among neighboring cells, but not beyond that. SMASH uses such a grid structure with a minimal cell size of (2.5\,fm)$^3$. The value of $d=2.5$ fm corresponds to a maximum cross section of $\sigma_\mathrm{tot}^\mathrm{max} = \pi d^2 \approx 200$ mb that can be handled. This maximum is reached in the $\Delta$ peak of the $\pi^+p$ cross section, see \cref{fig:xs_piN}.
The only exception of physical cross sections going above 200 mb are the elastic NN cross sections, which diverge at the threshold. Those are effectively being cut at 200 mb with our minimal cell size (at least without test particles). When larger numbers of test particles are used, the cross sections are scaled down accordingly and this limitation is lifted.

In the actual algorithm for iterating over the cells, a distinction is made between in-cell search and neighbor search. The in-cell search is used to find decays and collisions within a given cell.

\begin{figure}[h]
  \centering
 \includegraphics[width=0.48\textwidth]{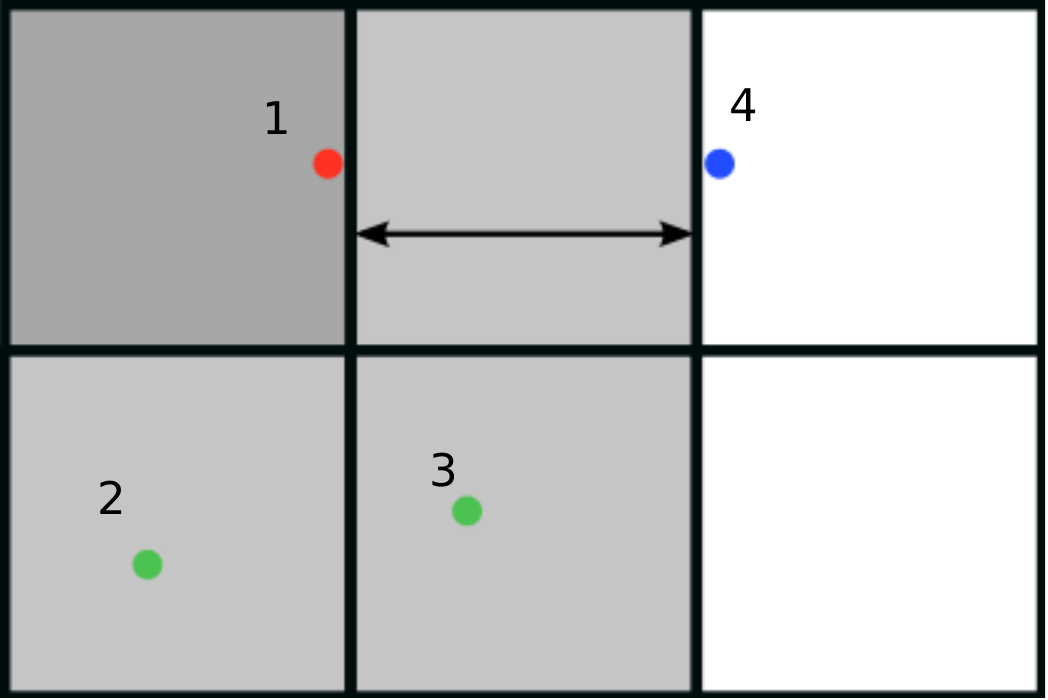}
  \caption{Two-dimensional schematic representation of the grid structure for finding
  collisions between particles. When searching for collisions, particle 1 is
checked with particles 2 and 3, but not with particle 4.}
\label{fig:grid}
\end{figure}

The neighbor search looks for actions between the particles in a given cell and its neighbors. To avoid finding duplicate actions, not all neighboring cells are used in the neighbor search. Consider the case depicted in \cref{fig:grid}: When starting the neighbor search from the dark gray cell, the actions between particle 1 and particles 2 and 3 will be found. Afterwards, when starting the search from a light gray cell, the dark gray cell can be omitted from the search, because there would be no new actions. After some analysis one comes to the conclusion that each cell needs to check only half of its neighbors (except for cells at the border which need to check even fewer).

The grid is also used to realize periodic boundary conditions. With periodic boundaries, particles that are on opposite sides of a fixed-sized box can interact. When this feature is activated, the neighbor search for a cell at the border does not only check the actual neighboring cells.  Instead, so-called ghost cells are added that contain the mirrored particles from the opposite side of the grid. Note that in 
this case it is important that no grid cell size at the boundaries is smaller 
than the minimal cell size. Therefore, the cell size is scaled up to fit the 
total volume with the minimal number of complete cells.

\subsubsection{Elastic Box Test}
To test the collision finding algorithm, we employ a simple setup, which we further call ''elastic box''. A box with periodic boundary conditions is uniformly filled with $N$ pions. The momenta are distributed according to a Boltzmann distribution
\begin{equation}
\frac{dN}{d^3 p} \sim \exp(-\sqrt{\vec{p}^{\,2} + m^2}/T),
\end{equation}
where the temperature $T$ is taken to be 0.13 GeV. The pions are only allowed to scatter elastically with a constant isotropic cross section $\sigma$. In this simple setup the scattering rate $s$ should be $s = n \sigma$, where it is taken into account that the relative velocity between particles is close to the speed of light (a calculation using Eq.~(52) of \cite{Cannoni:2013bza} gives $v_{\rm rel} \approx 0.98$ for our setup), and $n$ is the  particle density.

\begin{figure}
\centering
\includegraphics[width=0.48\textwidth]{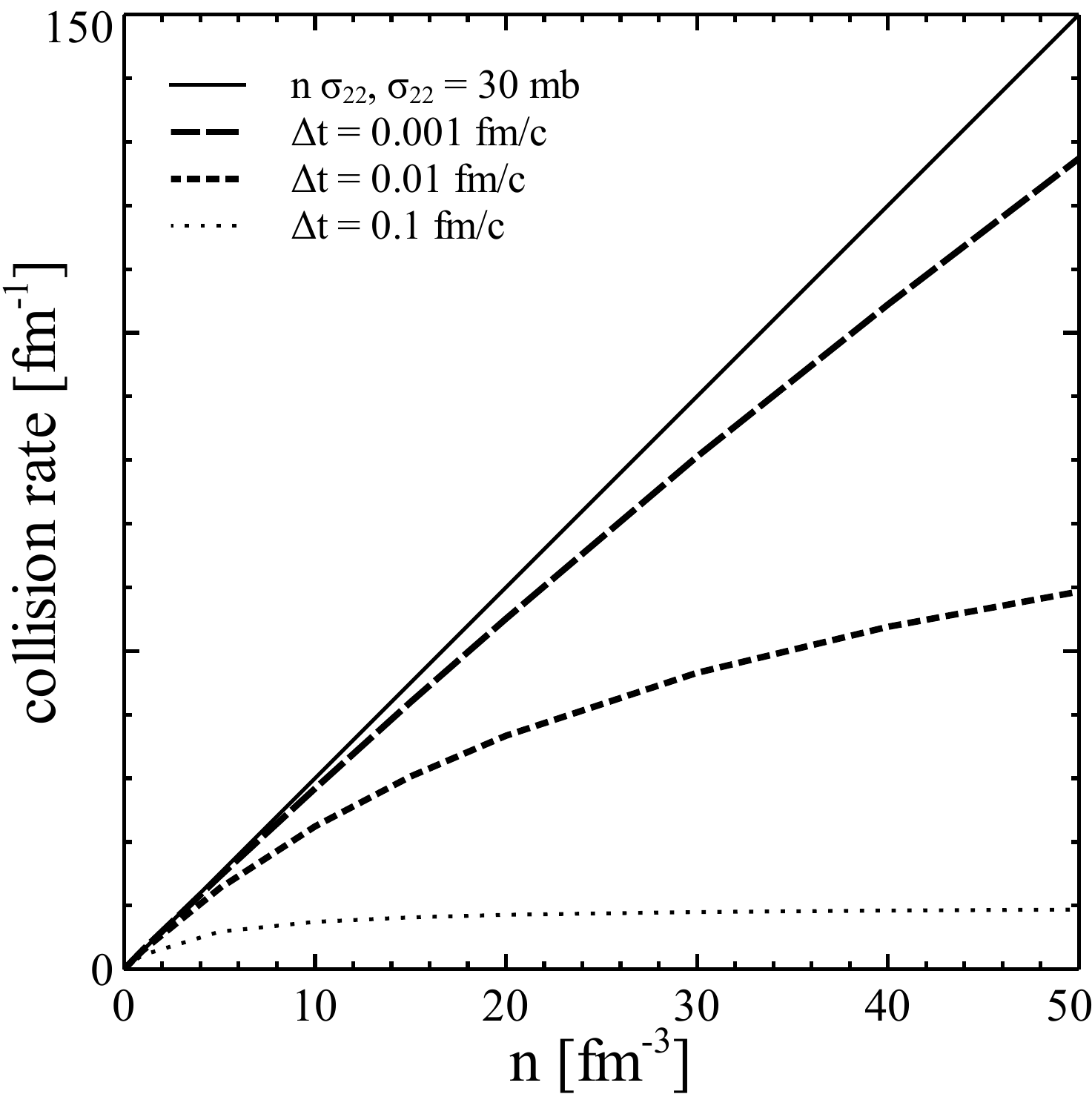} \\
\includegraphics[width=0.48\textwidth]{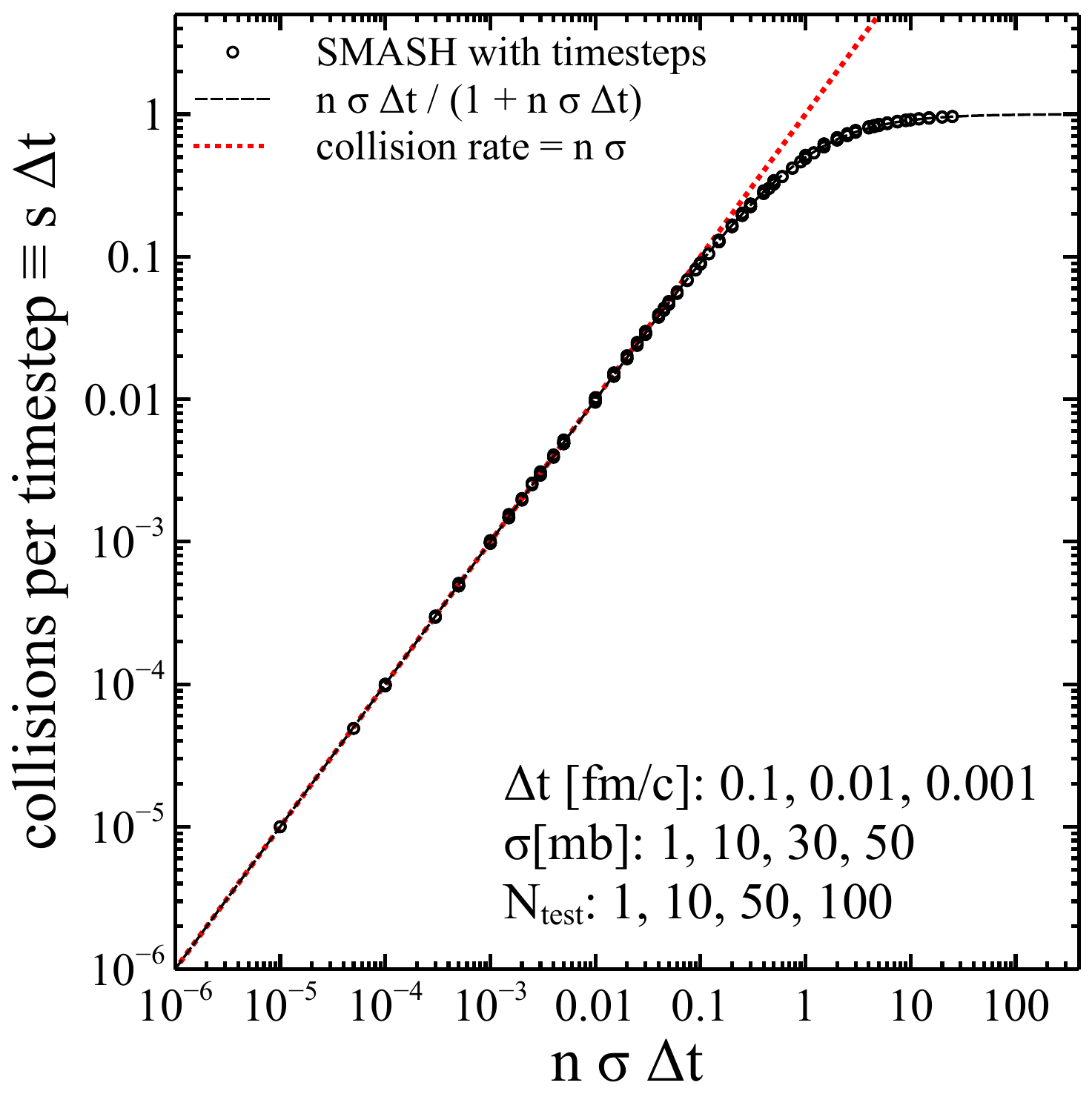}
\caption{''Elastic box'' (pion box with constant isotropic cross section). Upper panel: scattering rate versus density for different time steps. Lower panel: universal curve - number of collisions per timestep $s \Delta t$ versus $n \sigma \Delta t$ for 48 possible combinations of time step $\Delta t \in \{0.1, 0.01, 0.001 \}\,\text{fm}$, isotropic elastic cross section $\sigma \in \{1, 10, 30, 50\}\,\text{mb}$ and test particle numbers $N_{\rm test} \in \{1, 10, 50, 100\}$. }
\label{fig:rates}
\end{figure}

From \cref{fig:rates} one can see that $s=n \sigma$ is fulfilled for SMASH regardless of cross section $\sigma$ or test particle number $N_{\rm test}$, but only if the time step is sufficiently small. It is also interesting to observe that the scattering rate $s$ for all combinations of different $\sigma$, $N_{\rm test}$ and $\Delta t$ lies on one universal line
\begin{equation}
s \Delta t = \frac{n \sigma \Delta t}{1 + n \sigma \Delta t}.
\end{equation}
For small time steps $n \sigma \Delta t = \Delta t / \lambda \ll 1$ one retrieves the expected ideal gas behavior, while in the limit of large time steps there is one collision per particle per time step. This is expected, because more than one collision per particle per time step is prohibited by the SMASH algorithm. If one wants to have the full collision rate, the propagation from collision to collision without timesteps can be used. The universality of the curve can be explained in terms of dimensions. One can construct only three dimensionless quantities from $s$, $n$, $\sigma$ and $\Delta t$. Let us choose $s \Delta t$, $n \sigma \Delta t = \Delta t/\lambda$ and $n \sigma^{3/2}$, then
\begin{equation}
s \Delta t = f(n \sigma \Delta t, n \sigma^{3/2}).
\end{equation}
Assuming independence on the second argument one obtains our universal curve. We have additionally checked that these results depend only on the density $n = N/V$, but not on $N$ or $V$ separately. In other words, one can vary number of particles $N$, box volume $V$ or both, but the results are identical if $n = N/V$ is the same.

\subsection{Initial Conditions}
\subsubsection{Nuclear Collisions}
\label{sec:nuclear_collisions}

\paragraph{Nucleon Distribution in coordinate space}

To generate initial conditions for heavy-ion collisions the whole phase-space distribution of the initial nucleons needs to be sampled. In coordinate space, 'round' nuclei like gold or lead can be described by Woods-Saxon distributions

\begin{equation}
\frac{dN}{d^3r} = \frac{\rho_0}{\exp\left(\frac{r-r_0}{d}\right) + 1}
\end{equation}
where $d$ is the diffusiveness of the nucleus which controls the quick fall-off of the distribution. For $d\rightarrow 0$, the nucleus is a hard sphere. $\rho_0=0.168$ fm$^{-3}$ and $r_0$ are, in this limit, the nuclear ground state density and the nuclear radius. The default value for the diffusiveness is $d=0.545$ fm, where more specific values are used for Au, Pb, Cu and  U (see \cref{tab:nuclei}).

\begin{figure}
\centering
\includegraphics[width=0.48\textwidth]{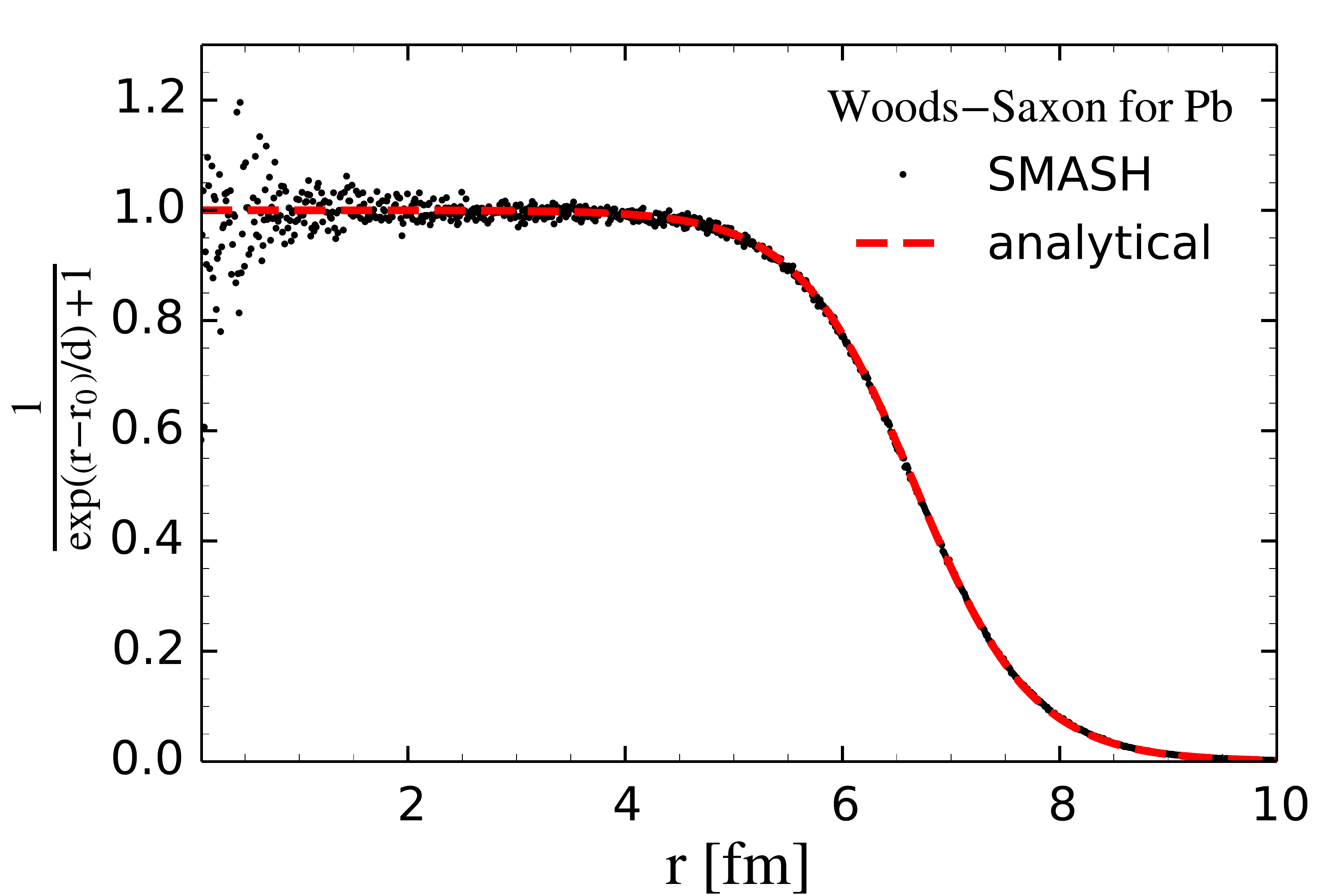}
\caption{Coordinate space distribution of 208 nucleons compared to the Woods-Saxon distribution with the parameters for a lead nucleus.}
\label{fig:woods-saxon}
\end{figure}

\begin{table}
\caption{\label{tab:nuclei} This table summarizes the specific parameters used in the Woods-Saxon initialization for some nuclei.}
\begin{tabular}{lccc}
\toprule
Nucleus & $A$ & $r_0$ [fm] & $d$ [fm] \\
\midrule
U &  238 & 6.86 & 0.556  \\
Pb & 208 & 6.67 & 0.54 \\
Au & 197 & 6.38 & 0.535 \\
Cu & 63 & 4.20641 & 0.597  \\
\bottomrule
\end{tabular}
\end{table}
Other values can be provided via the corresponding parameters in the configuration input file if necessary. Within the sampling procedure the finite size of the nucleons and nucleon-nucleon correlations are neglected for simplicity \cite{Alvioli:2009ab}. In \cref{fig:woods-saxon} it is shown that the sampling in coordinate space for a lead nucleus works as expected.

\begin{figure}
\centering
\includegraphics[width=0.48\textwidth]{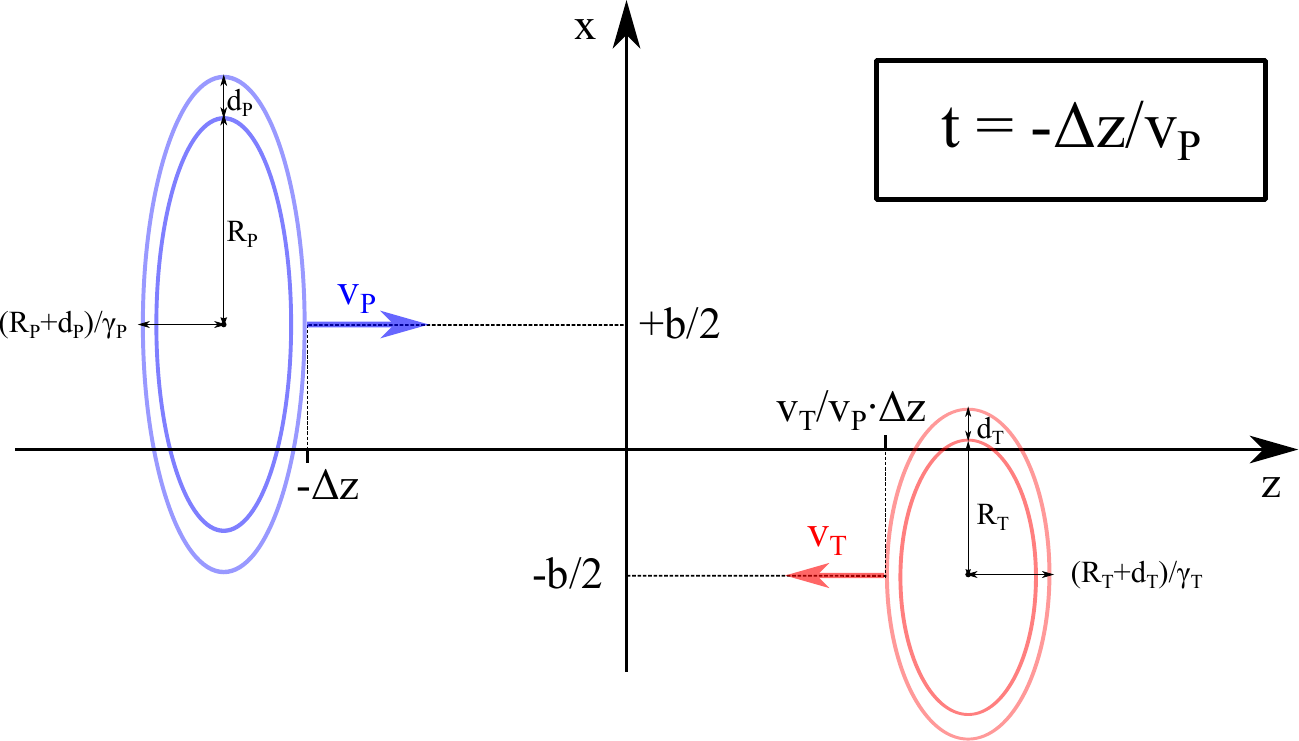}
\caption{Initial positions of nuclei such that contracted spheres of radii $(R + d)_{P,T}$ will touch at $t = 0$ in a central collision.}
\label{fig:init_coord_time}
\end{figure}

The initial positions of nuclei and the time of initialization are chosen as shown on \cref{fig:init_coord_time}.
We use Cartesian coordinates, where the $z$-direction corresponds to the beam direction and $x$ is the impact parameter direction. At the initialization the projectile center is at $xz$-coordinates $(b/2, - \Delta z- \gamma_P^{-1} (R_P + d_P))$ and the target center is at $(-b/2, \frac{v_T}{v_P}\Delta z + \gamma_T^{-1} (R_T + d_T))$. Here $R_{P,T}$ are the projectile/target radii and $d_{P,T}$ are the corresponding diffusiveness parameters from the Woods-Saxon distribution. By $v_{T,P}$ we denote absolute values of the velocities, while  $\gamma_{P,T} = (1 - v_{P,T}^2)^{-1/2}$ are the associated gamma-factors. The separation of the centers of the nuclei in $x$-direction equals the impact parameter $b$. For deformed nuclei an additional rotation along all three angles is applied. In this way, the simulation is started at such an initial separation that the potential of one nucleus does not influence the other one yet, otherwise initialization in the ground state would not be justified.
The initial coordinates and time are chosen in such a way that Lorentz-contracted spheres of radii $(R + d)_{P,T}$ will touch at $t = 0$ in a central collision. An alternative definition would be that $t=0$ fm corresponds to the maximal overlap of the two nuclei. The additional distance $\Delta z= 2$ fm is added to avoid missing any nucleon-nucleon collisions. Since the nucleons are distributed according to Woods-Saxon distributions, there is a small, but non-zero probability to position a nucleon at a large distance from the nucleus center. The initial separation distance $\Delta z$ is chosen such that all collisions are taken into account. The initial time is $t_0 = \Delta z/v_P$, which implies that the projectile is always moving, $v_P > 0$, while the target can be at rest depending on the reference frame for the calculation.

\paragraph{Fermi Motion}

In momentum space nucleons optionally get Fermi momenta, then target and projectile are boosted in $z$ direction according to the chosen energy of the reaction and computational frame. The gamma-factor of the boost is $\gamma = E_A/M_A$, where $E_A$ is the energy of the nucleus and $M_A$ is its mass. The velocity of the boost is $\beta = p_A/E_A$. Note that in $E_A$ and $M_A$ one has to account for the binding energy of the nucleus. For this we adopt an approximation used in the JAM transport code \citep{Nara:1999dz}, which assumes that all nucleons are equally bound. Thus, the energy of each nucleon in the rest frame of the nucleus is $E_i = M_A/A$, where $A$ is the number of nucleons. With this assumption the boost of the longitudinal momenta $p'_{iz}$ to the computational frame becomes
\begin{align}
p'_{iz} = \gamma (p_{iz} + \beta E_i) = \gamma p_{iz} + \frac{p_A}{M_A} \frac{M_A}{A} = p_\text{beam} + \gamma p_{iz} \,,
\end{align}
where $p_\text{beam}$ is the beam momentum per nucleon and $p_{iz}$ are the momenta of nucleons in the rest frame of the nucleus. In our implementation $p_\text{beam}$ and $\gamma$ themselves are computed without accounting for binding energy. We note that there is no well-established procedure of boosting nuclei with the account of their binding energy. Codes like UrQMD \citep{Bass:1998ca}, JAM \citep{Nara:1999dz} and GiBUU \citep{Buss:2011mx} apply different methods. Though the typical binding energy per nucleon is much smaller than the nucleon mass $(\simeq 8 \text{ MeV}/938 \text{ MeV} \approx 1\%)$, we found that the different methods of accounting for the binding energy produce small but noticeable differences in pion multiplicities and mean transverse momentum at low collision energies of $E_\text{kin} = 0.4-2A$ GeV.

The momentum distribution of nucleons in the ground state nucleus is a uniformly filled sphere in momentum space, known as the Fermi sphere. The radius of the Fermi sphere is given by the formula
\begin{equation}
p_F (\vec{r})= \hbar c (3 \pi^2 \rho (\vec{r}))^{1/3}
\end{equation}
where $\rho(\vec{r})$ is the density of nucleons at the point $\vec{r}$. A more detailed description of the density calculation is given in \cref{sec:thermodynamics}. A typical value of $p_F \approx$ 300 MeV corresponds to an energy of $p_F^2/(2m_N) \approx$ 45 MeV.

\begin{figure}
\centering
\includegraphics[width=0.3\textwidth]{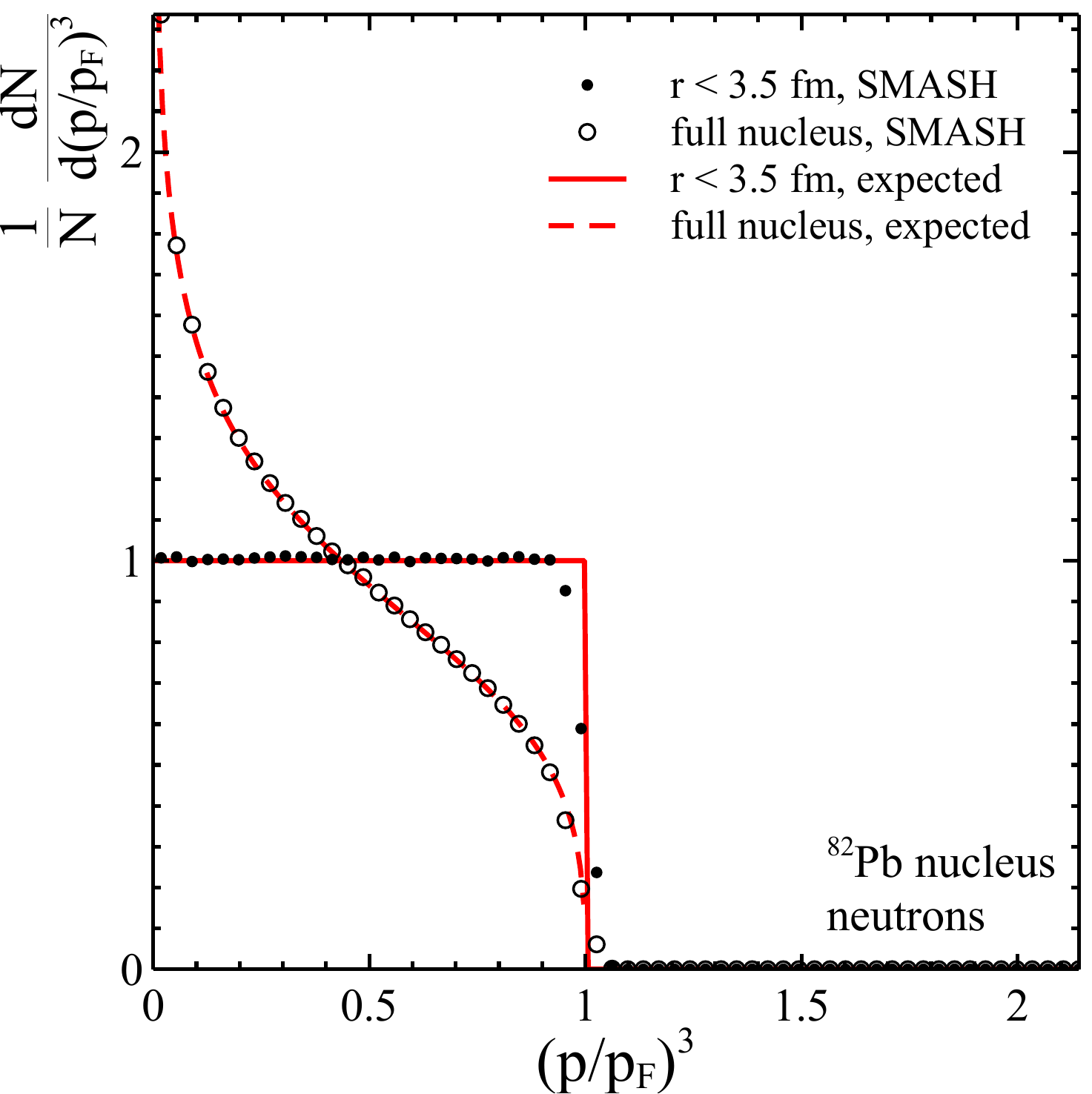}
\caption{Momentum space distribution of neutrons compared to the analytical expectation for a lead nucleus. }
\label{fig:fermi}
\end{figure}

To make sure that Fermi momenta are generated correctly, in \cref{fig:fermi} we plot the momentum distribution of the neutrons in a lead nucleus from SMASH and compare it to theoretical expectation computed as follows.
In the central part of the nucleus, where the  density is uniform, the expected normalized distribution $1/N\, dN/d(p/p_F^0)^3 = 1$, where $p_F^0 \equiv p_F (0)$, while the analogous momentum distribution integrated over the whole nucleus with $\rho(r) \sim 1/\left(1 + e^{(r-R)/d}\right)$ is
\begin{equation}
\frac{1}{N} \frac{dN}{dx^3} = \left[1 + \pi^2 \alpha^2 \right]^{-1} \big( 1 + \alpha \ln( x^{-3} - 1 )\big)^3 \theta(p_F^0-p) \,,
\end{equation}
where $x = p/p_F^0$ and $\alpha = d/R$.

Including the Fermi motion is only sensible if potentials are turned on simultaneously. Otherwise, the nucleus will fly apart due to the finite transverse momenta of the nucleons that need to be compensated by the attractive mean field interaction. Alternatively, one may employ the so-called frozen Fermi approximation: Fermi momenta are used for collisions, but not for propagation. This option is currently not implemented, but will be considered in the future. 

\begin{figure}
\centering
\includegraphics[width=0.23\textwidth]{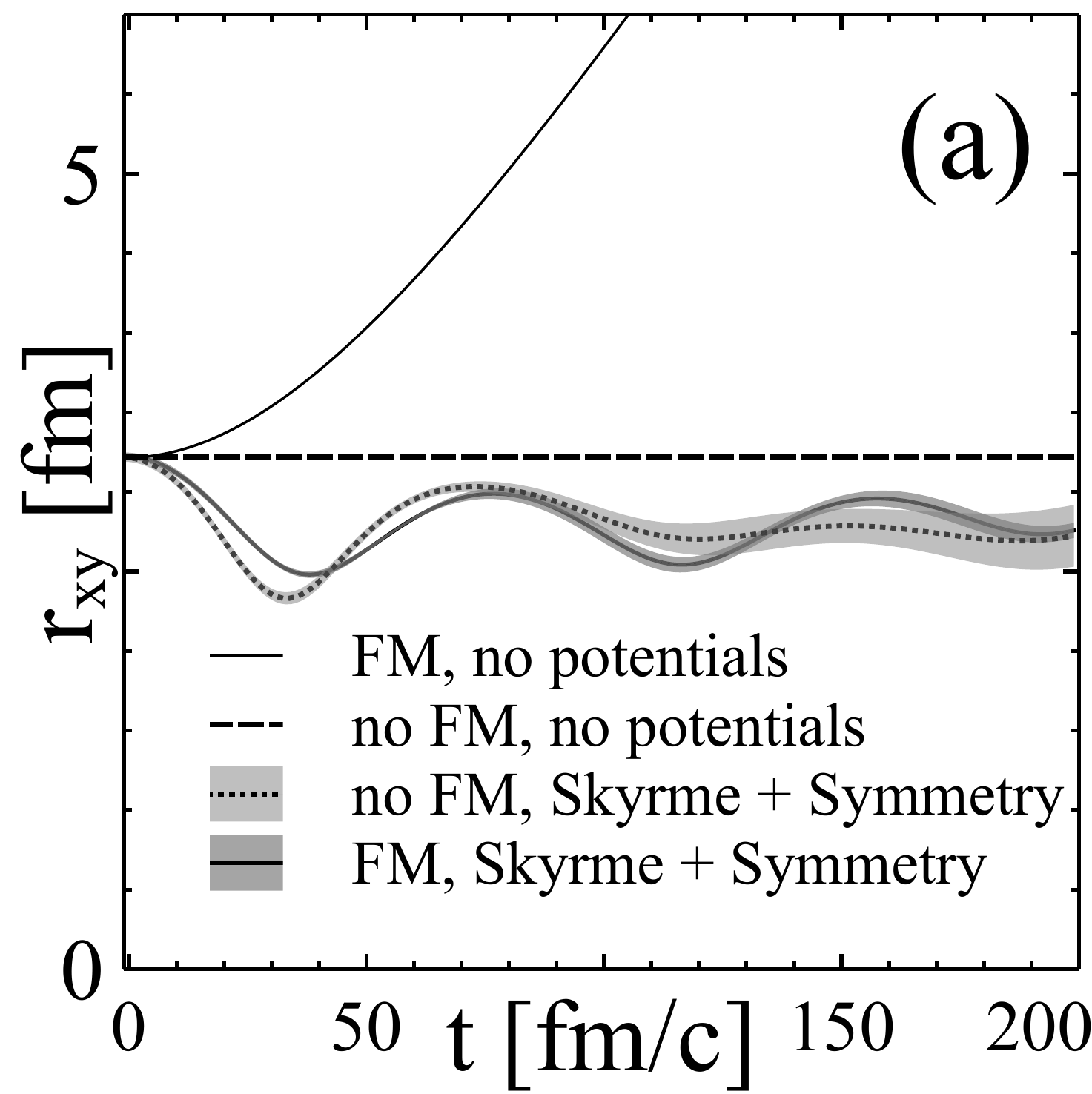}
\includegraphics[width=0.23\textwidth]{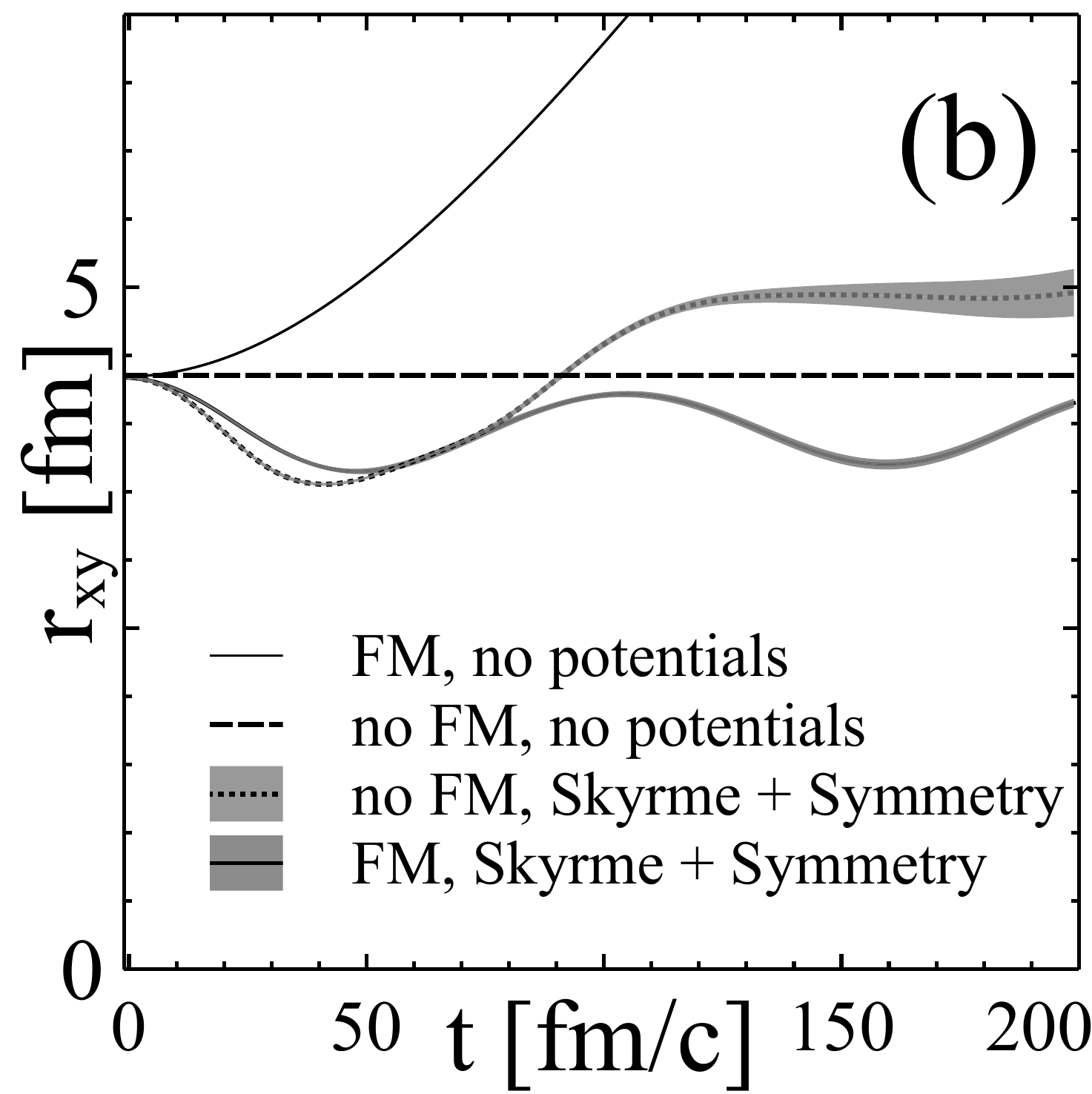}
\caption{Evolution of an average transverse radius $r_{xy} = \sqrt{\langle x^2 + y^2 \rangle}$ of a nucleus over 200 fm/c with different combinations of Fermi motion (FM) and potentials, $^{29}$Cu nucleus (a) and $^{79}$Au nucleus (b).}
\label{fig:stability}
\end{figure}

\cref{fig:stability} shows the nuclear stability over a large time range, much larger than what is actually relevant for a nucleus-nucleus collision. The nucleons fly apart as expected, if only Fermi motion without potentials to stabilize the nucleus are included. With potentials there is the expected oscillatory behavior: The nucleons drift apart due to Fermi motion and the potentials counteract and push them closer together again.

Computations with potentials require that time step is small enough - the energy change per timestep should be much smaller than the energy of the particle:
\begin{equation}
  \frac{\Delta E}{E} \simeq \frac{|\partial U/ \partial r| \Delta t}{E} \ll 1 \,.
\end{equation}
As an estimate for the maximal $|\partial U/ \partial r|_{\rm max}$ let us take two nucleons at the same point and consider $|\partial U/ \partial r|_{\rm max} = 2 m_N/\sigma$, where $\sigma$ is the width of the Gaussian smearing (as defined in \cref{eqn:smearing_kernel})
\begin{equation}
  \Delta t \ll \sigma/2 \,.
\end{equation}
Assuming the default value of $\sigma = 1$ fm, a time-step size of $\Delta t = 0.1$ fm is reasonable for physically relevant cases. Since the  potential becomes smoother with higher number of test particles $N_{\rm test}$, the estimate becomes in this case
\begin{equation}
  \Delta t \ll \sigma \sqrt{N_{\rm test}}/2 \,.
\end{equation}

\paragraph{Deformed Nuclei}

Despite the rather symmetric nuclei that are most often used for heavy-ion collisions, sometimes it is of interest to study deformed nuclei as well. For example, uranium has a prolate shape according to its nuclear many-body wave function. At RHIC U+U collisions at $\sqrt{s_{\rm NN}}=193$ GeV have been studied to evaluate the multiplicities and anisotropic flow as a function of geometry. Especially the case of tip-tip collisions, where the multiplicity is very high but the elliptic flow is close to zero, and the case of body-body collisions, where the elliptic flow is maximal, are of great interest. To differentiate between the different geometries, Monte Carlo event generators that yield the correct trends for the observables are needed that take into account the more involved geometry of deformed nuclei \cite{Goldschmidt:2015kpa}.

In SMASH, the Woods-Saxon distribution is enhanced with an angular dependent radius $r(\theta, \varphi)$
\begin{equation}
    \rho(r, \theta, \varphi) = \frac{\rho_0}{1 + \exp\Big(\frac{r - r(\theta, \varphi)}{d}\Big)}
\end{equation}
The deformation dependent nuclear radius $r(\theta, \varphi)$ can be described using the $\beta$ parameterization \cite{Moller:1993ed}
\begin{equation}
r(\theta ,\varphi ) = r_{ 0 }\left( 1+\sum_{ l=1 }^{ \infty }{ \sum_{ m=-l }^{ l } \beta_{ lm }Y_{ l }^{ m } }\right)
\end{equation}
Here $r_0$ is the initial nuclear radius and the coefficients in front of the spherical harmonics $Y_l^m$ are called the $\beta$ shape parameters (or deformation parameters). Note that our deformed nuclei are azimuthally symmetric and hence all terms with nonzero magnetic quantum number will vanish.

In SMASH, the deformed Woods-Saxon has been implemented by a rejection sampling routine. For the deformation we use the $\beta$ shape parameters up to angular momentum quantum number $l=4$ from \cite{Moller:1993ed}. For the initial nuclear radius $r_0$, we use values from \cite{DeJager:1987qc} (see Two-Parameter Fermi Model, abbreviated 2pF).  We also have a default initial radius that uses the empirical relation $1.2 A^{1/3}$. The diffusiveness parameter $d$ is on average given by 0.54 fm \cite{Chamon:2002mx}. Adjustments for specific nuclei come from \cite{DeJager:1987qc}. We sample a polar angle from the uniform solid angle distribution, and for the radius we set our maximum sampled value to be $r_{\rm max}=r_0/d + r_0 d$. Finally, the saturation density $\rho_0$ represents our normalization condition:
\begin{equation}
2 \pi \int_{0}^{\infty} {\int_{0}^{\pi} {\rho(r,\theta) r^2 dr } d\theta }= 1
\end{equation}

A deformed nucleus is no longer invariant to rotation. We therefore need to rotate the nucleus during the initialization phase. To do so, we treat the system of nucleons like a rigid body. A set of Euler angles is uniformly sampled that determines the rotation of the specific nucleus, for which we use the notation convention $(\varphi, \theta, \psi)$. 

\begin{figure}
\centering
\includegraphics[width=0.48\textwidth]{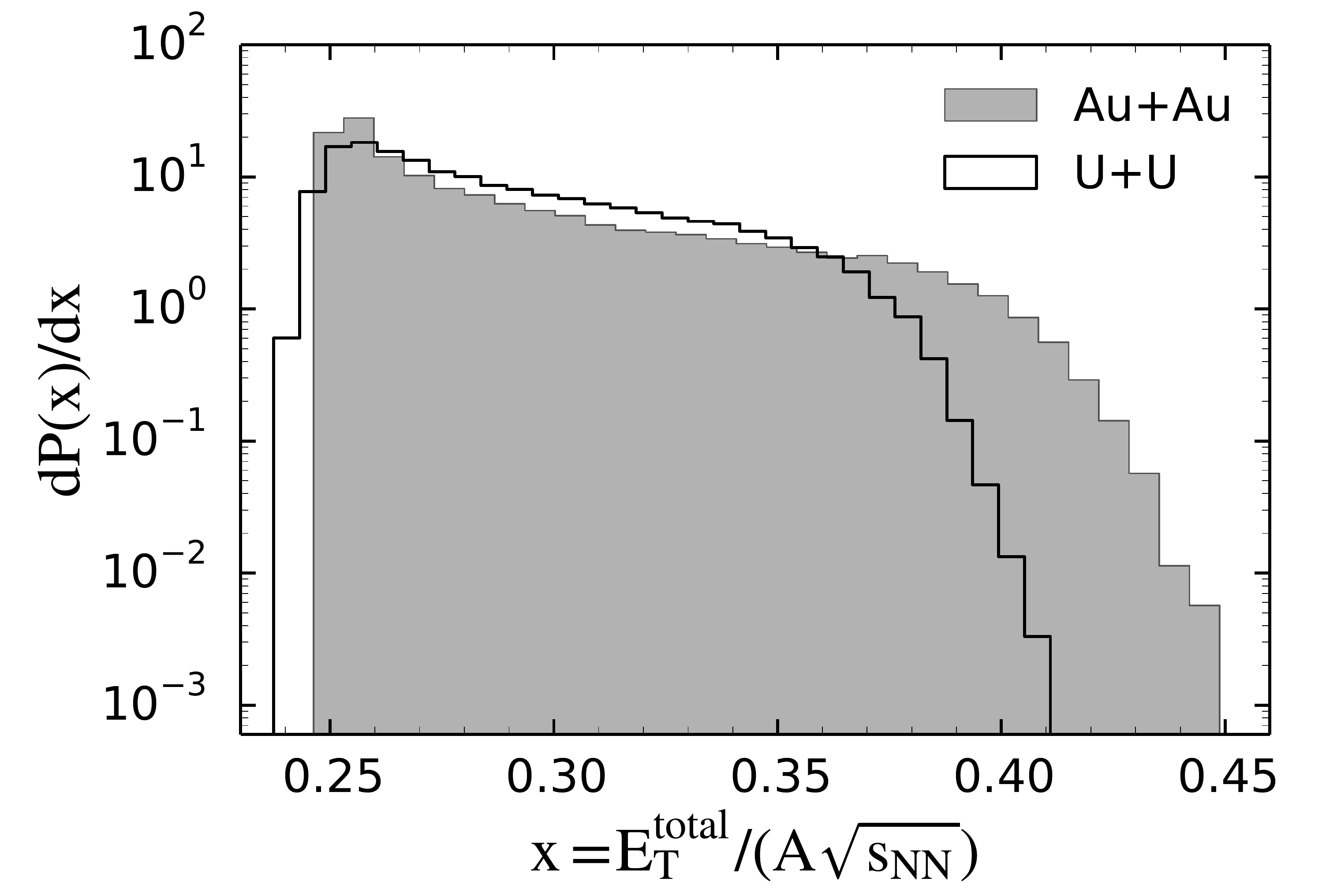}
\caption{Distribution of the ratio of transverse energy of charged particles and the beam energy scaled with the number of nucleons in minimum bias Au+Au and U+U collisions $\sqrt{s_{\rm NN}}=3$ GeV.}
\label{fig:mul_cen_uu}
\end{figure}

To visualize the differences between collisions of symmetric and deformed nuclei in \cref{fig:mul_cen_uu}, the ratio between total transverse energy of charged particles and the beam energy scaled with the number of nucleons (which is used to determine the centrality classes in low energy collisions), is compared for Au+Au and U+U collisions at $\sqrt{s_{\rm NN}}=3$ GeV. It can be seen that there are less high multiplicity events and more intermediate multiplicity events in U+U than Au+Au collisions. The difference comes from the fact that many tip-body and non-overlapped body-body collisions (with impact parameter $b=0$ fm) do not produce as many new particles as in Au+Au most central collisions. Those events are selected as semi-central collisions at experiments but provide much smaller elliptic flow. For most peripheral collisions, there are more non-empty events in U+U than Au+Au collisions, since the uranium can touch each other with much larger impact parameter along their long axes.

\paragraph{Frame Invariance}

To define the kinematics of the heavy-ion collision, different variables are commonly used. At lower beam energies often the kinetic energy per nucleon $E_{\rm kin}$ or the momentum per nucleon $p_{\rm lab}$ is given. Up to moderate beam energies of around 160 $A$GeV per nucleon, most experiments are fixed target experiments to increase the luminosity. Only if the whole available energy out of the accelerated bunches is needed to reach higher energies for the collision, the center-of-velocity frame of the two nuclei equals the laboratory frame in a collider setup. In this case, the center of mass energy for binary nucleon-nucleon collisions is usually specified $\sqrt{s_{\rm NN}}$ to characterize the collision energy. This frame is the standard computational frame for SMASH calculations, which is equal to the center-of-mass frame for symmetric systems.

\begin{figure}
\centering
\includegraphics[width=0.48\textwidth]{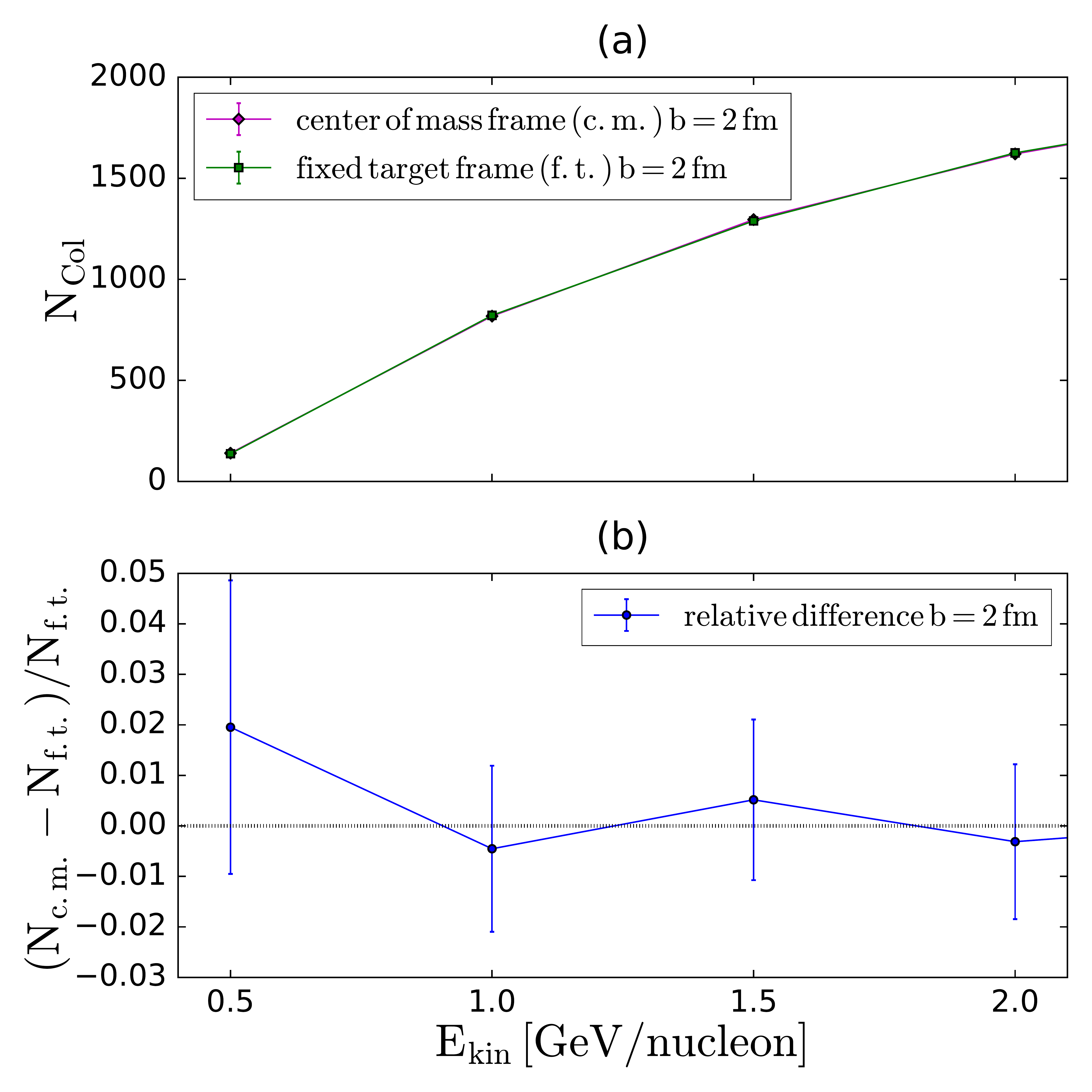}
\caption{Comparison of the number of interactions in the same heavy-ion reaction (60 central Au+Au collisions at different beam energies) calculated in the center-of-mass and the fixed-target frame. The upper plot~(a) shows the absolute numbers of interactions above a cut-off of $\sqrt{s} = 2$ GeV per binary collision, while the lower plot~(b) indicates the relative difference.}
\label{fig:frames}
\end{figure}

To give an estimate on how much the Lorentz invariance is violated by the non-local collision criterion, the number of interactions in one physically identical heavy-ion reaction is counted for calculations in different reference frames (see \cref{fig:frames}). The calculations have been performed as a function of beam energy. All interactions above a cut-off of $\sqrt{s} = 2$ GeV per binary collision are counted until the particles freeze out, therefore no Lorentz transformation of the runtime is necessary. The cut-off is necessary to exclude collisions within the nuclei at low momenta that are not relevant for the actual heavy-ion collision. These calculations do not assume specific time-steps, but all particles are propagated to the next interaction. Apart from the general trend that there are more collisions at higher beam energies, the relative difference between the reference frames is very small at all beam energies. The two calculations coincide within the statistical error bars at all energies.

\subsubsection{Infinite matter calculations}

To simulate infinite hadronic matter or other simple systems like an ideal massless or massive gas and investigate its thermodynamic properties, box calculations are performed. This section describes the initialization of $N$ particles of species $i$ in such a box. In general, every particle $j$ is characterized by coordinates $(x_j, y_j, z_j)$, the four-momentum $(E_j, \vec{p_j})$ and a spectral function, therefore the particle mass is given as $m = \sqrt{E^2 - p^2}$ and is not necessarily equal to its pole mass. The coordinates of the $N$ particles $(x_j, y_j, z_j)$ are sampled uniformly in the box: $x_j = U(0,L)$, $y_j = U(0,L)$, $z_j = U(0,L)$, where $U$ denotes the uniform distribution and $L$ is the length of the box. The momenta of the particles are sampled using the thermal Boltzmann distribution with temperature $T$:
\begin{align}
w(\vec{p}) &= N \exp(-\sqrt{\vec{p}^{\,2} + m^2}/T)\, p^2 dp \sin \theta \, d\theta \, d\varphi \; ,
\end{align}

where $w(p)$ is a probability to generate momentum $\vec{p}$, $\theta$ and $\varphi$ are angles in spherical coordinates and $N$ is a normalization factor. In other words, momentum directions are sampled uniformly in the solid angle $d\Omega = \sin \theta d\theta d\varphi$. Let us denote the total momentum of $N$ particles sampled from this distribution $p_{\rm tot}$. One can see that the ensemble average of $p_{\rm tot}$ is zero,
\begin{equation}
\int \exp(-\sqrt{\vec{p}^{\,2} + m^2}/T)\, d^3p \, p_{x,y,z} = 0
\; ,
\end{equation}
because it involves an integral over an odd function. However, in each single event $p_{\rm tot} \neq 0$, which is corrected by changing the momentum of every particle $p_j \to p_j - p_{\rm tot}/N$. After this procedure the thermal distribution is slightly spoiled, the total energy is changed and angle uniformity is disturbed. This is a small effect for large numbers of particles $N \gg 1$. After letting the system thermalize, the temperature differs by 1-2\% from the initialization temperature. One also has to note that
the total energy is not the same from event to event, it is fluctuating, even without this momentum shift. Fixed are the volume $V$, the number of particles $N$ and the temperature $T$. This picture corresponds to the canonical ensemble (CE) with the temperature and the particle number as independent parameters.

After initialization particles propagate along straight lines with velocities $\vec{v_i} = \vec{p_i}/E_i$ and collide with each other. The simulation is time-step-based and uses a grid to increase the performance of the collision finder as described above in \cref{sec:grid}. The box has per default periodic boundary conditions: at the end of each time step, particles outside of the box with coordinates $\vec{r}$ are returned to the coordinate $\vec{r} \mod (L,L,L)$.

\subsubsection{Expanding sphere}

A simplified scenario including expansion can be initialized using a three-dimensional sphere. For this purpose, $N$ particles of different species $i$ are uniformly distributed in a sphere with radius $R$. The momenta are sampled from a thermal distribution analogously to the box initialization. Then, the system expands freely. This setup provides the opportunity to analyze the numerical stability of the code by comparing to analytic solutions like \cite{Bazow:2015dha}, which is left for future work.

\subsubsection{Afterburner for hydrodynamic simulations}

The fourth method for initializing SMASH is to provide an externally generated particle list based on which the calculation is started. In heavy-ion collisions at high beam energies the hydrodynamic hot and dense stage is followed by a dilute phase that is dominated by hadronic rescattering and resonance decays. To include this late stage dynamically, a hadron transport approach like SMASH needs to be run for each particle configuration that is provided by sampling on the Cooper-Frye hypersurface. The hadronic transport calculation can be coupled in a similar way to other approaches than hydrodynamics, if necessary. Note that input particles in the list are not required to be at the same time $t$ in the computational frame, successive appearance of particles are implemented by setting non-zero formation times. These particles are propagated back to the earliest time in the list and free stream before their formation time.

\subsection{Particle Properties}
\label{sec:particles}

\begin{table}
\caption{All particles implemented in SMASH with their properties and PDG codes (see~\cite{Agashe:2014kda} for the definition). The corresponding antiparticles carry a minus sign and have identical properties.}
\label{tab:particles}
\begin{tabular}{cccc}
\toprule
\bf{Type} & \bf{Mass} & \bf{Width} & \bf{PDG codes} \\
          & [GeV]     & [GeV]      &               \\
\midrule
$\pi$    & 0.138 & 0      &     211, 111, -211 \\
$\rho$   & 0.776 & 0.149  &     213, 113, -213 \\
$\eta$   & 0.548 & 1.3e-6 &     221 \\
$\omega$ & 0.783 & 0.0085 &     223 \\
$\eta'$  & 0.958 & 0.198  &     331 \\
$\phi$   & 1.019 & 0.0043 &     333 \\
$\sigma$ & 0.800 & 0.400  & 9000221 \\
$f_2$    & 1.275 & 0.185  &     225 \\
\midrule
$K$         & 0.494 & 0      & 321, 311 \\
$K^*(892)$  & 0.892 & 0.0508 & 323, 313 \\
$K^*(1410)$ & 1.414 & 0.232  & 100323, 100313 \\
\midrule
$N$       & 0.938 & 0     & 2212, 2112 \\
$N(1440)$ & 1.462 & 0.350 & 202212, 202112 \\
$N(1520)$ & 1.515 & 0.115 & 102214, 102114 \\
$N(1535)$ & 1.535 & 0.150 & 102212, 102112 \\
$N(1650)$ & 1.655 & 0.140 & 122212, 122112 \\
$N(1675)$ & 1.675 & 0.150 & 102216, 102116 \\
$N(1680)$ & 1.685 & 0.130 & 202216, 202116 \\
$N(1700)$ & 1.700 & 0.150 & 112214, 112114 \\
$N(1710)$ & 1.710 & 0.100 & 212212, 212112 \\
$N(1720)$ & 1.720 & 0.250 & 212214, 212114 \\
$N(1875)$ & 1.875 & 0.250 & 9002214, 9002114 \\
$N(1900)$ & 1.900 & 0.200 & 9012214, 9012114 \\
$N(1990)$ & 1.990 & 0.500 & 9002218, 9002118 \\
$N(2080)$ & 2.000 & 0.350 & 9022214, 9022114 \\
$N(2190)$ & 2.150 & 0.500 & 9012218, 9012118 \\
$N(2220)$ & 2.220 & 0.400 & 9022218, 9022118 \\
$N(2250)$ & 2.250 & 0.470 & 9032218, 9032118 \\
\midrule
$\Delta$       & 1.232 & 0.117 & 2224, 2214, 2114, 1114 \\
$\Delta(1620)$ & 1.630 & 0.140 & 112222, 112212, 112112, 111112 \\
$\Delta(1700)$ & 1.700 & 0.300 & 122224, 122214, 122114, 121114 \\
$\Delta(1905)$ & 1.880 & 0.330 & 212226, 212216, 212116, 211116 \\
$\Delta(1910)$ & 1.890 & 0.280 & 222222, 222212, 222112, 221112 \\
$\Delta(1920)$ & 1.920 & 0.260 & 222224, 222214, 222114, 221114 \\
$\Delta(1930)$ & 1.950 & 0.350 & 9002226,9002216,9002116,9001116 \\
$\Delta(1950)$ & 1.930 & 0.285 & 202228, 202218, 202118, 201118 \\
\midrule
$\Lambda$       & 1.116 & 0      & 3122\\
$\Lambda(1405)$ & 1.405 & 0.0505 & 13122\\
$\Lambda(1520)$ & 1.520 & 0.0156 & 3124\\
$\Lambda(1670)$ & 1.670 & 0.0350 & 33122\\
$\Lambda(1690)$ & 1.690 & 0.0600 & 13124\\
$\Lambda(1820)$ & 1.820 & 0.0800 & 3126\\
$\Lambda(1830)$ & 1.830 & 0.0950 & 13126\\
$\Lambda(1890)$ & 1.890 & 0.1000 & 23124\\
\midrule
$\Sigma$       & 1.189 & 0     & 3222, 3212, 3112 \\
$\Sigma(1385)$ & 1.385 & 0.036 & 3224, 3214, 3114 \\
$\Sigma(1670)$ & 1.670 & 0.060 & 13224, 13214, 13114 \\
$\Sigma(1775)$ & 1.775 & 0.120 & 3226, 3216, 3116 \\
$\Sigma(1915)$ & 1.915 & 0.120 & 13226, 13216, 13116 \\
\midrule
$\Xi$       & 1.321 & 0     & 3322, 3312 \\
$\Xi(1530)$ & 1.532 & 0.009 & 3324, 3314 \\
\midrule
$\Omega$ & 1.672 & 0 & 3334 \\
\midrule
$e$      & 0.000511 & 0 & 11, -11 \\
$\mu$    & 0.105    & 0 & 13, -13 \\
$\gamma$ & 0        & 0 & 22 \\
\bottomrule
\end{tabular}
\end{table}

\subsubsection{Particle species}

We implement the most well-established hadronic states from the Review of Particle Properties \cite{Agashe:2014kda} with their corresponding decays and cross sections as detailed below. These particles and their properties are summarized in \cref{tab:particles}.

To simplify the extrapolation of cross sections and particle properties, full isospin symmetry is assumed. Therefore, small differences in the masses between isospin partners have been neglected. However, it should be noted that the cross sections for certain processes can indeed depend on isospin (thus breaking isospin symmetry, e.g. in channels like $NN\rightarrow NN^*$, see \cref{sec:coll_term}).

We treat all particles as stable which have a width below 10 keV (such as the $\pi$, $\eta$, $K$, N, $\Lambda$, $\Sigma$, $\Xi$, $\Omega$). All unstable particles (``resonances'') are assumed to have a Breit-Wigner shape. We note that this approximation is known to be questionable for the $\sigma$ meson, our parameters are adjusted to reproduce the $\pi \pi$ elastic cross section. 

\subsubsection{Spectral functions}

In general, the spectral function encodes the dispersion relation for a particle and can depend on the temperature and the density of the system. Medium modifications are currently neglected in SMASH and all spectral functions are described by relativistic Breit-Wigner distributions:

\begin{equation}
\mathcal{A}(m) = \frac{2\mathcal{N}}{\pi} \frac{m^2 \Gamma(m)}{(m^2 - M_0^2)^2 + m^2 \Gamma(m)^2}
\end{equation}

Here $m$ is the actual off-shell mass of the resonance and $M_0$ is the pole mass (i.e. a constant given in \cref{tab:particles}). However, the total width $\Gamma$ is not constant, but given by the mass-dependent width function $\Gamma(m)$. Each resonance has a minimum mass $m_\text{min}$ (corresponding to the sum of masses of the lightest decay channels), below which the width, and thus also the spectral function, vanishes. The total width is computed as the sum of all partial widths:
\begin{equation}
\Gamma(m)=\sum_i \Gamma_i(m)
\end{equation}

Note that the width given in \cref{tab:particles} is the total on-shell width, i.e. $\Gamma_0 = \Gamma(M_0)$.
The spectral function in relativistic Breit-Wigner form is normalized to one, when integrated from zero to infinity:

\begin{equation}
\int\displaylimits_0^\infty \mathcal{A}(m)dm = \int\displaylimits_{m_{\rm min}}^\infty \mathcal{A}(m)dm = 1
\end{equation}

In practice the integration can start from $m_{\rm min}$, since the spectral function vanishes below that value. Under the assumption of a constant width $\Gamma$, the normalization factor is exactly $\mathcal{N}=1$. As soon as the width becomes mass-dependent (as it is the case in SMASH), the normalization factor $\mathcal{N}$ can deviate from one and needs to be determined numerically. Practically all the normalization constants in SMASH are still rather close to one (within 25\%).

\subsubsection{Decay widths}
\label{sec:widths}

All the decay widths in SMASH are currently calculated following the treatment of Manley et al.~\cite{Manley:1992yb}, where in general the width of a two-body decay $R\rightarrow ab$ is written as

\begin{equation}
\Gamma_{R\rightarrow ab} = \Gamma^0_{R\rightarrow ab} \frac{\rho_{ab}(m)}{\rho_{ab}(M_0)}.
\end{equation}

Here $m$ is the actual off-shell mass of the resonance R, $M_0$ is its pole mass, $\Gamma^0_{R\rightarrow ab}=\Gamma_{R\rightarrow ab}(M_0)$ is the partial width at the pole mass and the function $\rho_{ab}$ is defined as
\begin{align}
\rho_{ab}(m) = \int & dm_a dm_b \mathcal{A}_a(m_a)\mathcal{A}_b(m_b) \nonumber \\
             \times & \frac{|\vec{p}_f|}{m} B_L^2(|\vec{p}_f|R) \mathcal{F}_{ab}^2(m).
\end{align}

In this formula, $m_a$ and $m_b$ denote the (off-shell) masses of the particles a and b (which are being integrated over), $\mathcal{A}_a$ and $\mathcal{A}_b$ are their spectral functions and $|\vec{p}_f|$ is the absolute value of the final-state momentum of a and b in the center-of-mass frame, which is given by:
\begin{align}
\vec{p}_f^{\,2} &= \vec{p}_{\rm cm}^{\,2}(m,m_a,m_b) \nonumber\\
                &= \frac{(m^2-(m_a+m_b)^2)(m^2-(m_a-m_b)^2)}{4m^2}
\end{align}
Finally, L is the orbital angular momentum of a and b in the final state and $B_L$ are the so-called 'Blatt-Weisskopf functions' \cite{BlaWei}. The parameter $R$ is usually called the 'interaction radius' and is assumed to have a universal value of $R=1$ fm for all processes. The form factor $\mathcal{F}_{ab}$ is only relevant for unstable decay products and will be discussed later.

The simplest case is that of a resonance R decaying into two stable daughter particles. Popular examples are $\Delta\rightarrow\pi N$ or $\rho\rightarrow\pi\pi$. In this case, the daughters have fixed masses (i.e. their spectral functions are just $\delta$ functions), so that the integrals collapse:

\begin{equation}
\rho_{ab}(m) = \frac{|\vec{p}_f|}{m} B_L^2(|\vec{p}_f|R)
\end{equation}

As an example, the width for the p-wave (L=1) decays of the $\rho$ and $\Delta$ (mentioned above) becomes
\begin{equation}
\Gamma(m) = \Gamma_0 \frac{M_0}{m} \left|\frac{\vec{p}_f}{\vec{p}_{f,0}}\right|^3 \frac{\vec{p}_{f,0}^{\,2}+\Lambda^2}{\vec{p}_f^{\,2}+\Lambda^2},
\end{equation}
using $B_1^2(x)=x^2/(1+x^2)$. Here $m$ and $M_0$ are the off-shell and pole mass, respectively, while $\vec{p}_f$ and $\vec{p}_{f,0}$ denote the final-state momenta in the center-of-mass frame for mass $m$ and $M_0$, respectively. $\Lambda=1/R$ can be viewed as a cut-off parameter. For an s-wave (L=0) decay like $\sigma\rightarrow\pi\pi$, the width simply becomes
\begin{equation}
\Gamma(m) = \Gamma_0 \frac{M_0}{m} \left|\frac{\vec{p}_f}{\vec{p}_{f,0}}\right|,
\end{equation}
since $B_0^2=1$.

In the case that one of the daughter particles is itself a resonance, the width calculation becomes more difficult, since the mass of this daughter resonance is not fixed and needs to be integrated over. Examples for this case are $N^*(1440)\rightarrow\pi\Delta$ or $\omega\rightarrow\pi\rho$. As one of the daughters is stable, at least one of the two integrals collapses:

\begin{equation}
 \label{eq:decay_semistable}
\rho_{ab}(m) = \int\displaylimits_{m_a^{\rm min}}^{m-m_b} dm_a \mathcal{A}_a(m_a) \frac{|\vec{p}_f|}{m} B_L^2(|\vec{p}_f|R) \mathcal{F}_{ab}^2(m)
\end{equation}

The remaining integral runs from the minimum allowed mass of particle a (i.e. the threshold of its lightest decay channel) up to the maximum possible mass of a in the decay process (given by $m-m_b$). The form factor $\mathcal{F}_{ab}$ (by M. Post \cite{Post:2003hu}) is used only if unstable decay products are involved and is defined as

\begin{equation}
\mathcal{F}_{ab}(m) = \frac{\lambda^4+1/4(s_0-M_0^2)^2}{\lambda^4+\big(m^2-1/2(s_0+M_0^2)\big)^2},
\end{equation}
where the cut-off factors given in \cref{tab:cut-off} are used.

\begin{table}
\caption{Cut-off parameter $\lambda$ for form factor in resonance decay widths.}
\label{tab:cut-off}
\begin{tabular}{lc}
\toprule
 decay & $\lambda$ [GeV] \\
\midrule
 $\pi\rho$                                   & 0.8 \\
 unstable mesons (e.g. $\rho N$, $\sigma N$) & 1.6 \\
 unstable baryons (e.g. $\pi\Delta$)         & 2.0 \\
 two unstable daughters (e.g. $\rho\rho$)    & 0.6 \\
\bottomrule
\end{tabular}
\end{table}

It is easy to see that $\mathcal{F}_{ab}(M_0)=\mathcal{F}_{ab}(\sqrt{s_0})=1$. Note that this form factor was not used by Manley originally, but was added only later in the GiBUU implementation. The effect of the form factor is that it suppresses the high-mass tail ($m>M_0$) and slightly enhances the low-mass tail ($m<M_0$). Both of these effects get stronger with decreasing $\lambda$ ($\mathcal{F}_{ab}\rightarrow 1$ for $\lambda\rightarrow\infty$). We have decided to follow the GiBUU framework for the width parametrization of resonances, since it has been proven to give a good description of experimental data \cite{Buss:2011mx}.

All the formulas described above are for the case of resonance decays. For the inverse process, i.e. resonance formation via $ab\rightarrow R$, the Breit-Wigner cross section involves the so-called 'in-width' $\Gamma_{ab\rightarrow R}$. For stable particles a and b it is identical to the 'out-width' $\Gamma_{R\rightarrow ab}$. However, the two differ if a or b are unstable. In the Manley formalism, the in-width for unstable particles becomes
\begin{equation}
\label{eq:in-width}
\Gamma_{ab\rightarrow R}(m) = \Gamma_{R\rightarrow ab}^0 \frac{|\vec{p}_{ab}|B_L^2(|\vec{p}_{ab}|R)\mathcal{F}_{ab}(m)}{m\rho_{ab}(M_0)},
\end{equation}
where~$m$ is the off-shell mass of the produced resonance~$R$ (i.e. the $\sqrt{s}$ in the process) and $\vec{p}_{ab}=\vec{p}_{cm}(m,m_a,m_b)$ is the momentum of~$a$ and~$b$ in the center-of-mass frame. The difference between the in- and the out-width is essentially due to the fact that for the out-width one integrates over the mass of the unstable particle, while in the in-width this mass is fixed.

\begin{figure}
\centering
\includegraphics[width=0.48\textwidth]{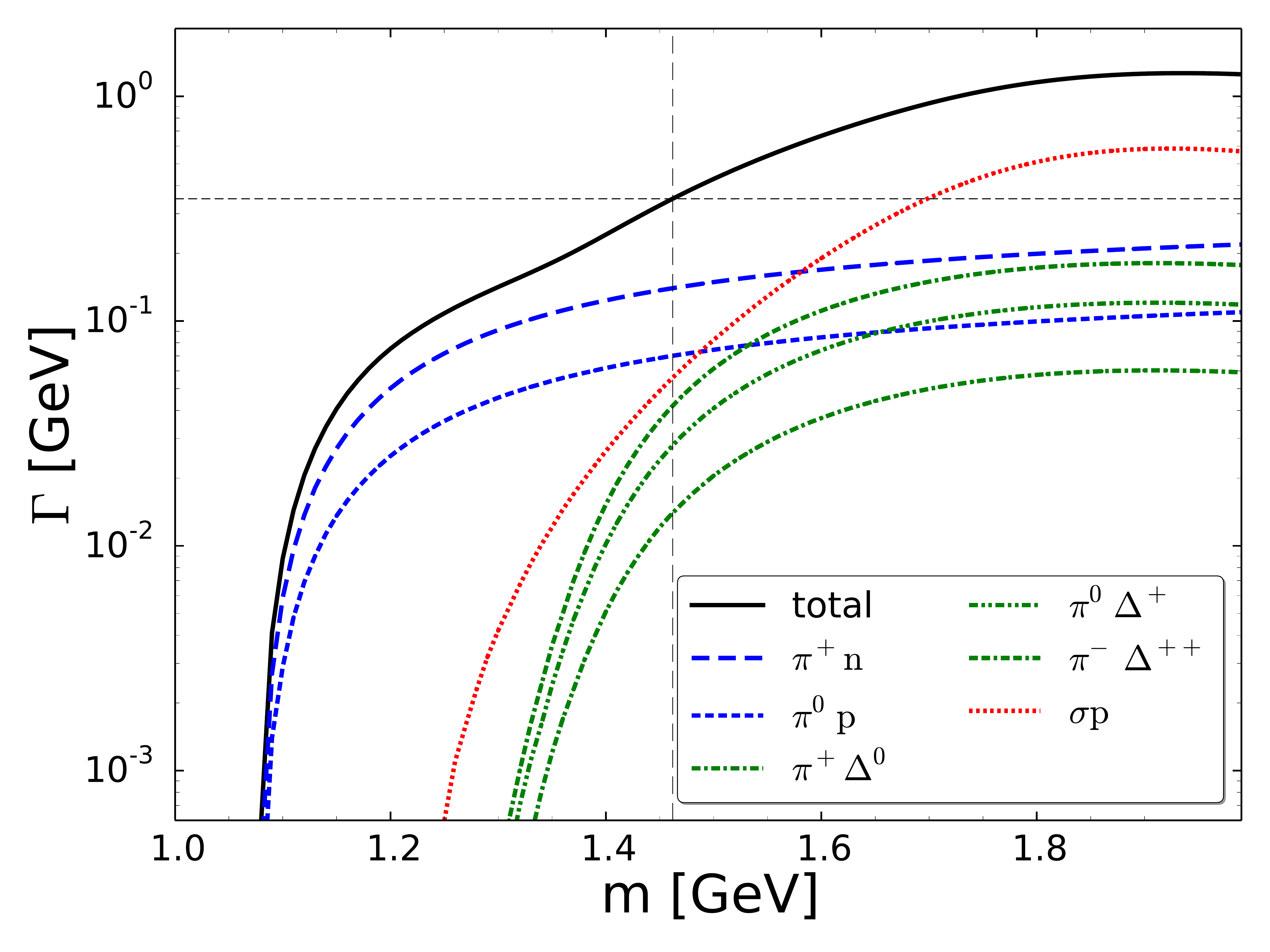}
\caption{Total and partial decay widths of the $N^*(1440)^+$ resonance as a function of mass. The vertical and horizontal dashed lines mark the pole mass and width.}
\label{fig:width}
\end{figure}

In \cref{fig:width} the theoretical decay width of the $N^*(1440)$ resonance is shown as a function of mass. The total width is given as the sum of all partial widths. Each partial width has a threshold that is given by the sum of the minimal masses of the decay products. The branching ratios are fixed at the pole mass. One can see that all partial widths increase as a function of mass, since more phase space is available for heavier resonances. The lifetime correspondingly has an opposite trend and heavy particles decay faster than low-mass resonances. Since the width also enters in the production cross section (\cref{eq:sigma21}), the production of such low-mass resonances becomes more unlikely.

\subsection{Collision Term}
\label{sec:coll_term}

The collision term includes all different processes (decays and collisions) that can happen to particles within this hadronic transport approach. At this point, unstable particles can decay, 2 particles scatter in-/elastically or excite a resonance. 
Weak decays are neglected since they have significantly longer lifetimes than processes associated with the strong interaction. Electromagnetic processes are treated perturbatively and will be discussed in detail in a forthcoming publication \cite{dileptons_to_be}.

We note that the current implementation is limited to the energy regime of a few GeV, where all hadronic cross sections are expected to be dominated by the excitation and decay of resonances. Since the model at present is lacking a string fragmentation mechanism, the cross sections are not sufficient at higher energies. In the following, a detailed description of all implemented processes is given.

\subsubsection{Decays}

The lifetime of a resonance is defined as $\tau = 1/\Gamma(m)$, where $\Gamma(m)$ is the mass-dependent total decay width. The probability to decay in a sufficiently small time interval $\Delta t$ is
\begin{equation}
P(\text{decay at }\Delta t) = \frac{\Delta t}{\tau} = \Gamma(m) \Delta t
\end{equation}
This leads to exponential decay, as the survival probability after $n$ time steps is

\begin{align*}
P(\text{alive after n steps}) &= (1 - \Gamma(m) \Delta t)^n \\
                              &= (1 - \Gamma(m) \Delta t)^{t / \Delta t}\\
                              &\rightarrow \exp(-\Gamma(m) t)
\end{align*}
when $\Delta t \rightarrow 0$. As noted above, the total width $\Gamma(m)$ is computed as the sum of all partial widths.

When a resonance decays in SMASH, the decay channel is randomly chosen from the list of allowed channels for this particle, based on the off-shell branching ratios $\Gamma_i(m)/\Gamma(m)$.
The decay channels and their on-shell ratios $\Gamma_i(M_0)/\Gamma(M_0)$ are listed in an input file and can be turned on and off separately.

\subsubsection{\texorpdfstring{2 $\rightarrow$ 1 processes}{2 -> 1 processes}}

The cross section formula for 2 $\rightarrow$ 1 resonance production is based on Eq. (176) in \cite{Buss:2011mx}:

\begin{equation} \label{eq:sigma21}
\sigma_{ab\rightarrow R}(s) = \frac{2J_R+1}{(2J_a+1)(2J_b+1)}\mathcal{S}_{ab} \frac{2\pi^2}{\vec{p}_i^{\,2}} \Gamma_{ab\rightarrow R}(s) \mathcal{A}_R(\sqrt{s})
\end{equation}
where:
\begin{itemize}
\item $J$ is the spin of the particle
\item $\Gamma_{ab\rightarrow R}$ is the partial in-width for the process
\item $\mathcal{A}_R$ is the spectral function of the resonance
\item $\mathcal{S}_{ab}$ is a symmetry factor, which is 2 if $a$ and $b$ are identical, and 1 otherwise
\item $\vec{p}_i=\vec{p}_{cm}(\sqrt{s},m_a,m_b)$ is the center-of-mass momentum of the initial state
\end{itemize}
Note that in the above, $\Gamma_{ab\rightarrow R}(s)$ refers to the isospin-specific channel instead of the isospin-generic channel. Hence there is no need for isospin factors in the cross section formula.

The so-called ``in-width'' $\Gamma_{ab\rightarrow R}$ simply equals the usual decay width $\Gamma_{R\rightarrow ab}$ for the case of stable particles $a$ and $b$, see \cref{sec:widths}. For unstable particles however, it is given by \cref{eq:in-width}, which differs from the decay width.

\subsubsection{Elastic collisions}

There are different cases of elastic collisions in SMASH. For the meson-baryon and meson-meson collisions, one assumes that the elastic cross sections are fully determined by resonance excitation and decay, e.g. $\pi N \to \Delta \to \pi N$ or $\pi\pi \to \rho \to \pi\pi$. For baryon-baryon collisions on the other hand, one typically uses parametrized cross sections. The parametrizations of the elastic pp and pn cross sections in particular are taken from \cite{Weil:2013mya}, eq. (44) and (45).

\subsubsection{\texorpdfstring{2 $\rightarrow$ 2 processes with one resonance in final state}{2 -> 2 processes with one resonance in final state}}
\label{sec:2_to_2}

When there is another particle in the final state, the resonance mass must be integrated over the allowed range:
\begin{align} \label{eq:sigma22_1res}
\sigma_{ab\rightarrow Rc}(s) &= \frac{(2J_R+1)(2J_c+1)}{s|\vec{p}_i|} \nonumber\\
                             &\times \sum_I \left(C_{ab}^IC_{Rc}^I\right)^2 \frac{|\mathcal{M}|^2_{ab\leftrightarrow Rc}(s,I)}{16\pi} \nonumber \\
                             &\times \int\displaylimits_{m_R^\textrm{min}}^{\sqrt{s}-m_c}dm\, \mathcal{A}_R(m) \, |\vec{p}_f|(\sqrt{s},m,m_c),
\end{align}
where $\vec{p}_i$ and $\vec{p}_f$ are the center-of-mass momenta of the initial and the final state and $\mathcal{A}_R$ is the spectral function of the resonance R. The symbol $C$ refers to isospin Clebsch-Gordan factors, which couple the initial and final state to a total isospin $I$.
Here it is assumed that the matrix element $|\mathcal{M}|^2$ is a constant (or only depends on $s$) without angular dependence, resulting in the factor $4\pi$ from the trivial angle integration. If the matrix element does depend on the angle, the factor $4\pi$ must be replaced with the proper integration of $|\mathcal{M}|^2$ over the phase space. The lower mass limit for the resonance, $m_R^{\rm min}$, is defined as the sum of the particle masses in the lightest decay channel. This is the lowest mass the resonance can have and still be able to decay into one of the implemented channels.

For the process $NN\rightarrow N\Delta$, the parametrized energy dependence
\begin{equation}
 \frac{|\mathcal{M}|^2(s)}{16\pi}=\frac{A}{(\sqrt{s} - b)^c}
\end{equation}
(with the parameters $A=68$, $b=1.104$ GeV and $c=1.951$) is based on a fit to the Dmitriev one-boson-exchange (OBE) model \cite{Dmitriev:1986st}. For other resonance production processes (i.e. $NN\rightarrow NR$ and $NN\rightarrow \Delta R$, with $R=N^*,\Delta^*$), the matrix element is assumed to be a constant (independent of $s$), but can depend on the total isospin and the pole masses $m_a$ and $m_b$ of the outgoing particles. It is parametrized as
\begin{equation} \label{eq:matrix}
 \frac{|\mathcal{M}|^2}{16\pi}=\frac{A_I}{2(m_a^2+m_b^2)}
\end{equation}
with parameters $A_I$ as given in \cref{tab:matrix}.

\begin{table}
\caption{Parameters for matrix elements in baryonic $2\to2$ processes (in units of $\mathrm{mbGeV}^4$).}
\label{tab:matrix}
\begin{tabular}{lcc}
\toprule
 process & $A_{I=1}$ & $A_{I=0}$ \\
\midrule
 $NN\to NN^*$          &  7 &  14 \\
 $NN\to N\Delta^*$     & 15 &   - \\
 $NN\to\Delta\Delta$   & 45 & 120 \\
 $NN\to\Delta N^*$     &  7 &   - \\
 $NN\to\Delta\Delta^*$ & 15 &  25 \\
\bottomrule
\end{tabular}
\end{table}

\subsubsection{\texorpdfstring{2 $\rightarrow$ 2 processes with two resonances in final state}{2 -> 2 processes with two resonances in final state}}

Analogously to \cref{eq:sigma22_1res}, one can write down the cross section for a process with two resonances in the final state. In this case both their masses must be integrated over the allowed range:

\begin{align} \label{eq:sigma22_2res}
\sigma_{ab\rightarrow R_1R_2}(s) &= \frac{(2J_{R_1}+1)(2J_{R_2}+1)}{s|\vec{p}_i|}  \\
&\times \sum_I \left(C_{ab}^IC_{R_1R_2}^I\right)^2 \frac{|\mathcal{M}|^2_{ab\leftrightarrow R_1R_2}(s,I)}{16\pi} \nonumber \\
&\times \int\displaylimits_{m_1^\textrm{min}}^{\sqrt{s}-m_2^\textrm{min}} dm_1\, \mathcal{A}_1(m_1) \nonumber \\
&\times \int\displaylimits_{m_2^\textrm{min}}^{\sqrt{s}-m_1^\textrm{min}} dm_2\, \mathcal{A}_2(m_2) \, |\vec{p}_f|(\sqrt{s},m_1,m_2). \nonumber
\end{align}
The double-resonance production processes that are currently implemented in SMASH are $NN\rightarrow\Delta\Delta$, $\Delta N^*$ and $\Delta\Delta^*$. The matrix elements are parametrized in the same way as for single-resonance production, see \cref{eq:matrix,tab:matrix}.

\subsubsection{Detailed balance}

The cross sections for the inverse resonance-absorption processes are derived from the production cross section by imposing the principle of detailed balance (see Eqs. (B.6), (B.9) and (181) in \cite{Buss:2011mx}):
\begin{align}
\sigma_{cd\rightarrow ab} (s) &= (2J_a+1)(2J_b+1) \frac{\mathcal{S}_{cd}}{\mathcal{S}_{ab}} \left|\frac{\vec{p}_f}{\vec{p}_i}\right| \frac{1}{s} \nonumber \\
                              &\times \sum_I \left(C_{ab}^IC_{cd}^I\right)^2 \frac{|\mathcal{M}|^2_{ab\leftrightarrow cd}(s,I)}{16\pi}
\end{align}
This equation holds for both single- and double-resonance absorption, i.e.~$c$ and $d$ can be either two resonances or a resonance and a stable particle.
The symmetry factors $\mathcal{S}_{xy}$ here are defined such that they are 2 if $x$ and $y$ are in the same isospin multiplet and 1 otherwise.
In SMASH all processes are following explicit detailed balance in the whole phase space, as will be demonstrated in \cref{sec:detailed_balance} below.

\subsubsection{Mass sampling}

In any process where a resonance is produced in the final state, its mass needs to be sampled according to the spectral function and the available phase space. The simplest case is that a single resonance is produced in a $2\to2$ collision together with a stable particle. Then the mass of the resonance is sampled from the integrand of \cref{eq:sigma22_1res}:

\begin{equation}
F(m) = \mathcal{A}(m) \, |\vec{p}_f|(\sqrt{s},m,m_{\rm stable})
\end{equation}

The allowed mass range is from $m_R^{\rm min}$ to $\sqrt{s}-m_{\rm stable}$, where $s$ is the Mandelstam s of the process and $m_{\rm stable}$ is the mass of the stable final-state particle.

The mass sampling is slightly more complicated for the case of a resonance decay ($1\to2$) with one resonance and one stable particle in the final state. In this case an additional Blatt-Weisskopf factor appears, which takes into account the angular momentum in the decay, cf.~\cref{eq:decay_semistable}:

\begin{equation}
F(m) = \mathcal{A}(m) \, |\vec{p}_f|(\sqrt{s},m,m_{\rm stable}) \, B_L^2(|\vec{p}_f|R)
\end{equation}

For a scattering process with two resonances in the final state, the masses of both resonances have to be chosen according to the function

\begin{equation}
F(m_1,m_2) = \mathcal{A}_1(m_1) \, \mathcal{A}_2(m_2) \, |\vec{p}_f|(\sqrt{s},m_1,m_2),
\end{equation}

which is the integrand of \cref{eq:sigma22_2res}. It is important to note that both masses cannot be determined independently, but have to be chosen simultaneously according to a common sampling function.

Analogously to the single-resonance case, a decay into two resonances also includes an additional Blatt-Weisskopf factor:

\begin{align}
F(m_1,m_2) = {} & \mathcal{A}_1(m_1) \, \mathcal{A}_2(m_2) \nonumber \\
                & \, |\vec{p}_f|(\sqrt{s},m_1,m_2)  \, B_L^2(|\vec{p}_f|R)
\end{align}

Drawing random numbers from these distribution functions is numerically non-trivial. We first draw from a Cauchy distribution which approximates the spectral function and handle the remaining factors by rejection sampling (where the unknown maximum value is determined adaptively).

\subsubsection{Angular distributions}
\label{sec:angular}

We currently have anisotropic angular distributions implemented for $NN\rightarrow NN$, $NN\rightarrow N\Delta$ and $NN\rightarrow NR$ (with $R=N^*,\Delta^*$). For elastic nucleon-nucleon collisions we follow the prescription by Cugnon et al.~\cite{Cugnon:1996kh}, using an exponential ansatz $d\sigma/dt\propto e^{-bt}$, with an energy-dependent parameter b which is fit to data. In the second case we also follow Cugnon et al.~\cite{Cugnon:1996kh}, using the same ansatz as for elastic NN collisions. For the last case of $NN\rightarrow NR$ we use the ansatz $d\sigma/dt\propto t^{-a}$, with parameters $a$ which have been fitted to HADES data \cite{Agakishiev:2014wqa}. In \cref{sec:angular_dist} a comparison to elementary data is shown. We note that in the present implementation all resonances decay isotropically in SMASH.

\subsubsection{Pauli blocking}
\label{sec:pauli_blocking}

Pauli blocking is an effective way to obtain the solution of the quantum BUU (Boltzmann-Uehling-Uhlenbeck) equation from classical molecular dynamics.
To understand the way this is achieved one has to compare the classical Boltzmann equation
\begin{align}
p^{\mu}\frac{\partial f}{\partial x^{\mu}} &= \frac{1}{2} \int \frac{d^3p_2}{E_2} \frac{d^3p'_1}{E_1} \frac{d^3p'_2}{E'_2}
          \times W(p_1,p_2 \to p'_1, p'_2) \nonumber\\
          &\times (f'_1 f'_2 - f f_2)
\end{align}
and the BUU equation - its quantum analog:
\begin{align}
p^{\mu}\frac{\partial f}{\partial x^{\mu}} &= \frac{1}{2} \int \frac{d^3p_2}{E_2} \frac{d^3p'_1}{E_1} \frac{d^3p'_2}{E'_2}
          \times W(p_1,p_2 \to p'_1, p'_2) \nonumber\\
          &\times (f'_1 f'_2(1 \pm f)(1 \pm f_2) - f f_2 (1 \pm f'_1) (1 \pm f'_2))
\end{align}
Here the plus sign is for bosons and the minus sign for fermions. One can see that quantum BUU equation differs from classical Boltzmann only in the Uehling-Uhlenbeck factors in the collision term. One can interpret this factors as a multiplication of the cross sections by $\prod_i (1 \pm f_i)$, where the product is taken over all final states in the reaction and $f_i \equiv f(r_i,p_i,t)$ is the phase-space density of final-state particle $i$. This means that for bosons cross sections are effectively increased and for fermions cross sections are effectively decreased. This is called Bose enhancement and Pauli-blocking respectively. While Bose enhancement has been attempted to implement recently in a parton cascade \cite{Xu:2014ega}, Pauli blocking is taken into account in many transport approaches. Since Pauli blocking is important in the energy range under consideration in this work, we describe in the following how it is taken into account in a Monte-Carlo model.

The implementation of Pauli blocking consists of two parts: the calculation of the phase-space density and the rejection of reactions with probability $1 - \prod_i (1-f_i)$. For the latter SMASH loops over all baryons in the final state after a collision has taken place and returns 'true' for blocking, if a uniformly distributed random number $r > f_i$. This means that the reaction is not blocked with probability $\prod_i (1-f_i)$. In this way, no fermion can be produced or scatter into a phase space bin that is already occupied by another fermion. 

The implementation of the phase space density calculation basically follows the method used in the GiBUU model, see section D.4.3 in \cite{Buss:2011mx}. By definition $N(\Delta V_r, \Delta V_p) = g f(r,p) \Delta V_r  \Delta V_p$, where $N$ is the number of (test)particles in a given phase-space volume $\Delta V_r \Delta V_p$ and $g$ is the degeneracy. Theoretically, the size of the phase-space goes to zero $\Delta V_r, \Delta V_p \to 0$. In practice $\Delta V_r$, $\Delta V_p$ and the way of averaging are chosen to balance between the smoothness of the obtained distribution function and the resolution of coordinate and momentum space. This implementation relies on a large number of test particles ($N_{\rm test} \gtrsim 20$).

The phase-space density is calculated according to the following equations:
\begin{align}
f_i (r_j,p) &= \sum_{j: p_j \in V_p} \frac{1}{\kappa (2 \pi \sigma^2)^{3/2}} \int_{\Delta V_r, |r - r_j| < r_c} d^3r \, \nonumber\\
            &\times \exp \left( - \frac{(r-r_j)^2}{2 \sigma^2} \right)
\end{align}
with $\kappa$ given as
\begin{align}
\kappa &= \frac{2 \Delta V_r \Delta V_p N}{(2 \pi)^3} \frac{4 \pi}{(2 \pi \sigma^2)^{3/2}} \int_0^{r_c} dr \, \nonumber\\
       &\times r^2 \exp \left(- \frac{r^2}{2 \sigma^2}\right)
\end{align}

Here $\vec{r}_j$ is a vector connecting the point, where $f$ is calculated, and the position of the $j$-th particle. All these expressions can be analytically further evaluated for $r_c>r_r$. This is a reasonable assumption, because the Gaussian cut-off $r_c$ has to be large enough, so that the results do not depend on it. If $r_c<r_r$ the whole method is hardly applicable. In GiBUU these integrals are computed numerically, but we have found analytical expressions for them (see \cref{sec:paulibl_append}). For $V_p$ a sphere of radius 80 MeV is taken.

\begin{figure}
\centering
\includegraphics[width=0.48\textwidth]{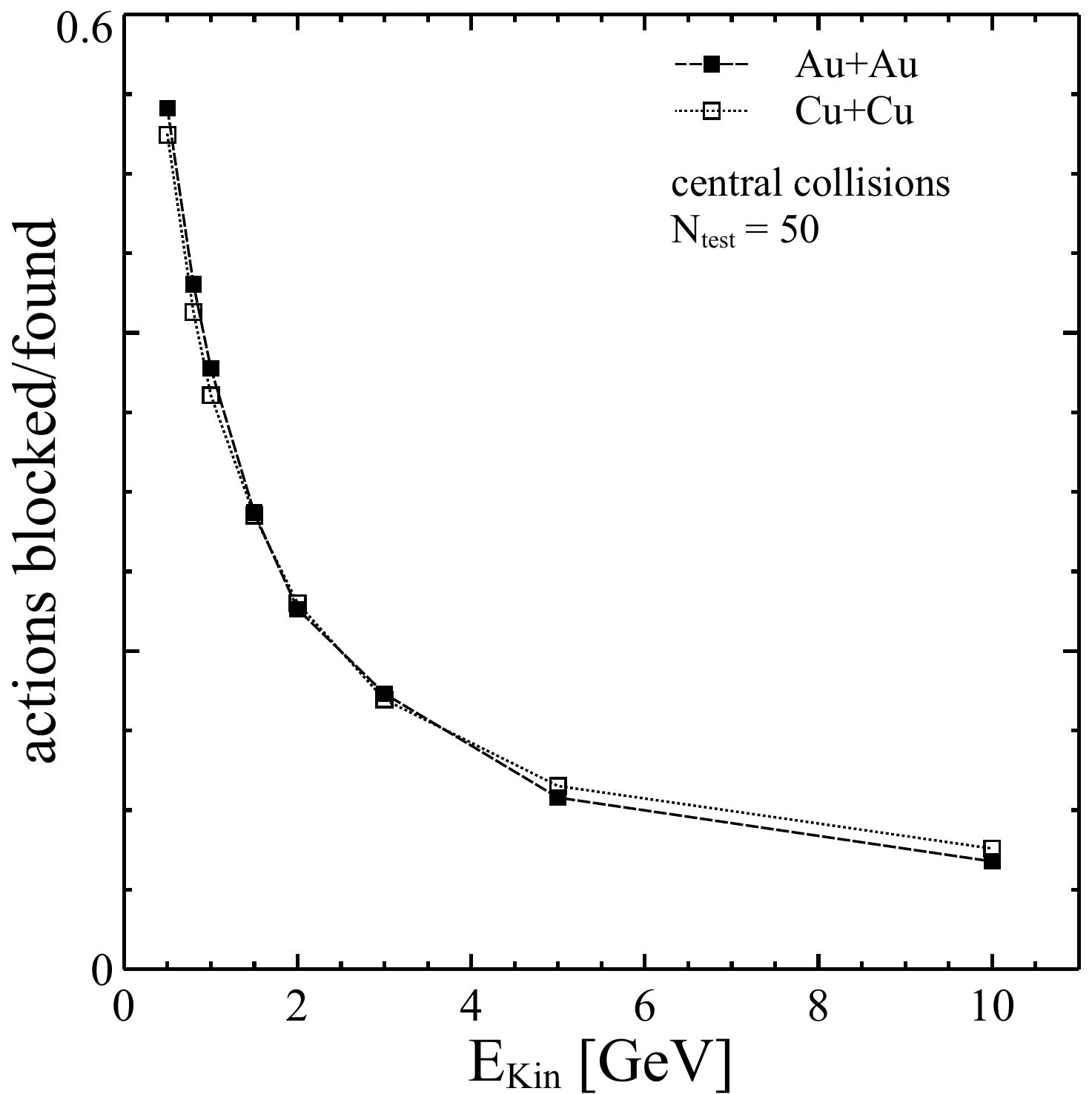}
\caption{Ratio of Pauli blocked to total found actions in Cu+Cu and Au+Au collisions at different beam energies. For reference, the total number of found actions per event (both blocked and performed) in an Au+Au collision at $E_\text{kin} = 0.5A\,\text{GeV}$ is $0.99 \times 10^5$, for $E_\text{kin} = 5A\,\text{GeV}$ it constitutes $1.32 \times 10^5$. The number of test particles used in the simulation is $N_\text{test} = 50$.}
\label{fig:pauli_blocking}
\end{figure}

\begin{figure}
\centering
\includegraphics[width=0.48\textwidth]{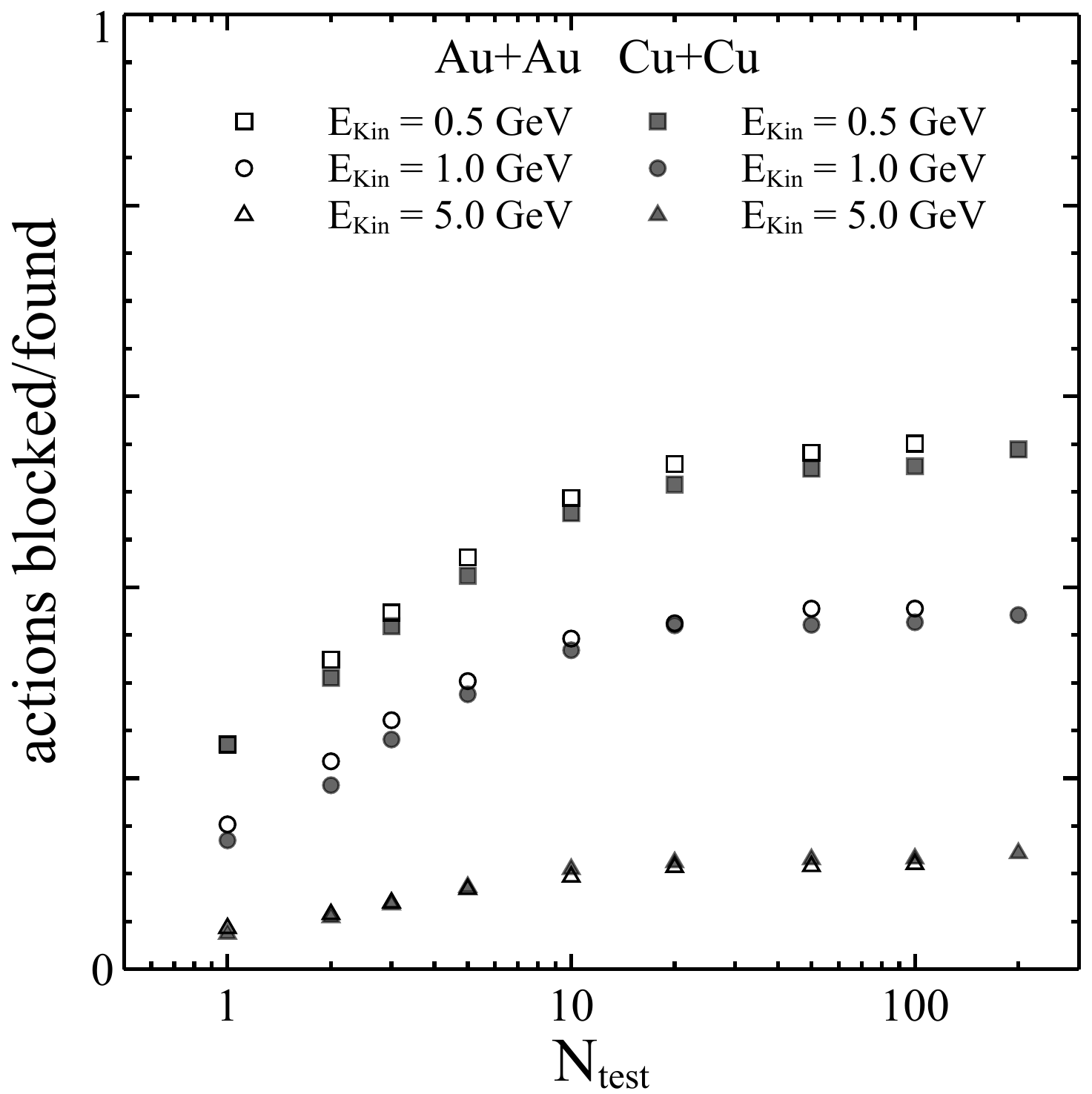}
\caption{Ratio of Pauli blocked to total found actions in Cu+Cu (filled symbols) and Au+Au (open symbols) collisions for different numbers of test particles.}
\label{fig:pauli_ntest}
\end{figure}

In \cref{fig:pauli_blocking} the number of collisions that is blocked due to prior phase space occupation has been calculated in central Cu+Cu and Au+Au collisions as a function of beam energy. One can see that at very low energies there are as many blocked collisions as collisions taking place. The ratio drops rather fast and around $E_{\rm kin} = 2A$ GeV only a quarter of the collisions are blocked. It then saturates around 10\% for higher beam energies.

\cref{fig:pauli_ntest} demonstrates the need for a decent number of test particles to obtain stable results. If the number of test particles is low the phase-space volume cannot be calculated with enough precision and therefore, there are too many collisions allowed. Saturation sets in around $N_{\rm test} =20$ and is very similar for Au+Au and Cu+Cu collisions.

\section{Validation}
\label{sec:validation}

\subsection{Elementary cross sections}
\label{sec:elementary_reactions}

The elementary hadron-hadron scattering cross sections are among the most
important ingredients of a transport model. The production mechanisms and cross
section formulae were discussed in detail in \cref{sec:coll_term}.

\begin{figure}
\centering
\includegraphics[width=0.48\textwidth]{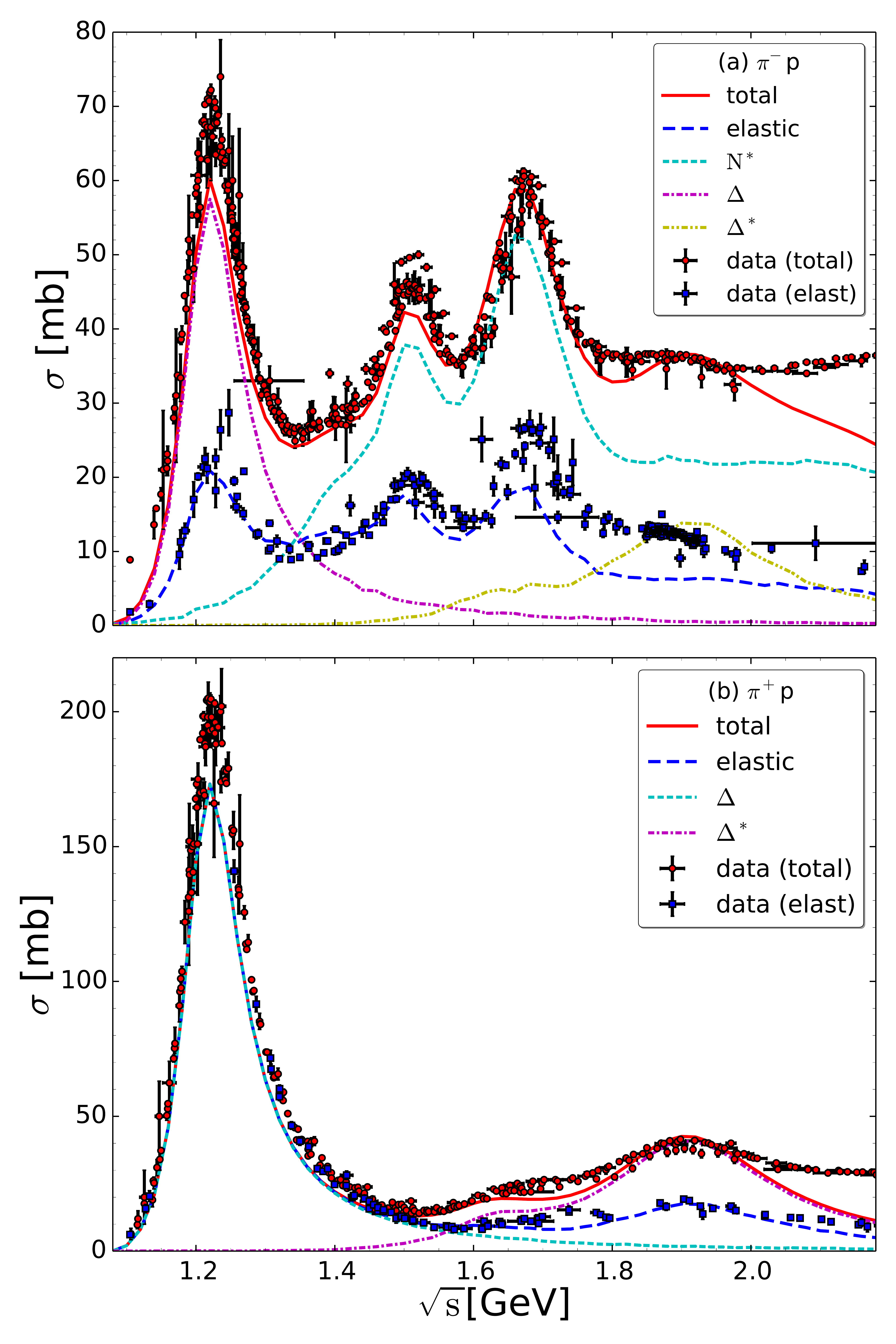}
\caption{$\pi^-$-proton (a) and $\pi^+$-proton (b) cross sections compared to data from \cite{Agashe:2014kda}.}
\label{fig:xs_piN}
\end{figure}

\begin{figure}
\centering
\includegraphics[width=0.48\textwidth]{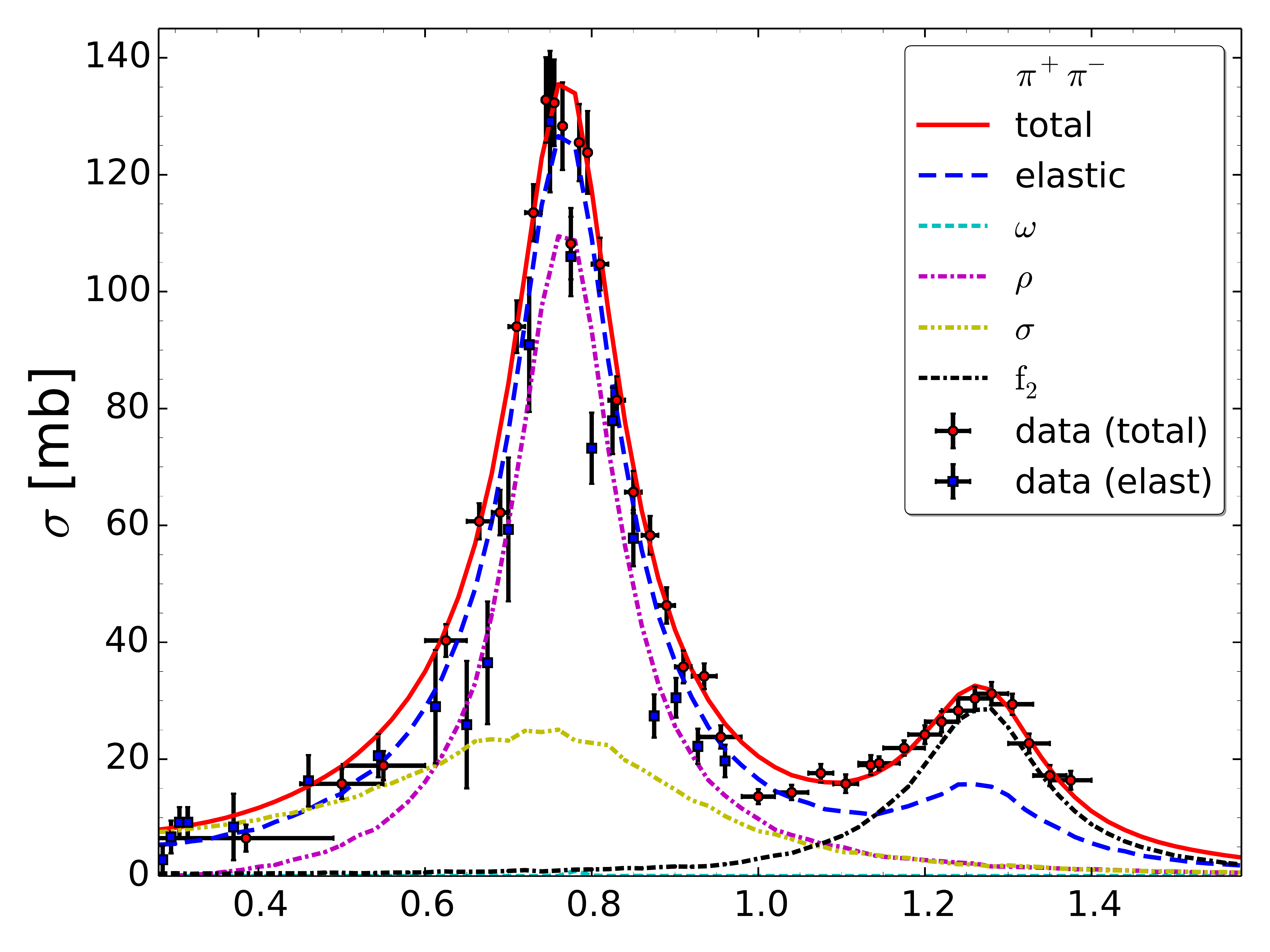}
\caption{Pion-pion cross section compared to data from \cite{Protopopescu:1973sh,Alekseeva:1982uy}.}
\label{fig:xs_pipi}
\end{figure}

\begin{figure}
\centering
\includegraphics[width=0.48\textwidth]{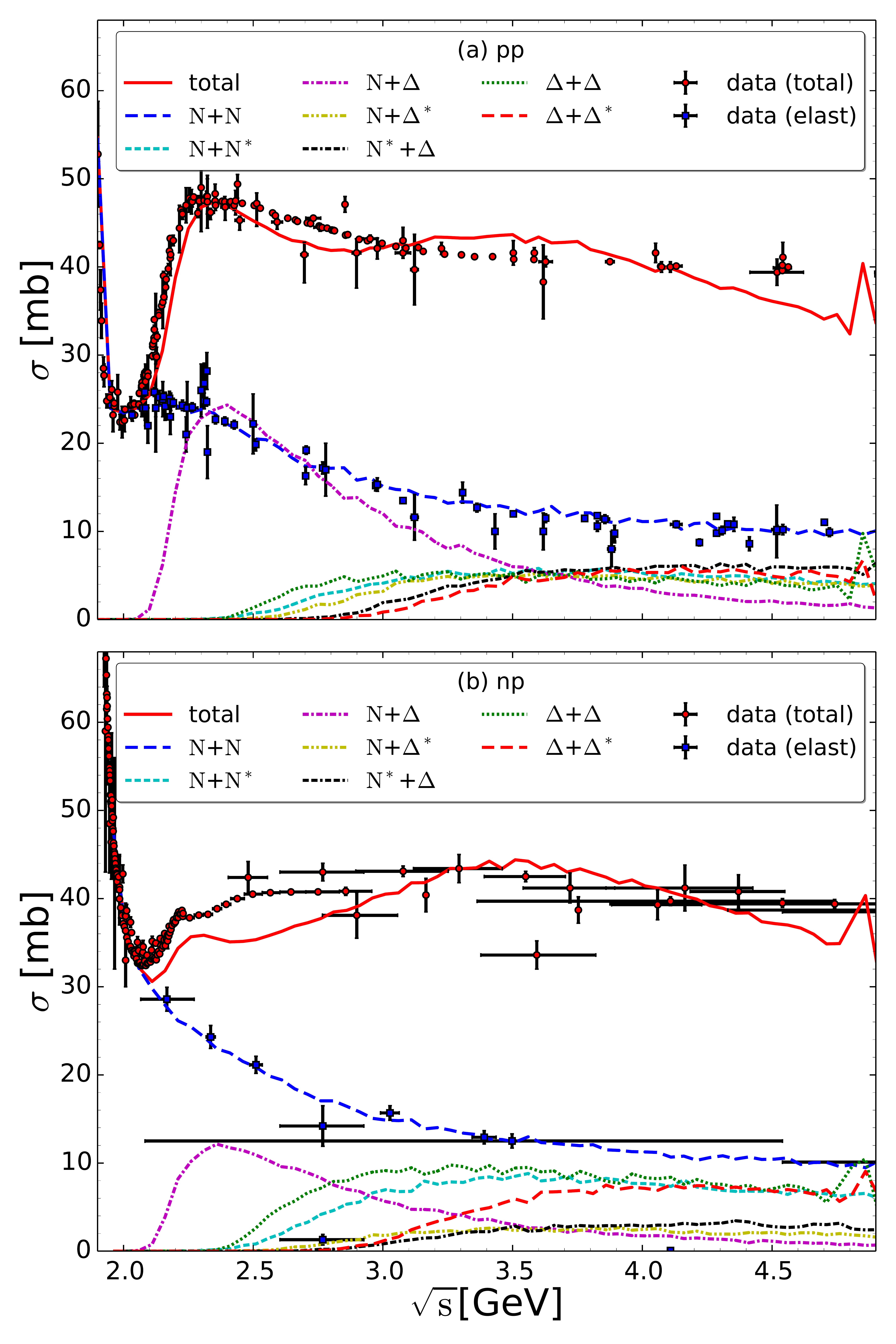}
\caption{Proton-proton (a) and proton-neutron (b) cross sections compared to data from \cite{Agashe:2014kda}.}
\label{fig:xs_NN}
\end{figure}

Since nucleons and pions are clearly
the most abundant particles in a heavy-ion collision, we show in
\cref{fig:xs_NN,fig:xs_piN,fig:xs_pipi} the cross sections for $NN$, $\pi N$ and
$\pi\pi$ collisions at energies of a few GeV, where the cross sections are
expected to be dominated by the excitation of hadronic resonances.

In particular the $\pi^-p$ cross section in the upper panel of \cref{fig:xs_piN}
shows some very clear resonance structures. The lowest excitation here is the
$\Delta(1232)$, followed by several $N^*$ resonances in the second and third
resonance region at around 1.5 and 1.7 GeV, respectively. $\Delta^*$ states only
play a significant role at higher energies of around 1.9 GeV. In fact SMASH
exclusively produces $s$-channel resonances in this case, which then decay into
different final states. In this way, we can saturate the total cross section up
to about 2 GeV with only minor deviations, which may be caused by the negligence of
non-resonant backgrounds and/or uncertainties regarding resonance parameters.

Also the elastic cross section only involves contributions from $s$-channel
resonances, which then decay back into $\pi^-p$, and is reasonably well
described over most of the displayed energy range. Only above energies of 2 GeV,
SMASH starts to underestimate the total and elastic cross section. Here further
production mechanisms, such as string fragmentation, will be necessary to
achieve agreement with the data.

The $\pi^+p$ cross section in the bottom panel of \cref{fig:xs_piN}, shows a
similar dominance of $s$-channel resonances. However it is limited to
$\Delta$-type excitations due to isospin arguments. The resonance contributions
in $\pi^-p$ and $\pi^+p$ are related by simple Clebsch-Gordan factors.

The purely mesonic case of the $\pi^+\pi^-$ cross section in
\cref{fig:xs_pipi} exhibits a similar resonance pattern. Here the dominant
resonances are the $\rho$ and $f_2$ states. There is also a contribution from
the scalar $\sigma$ (or $f_0$) meson. However, it should be noted that the
parameters (mass and width) of the $\sigma$ in SMASH differ significantly from
the PDG values \cite{Agashe:2014kda}, in order to achieve a reasonable agreement
with the $\pi\pi$ data. Presumably this discrepancy is due to our usage of the
Breit-Wigner approximation, which is known to be questionable for a state like
the $\sigma$ meson, for which the width is comparable to the mass.

For the nucleon-nucleon cross sections in \cref{fig:xs_NN}, the resonance
contributions are less apparent, simply because the resonances do not occur in
the $s$-channel. Instead the prevalent physical picture in this case is a
$t$-channel meson exchange, which may excite one or both of the scattered nucleons
into a resonance state that subsequently decays. Both the pp and pn cross
sections include a significant elastic contribution that rises towards the
threshold. We simply parametrize the $\sqrt{s}$ dependence in this case, cf.
\cref{sec:coll_term}. The first inelastic channel that opens up is the excitation
of a single $\Delta$ resonance. At higher energies it is followed by the
excitation of heavier resonance states ($N^*$ and $\Delta^*$) as well as
double-resonance excitations. For the nucleon-nucleon case, the resonance-based
mechanisms are able to saturate the total cross section up to energies of 4 to
4.5 GeV, above which they need to be supplemented by additional production
mechanisms (e.g. string fragmentation). Further it should be noted that the
$\sqrt{s}$ dependence of the total cross section is not described perfectly well
here, which may be caused by assuming matrix elements which are independent of
$s$, e.g. in \cref{eq:sigma22_1res}.

\begin{figure}
\centering
\includegraphics[width=0.48\textwidth]{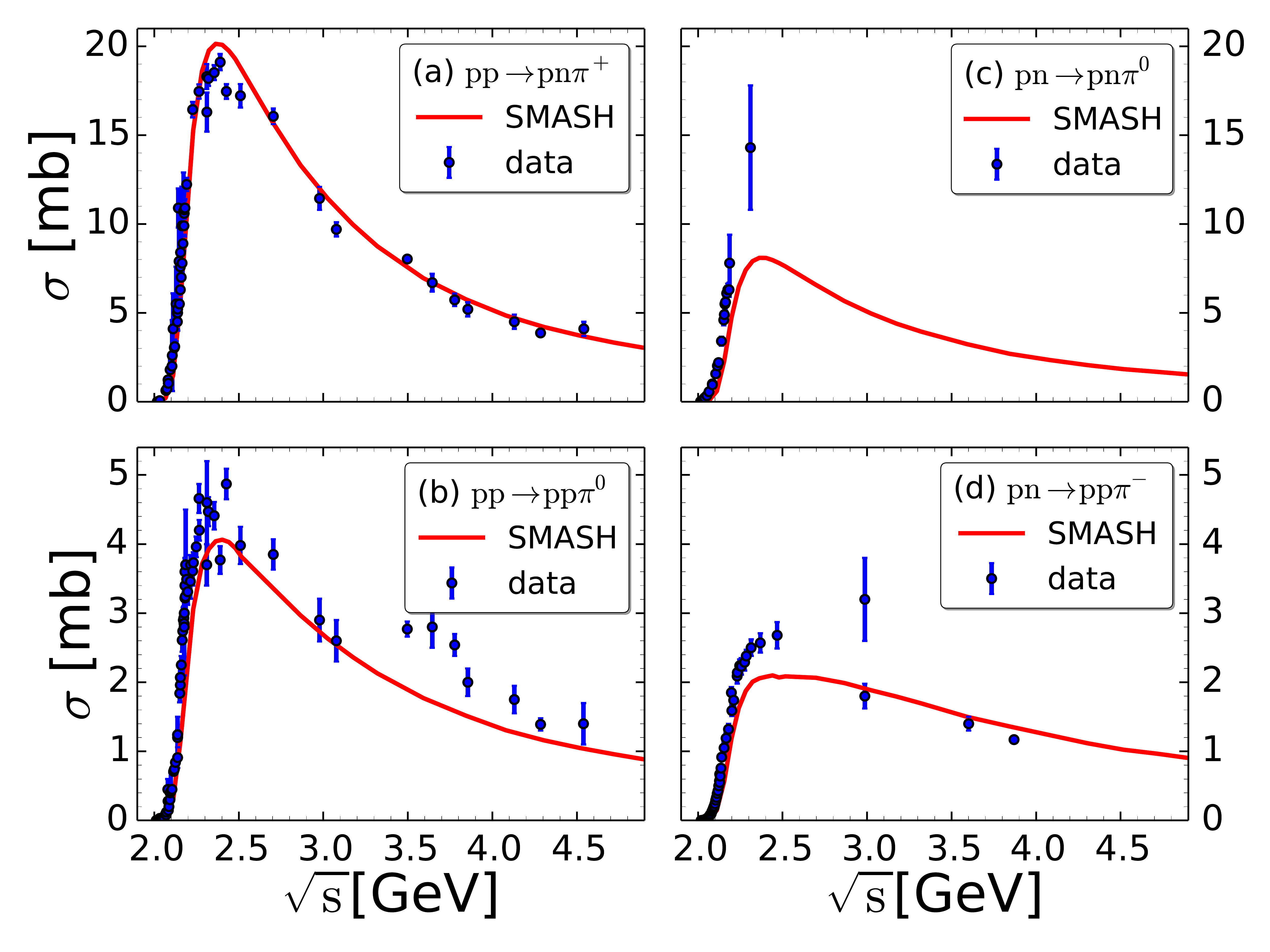}
\caption{Cross sections for single pion production from proton-proton (left) and proton-neutron (right) collisions compared to data from \cite{LaBoer}.}
\label{fig:xs_NN_NNpi}
\end{figure}

The exclusive cross section for single pion production in proton-proton collisions in \cref{fig:xs_NN_NNpi} shows an overall good agreement with the data. The dominant contribution for single pion production in nucleon-nucleon collisions is the $\Delta$ resonance (compare \cref{fig:xs_NN}). Above 2.5 GeV also additional contributions from excited resonance states ($N^*$ and $\Delta^*$) occur. A slight undershoot for the $\pi^0$ production at low energies in proton-proton collisions might come from non-resonant background terms that are not included in the model. \cref{fig:xs_NN_NNpi} also reveals a systematic undershooting for the single pion production in proton-neutron collisions, which could be due to an underestimation of the contributions with total isospin $I=0$, see \cref{sec:2_to_2}.

\subsection{Angular distributions}
\label{sec:angular_dist}

\begin{figure}
\centering
\includegraphics[width=0.48\textwidth]{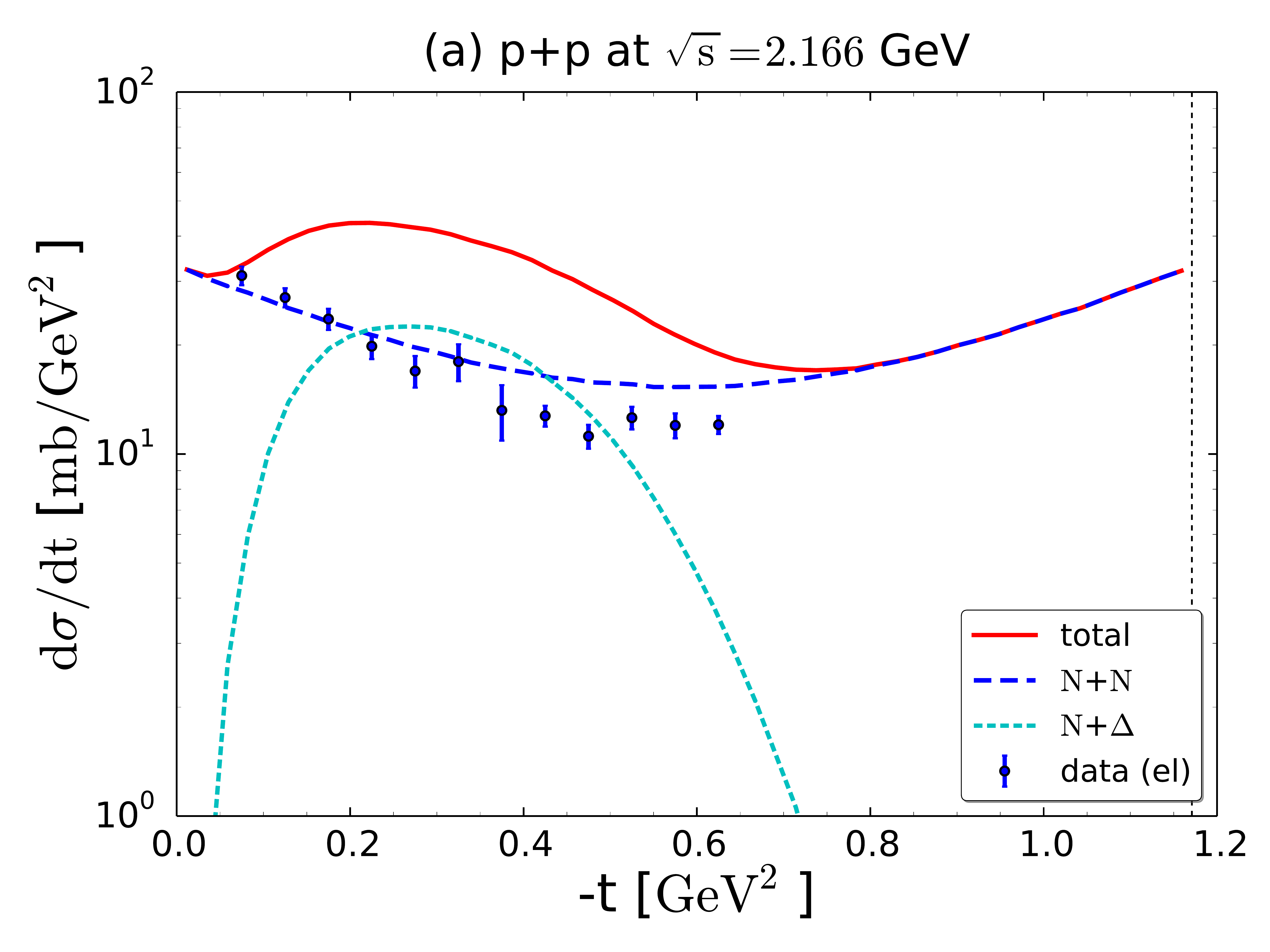}
\includegraphics[width=0.48\textwidth]{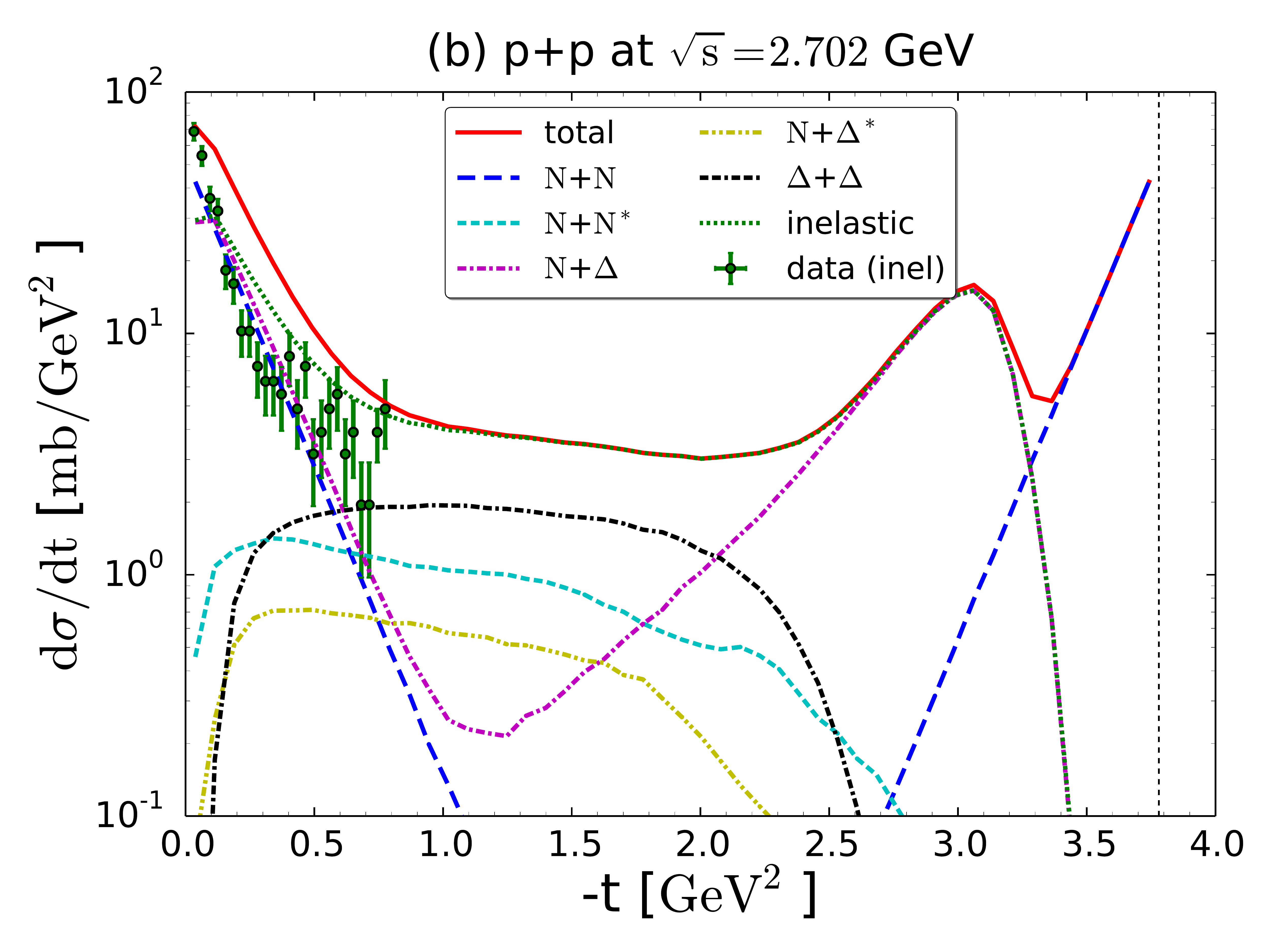}
\caption{Angular distributions for elastic and inelastic pp collisions at two different energies, compared to data from \cite{Bacon:1967zz,Ryan}.}
\label{fig:angular}
\end{figure}

In \cref{fig:angular} we show two examples of angular distributions $d\sigma/dt$
in pp collisions, $t$ being the Mandelstam variable. The upper plot shows a
collision at a relatively low energy, where essentially only the elastic and
single-$\Delta$-production channels are open. The angular distribution of the
elastic channel is of course symmetric in the allowed $t$-range and matches the
data points rather well, even though the slope at this particular energy appears
to be slightly too flat.
The distribution for single-$\Delta$ production is not symmetric and restricted
to a smaller range in $t$, due to the larger mass of the $\Delta$ in the final
state. Unfortunately there is no inelastic data to compare to at this energy.

The lower plot in \cref{fig:angular} shows a pp collision at a somewhat higher
energy, where additional resonance production channels are open. In principle
the distributions for all these channels are forward/backward-peaked (either
exponential or power-law shaped), as mentioned in \cref{sec:angular}. This
forward/backward peaking is clearly visible for the $NN$ and $N\Delta$ final
states at least, while those final states with heavier resonances exhibit a more
plateau-like structure, due to the limited phase space and the mass
distributions of the resonances. Here the sum of all inelastic channels is
compared to data and indeed shows a reasonable agreement, again with a slight
tendency of being too flat.

\subsection{Detailed balance}
\label{sec:detailed_balance}

\begin{figure}
\includegraphics[width=0.48\textwidth]{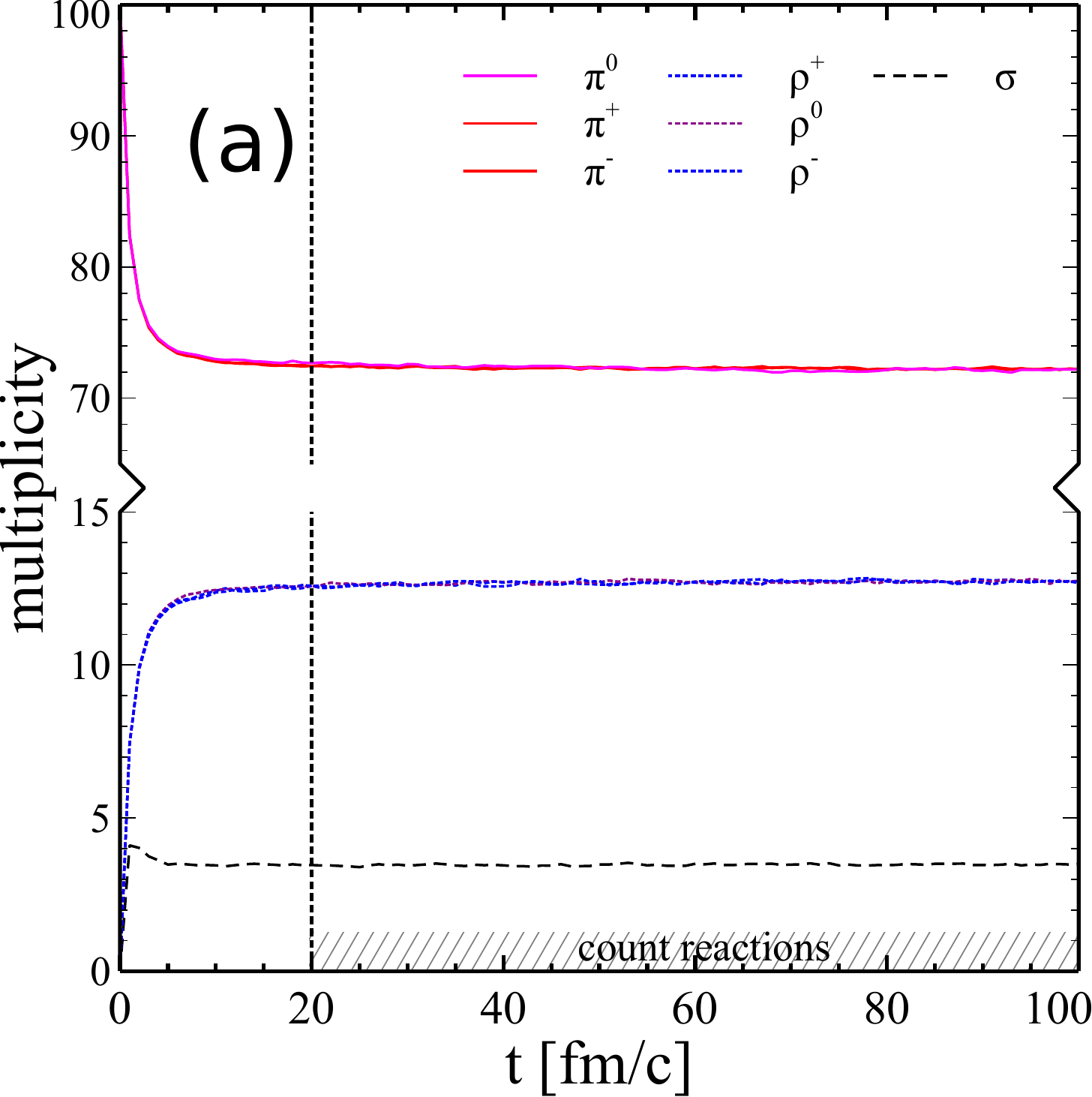} \\
\includegraphics[width=0.48\textwidth]{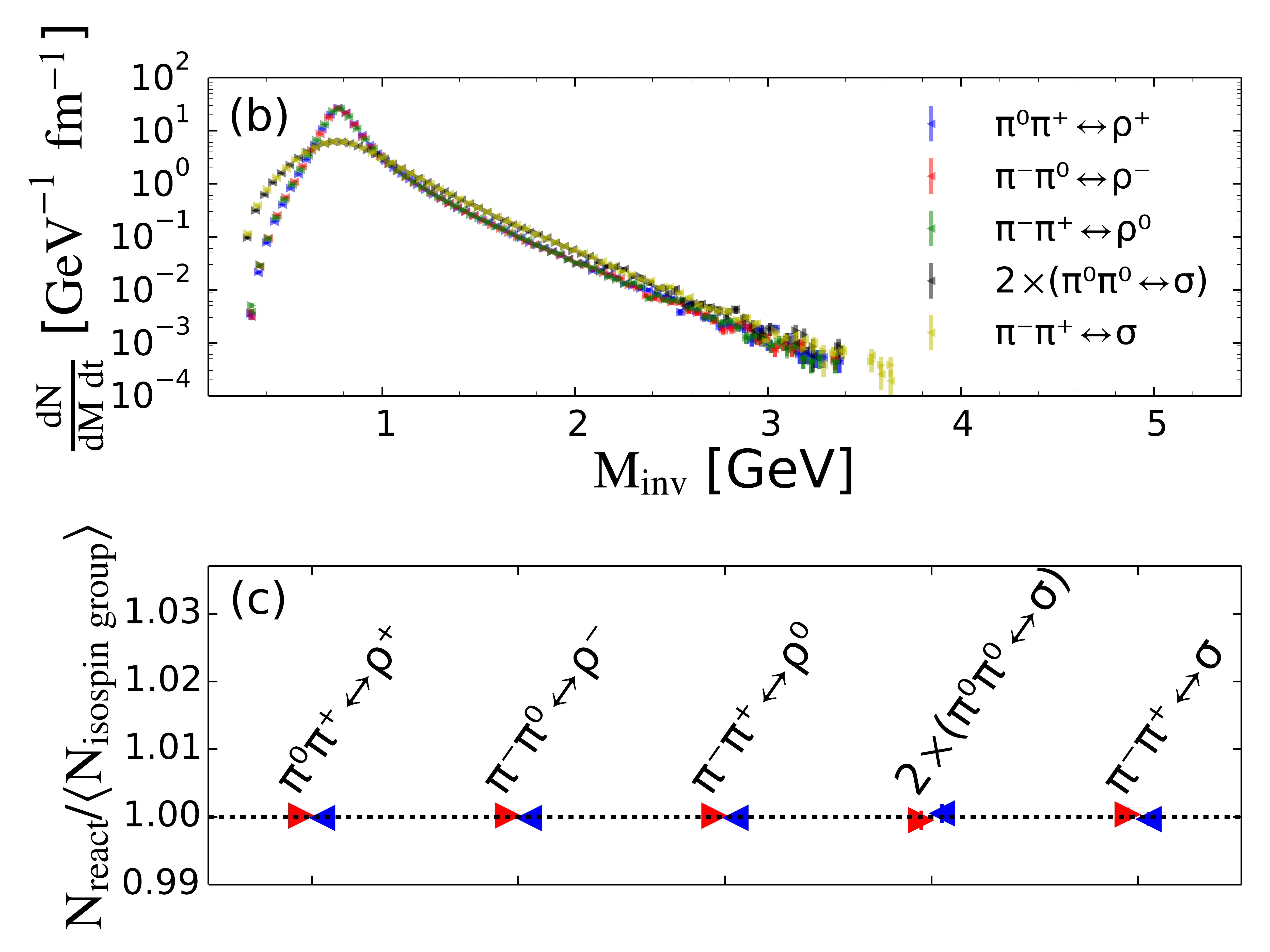}
\caption{Detailed balance for the $\pi$-$\rho$-$\sigma$ system in a box with periodic boundary conditions. Multiplicities versus time (a), scaled numbers of forward and backward reactions for $t>20$ fm/c (c), and the same differentially versus the invariant mass of the reaction, which is equal to the resonance mass in this case (b).}
\label{fig:pi_rho_sig_box}
\end{figure}

The strong interaction is invariant under time reversal, which implies that for any scattering or decay process the probability of transition $w(\Gamma_i, \Gamma_f)$ from the point in phase space $d\Gamma_i$ to $d\Gamma_f$ is equal to the probability of the reverse process.
\begin{align} \label{eq:prob_t_reversal}
w(\Gamma_i, \Gamma_f) = w(\Gamma_f, \Gamma_i)
\end{align}

\cref{eq:prob_t_reversal} is embodied in SMASH via the equality of matrix elements of the forward and backward reactions,
\begin{align}
|M_{\rightarrow}|^2 = |M_{\leftarrow}|^2 = |M|^2.
\end{align}
With this formula one can connect cross sections of the forward and backward $2\to 2$ reaction, or the width of the $1\to 2$ decay to the backward $2 \to 1$ reaction. For example, for $1 2\to 1' 2'$ scatterings
\begin{align}
d\sigma = (2\pi)^{-2} \delta^{(4)}(P_i-P_f) |M|^2 \frac{1}{4 I} \frac{d^3p_1'}{2E_1'} \frac{d^3p_2'}{2E_2'} \frac{1}{1+\delta_{1'2'}} \,,
\end{align}
where $I = \sqrt{(P_1 \cdot P_2)^2 - m_1^2 m_2^2}$ and the term $1/(1+\delta_{1'2'})$ accounts for identical particles in the final state. Integrating this over momenta one arrives at \cref{eq:sigma22_1res}. For resonances in the final state the transformation from \cite{Wolf:1992qja} is applied. The corresponding \cref{eq:sigma21} for decays is derived analogously.

Substituting \cref{eq:prob_t_reversal} back into the Boltzmann equation \ref{eq:boltzmann_equation} leads to the principle of detailed balance: In equilibrium the rate of forward reactions $d\Gamma_i \to d\Gamma_f$ is equal to the rate of backward reactions \cite{Lifsh10}.

To test, if detailed balance actually holds in our calculations, a periodic box is initialized with multiple particle species. After the matter reaches equilibrium, we check that the numbers of forward and backward reactions are identical. The fact that the box should reach equilibrium is granted by the H-theorem, which is derived assuming \cref{eq:prob_t_reversal} and the hypothesis of molecular chaos (two-particle distribution function $\mathit{f}_2(\Gamma_1, \Gamma_2) = \mathit{f} (\Gamma_1) \mathit{f}(\Gamma_2)$ or, in other words, participants of the reaction are uncorrelated). Strictly speaking, in a transport code both assumptions are valid only in the limit $N_\text{test} \to \infty$. At finite $N_\text{test}$ the interactions are non-local due to the geometrical cross sections.
In addition, while two particles with space coordinates $\vec{r}_1$ and $\vec{r}_2$ form a resonance at $(\vec{r}_1 + \vec{r}_2)/2$, the products of resonance decay gain the same position as the decaying resonance. This breaks \cref{eq:prob_t_reversal}, where for non-local interactions the phase space $\Gamma$ includes coordinate space. This leads to a small violation of detailed balance, which vanishes at large $N_\text{test}$ as we show in the following.

For the test we are using two configurations: a $\rho-\pi-\sigma$ box and a $N-\pi-\Delta$ box. The first one is initialized with a 100 $\pi^+$, 100 $\pi^-$ and 100 $\pi^0$ in a volume of $V = (10$ fm$)^3$. The reactions $\pi\pi \leftrightarrow \rho$ and $\pi\pi \leftrightarrow \sigma$ are allowed, while all the other possible reactions are switched off. From \cref{fig:pi_rho_sig_box} one observes that the system reaches chemical equilibrium, since the particle multiplicities in the box saturate after around $t=20$ fm/c. Starting from this time, forward and backward reactions are counted. The matrix elements of reactions in the same isospin group differ only by Clebsch-Gordan coefficients. Thus one expects, for example, that the number of reactions $N(\sigma \leftrightarrow \pi^+ \pi^-) = 2 N(\sigma \leftrightarrow \pi^0 \pi^0)$. Therefore, the reaction numbers in \cref{fig:pi_rho_sig_box} are scaled by the isospin and symmetry factors appropriately to make sure that this expectation is fulfilled.
Detailed balance is valid not only for the total number of reactions, but it also has to be fulfilled differentially in momentum space. We show in \cref{fig:pi_rho_sig_box} that detailed balance is indeed fulfilled differentially in each invariant mass bin of the reaction.
Let us note that for the $\rho-\pi-\sigma$ box detailed balance for the total (but not differential) number of reactions follows trivially from the multiplicity saturation. Indeed, denoting forward and backward reaction rates by $r^{\rightarrow}$ and $r^{\leftarrow}$, one arrives at
\begin{align}
\frac{dN_{\rho}}{dt} = - r^{\rightarrow}_{\rho\pi\pi} + r^{\leftarrow}_{\rho\pi\pi} = 0\\
\frac{dN_{\sigma}}{dt} = - r^{\rightarrow}_{\sigma\pi\pi} + r^{\leftarrow}_{\sigma\pi\pi} = 0
\end{align}

\begin{figure*}
\includegraphics[width=\textwidth]{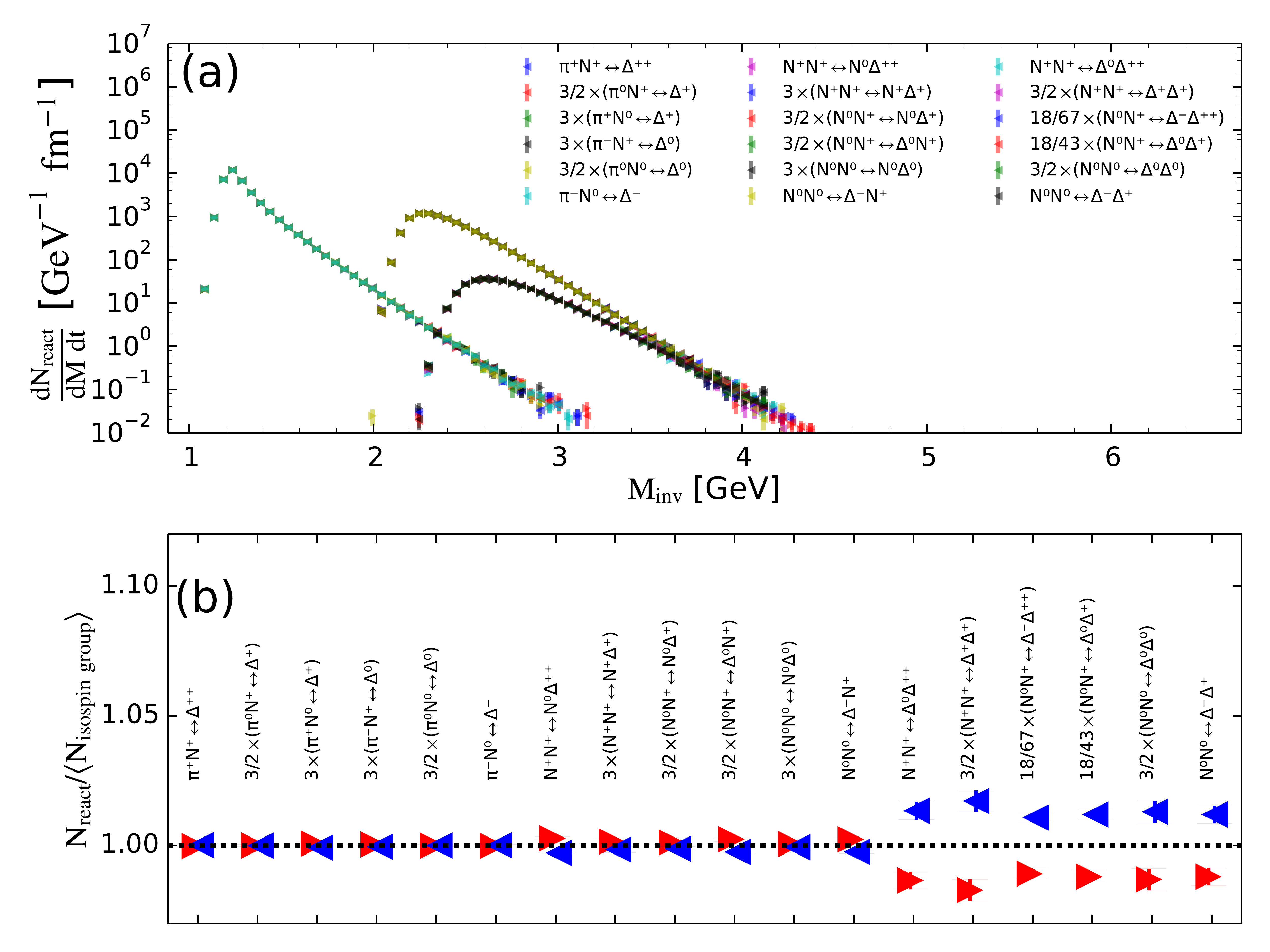}
\caption{Scaled numbers of forward (triangles right) and backward (triangles left) reactions for $t>80$ fm/c $\pi$-$N$-$\Delta$ (b) and the same differentially in the invariant mass of reaction (a).}
\label{fig:pi_N_D_reactions}
\end{figure*}

For the $N-\pi-\Delta$ box similar relations become less trivial. We initialize the $N-\pi-\Delta$ box with 100 neutrons and 100 protons and allow reactions $\Delta \leftrightarrow N\pi$ (1), $NN \leftrightarrow N\Delta$ (2) and $NN \leftrightarrow \Delta \Delta$ (3), with all the other reactions being forbidden. In chemical equilibrium the following equations are fulfilled:
\begin{align}
\frac{dN_{\pi}}{dt} &=& r^{\rightarrow}_1 - r^{\leftarrow}_1 = 0\\
\frac{dN_{N}}{dt} &=& -r^{\rightarrow}_2 + r^{\leftarrow}_2 - 2 (r^{\rightarrow}_3 - r^{\leftarrow}_3)  = 0 \\
\frac{dN_{\Delta}}{dt} &=& r^{\rightarrow}_2 - r^{\leftarrow}_2 + 2 (r^{\rightarrow}_3 - r^{\leftarrow}_3)  = 0
\end{align}
It can be observed that forward and backward rates for $NN \leftrightarrow N\Delta$ and $NN \leftrightarrow \Delta \Delta$ being equal does not necessarily follow from multiplicities being saturated. As one can see from \cref{fig:pi_N_D_reactions}, with $N_\text{test} = 100$ detailed balance is violated at maximum by 2\%. For $N_\text{test} = 1$ this violation can reach 10\% because of the non-locality effect described above.

To see if the numbers of reactions within one isospin group relate as expected from Clebsch-Gordan factors, we multiply every number of reactions $N_i$ by a factor $\alpha_i$ that compensates for the isospin factors of this reaction. Let us denote $\langle N_\text{isospin group} \rangle = \frac{1}{k} \sum_{i=1}^k \alpha_i N_i$, where $k$ is amount of reactions in the isospin group (forward + backward). If the SMASH result corresponds to the theoretical expectation, then $N_i/\langle N_\text{isospin group} \rangle$ should be strictly 1 for every reaction. One can make sure from
\cref{fig:pi_rho_sig_box} and from \cref{fig:pi_N_D_reactions} that SMASH matches this expectation. \cref{tab:clebsch} shows the origin of compensating coefficients $\alpha_i$. While most of the Clebsch-Gordan factors are simple, for $pn \leftrightarrow \Delta\Delta$ reactions they are less intuitive. The matrix element for $NN \leftrightarrow \Delta\Delta$ reaction is isospin dependent, namely $|M(I=0)|^2 = \kappa |M(I=1)|^2$, where $\kappa = \frac{8}{3}$. Here is one explicit example illustrating the calculation (where states beyond $I=1$ have been omitted, since they drop out):
\begin{align}
&|pn\rangle &=& \sqrt{\frac{1}{2}} | I=1 \rangle + \sqrt{\frac{1}{2}} | I=0 \rangle \\
&|\Delta^- \Delta^{++} \rangle &=& \dots + \sqrt{\frac{9}{20}} | I=1 \rangle - \sqrt{\frac{1}{4}} | I=0 \rangle \\
&\langle pn | \Delta^- \Delta^{++} \rangle ^2 &=& \frac{9}{40} |M(I=1)|^2 + \frac{5}{40} |M(I=0)|^2  \\
&\langle pn | \Delta^- \Delta^{++} \rangle ^2 &=& \frac{5\kappa +9}{40} |M(I=1)|^2
\end{align}

Thus, we have shown that the detailed balance in SMASH for a mesonic system and a more complex situation involving baryons and mesons is fulfilled.

\begin{figure}
\includegraphics[width=0.48\textwidth]{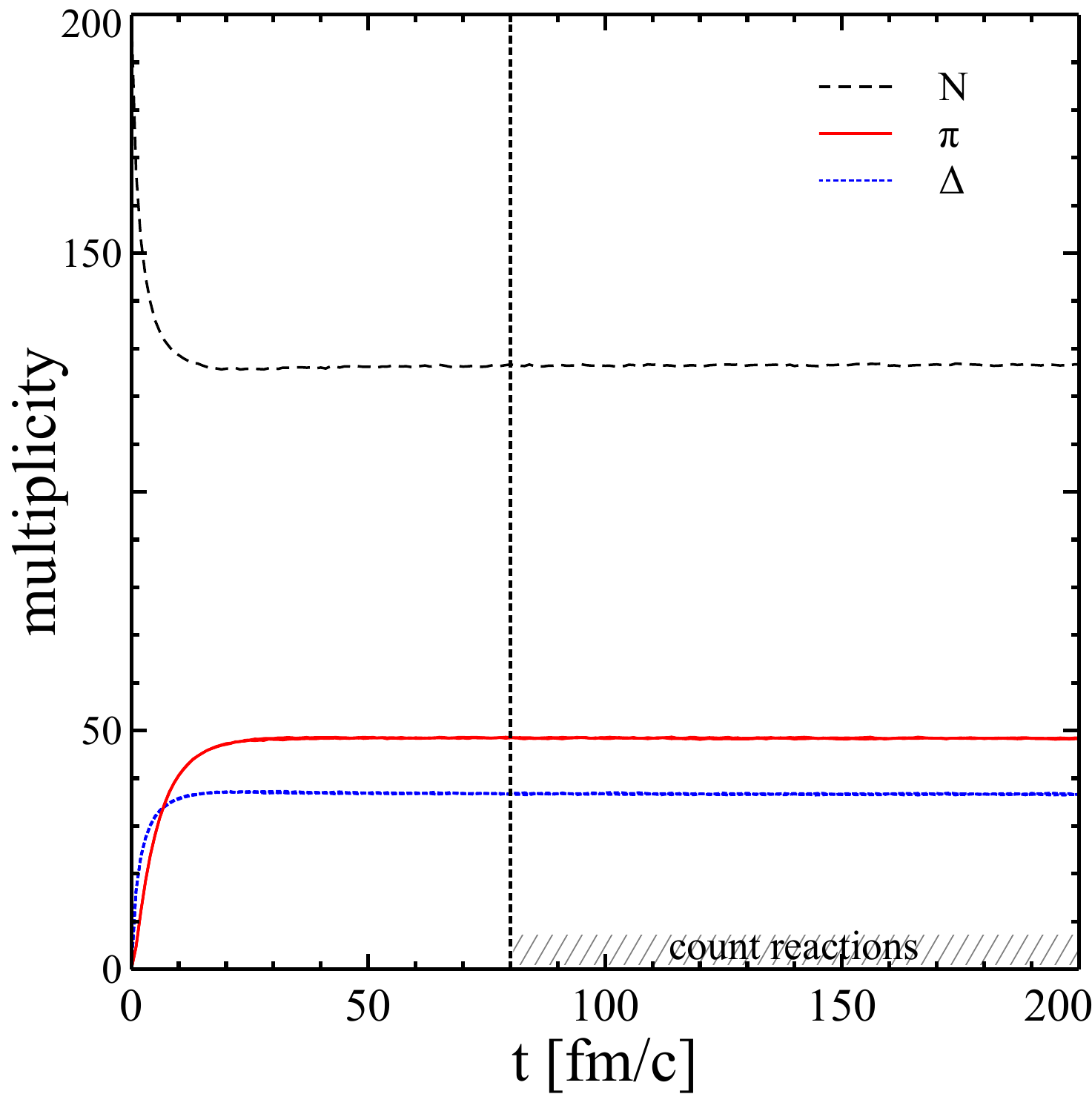}
\caption{Multiplicities versus time for $\pi$-$N$-$\Delta$ system in a box.}
\label{fig:pi_N_D_box}
\end{figure}

\begin{table}
\begin{tabular}{cccc}
\toprule
Reaction                           &   Clebsch       &  Symmetry  &   Total \\
\midrule
$\rho^+\to\pi^+\pi^0$              &  $      1/2$  &     1      &   1/2     \\
$\rho^-\to\pi^-\pi^0$              &  $      1/2$  &     1      &   1/2     \\
$\rho^0\to\pi^0\pi^0$              &          $0$  &   1/2      &   0       \\
$\rho^0\to\pi^+\pi^-$              &  $      1/2$  &     1      &   1/2     \\
\midrule
$\sigma\to\pi^+\pi^-$              &  $      1/3$  &     1      &   2/6     \\
$\sigma\to\pi^0\pi^0$              &  $      1/3$  &   1/2      &   1/6     \\
\midrule
$p\pi^+\to\Delta^{++}$             &          $1$  &     1      &   3/3     \\
$p\pi^0\to\Delta^{+}$              &  $      2/3$  &     1      &   2/3     \\
$p\pi^-\to\Delta^{0}$              &  $      1/3$  &     1      &   1/3     \\
$n\pi^+\to\Delta^{+}$              &  $      1/3$  &     1      &   1/3     \\
$n\pi^0\to\Delta^{0}$              &  $      2/3$  &     1      &   2/3     \\
$n\pi^-\to\Delta^{-}$              &          $1$  &     1      &   3/3     \\
\midrule
$pp\to p\Delta^{+}$                &  $      1/4$  &   1/2      &  1/8      \\
$pp\to n\Delta^{++}$               &  $      3/4$  &   1/2      &  3/8      \\
$pn\to n\Delta^{+}$                &  $      1/4$  &     1      &  2/8      \\
$pn\to p\Delta^{0}$                &  $      1/4$  &     1      &  2/8      \\
$nn\to p\Delta^{-}$                &  $      3/4$  &   1/2      &  3/8      \\
$nn\to n\Delta^{0}$                &  $      1/4$  &   1/2      &  1/8      \\
\midrule
$pp\to\Delta^{0}\Delta^{++}$       &  $      6/20$ &   1/2      &  18/120   \\
$pp\to\Delta^{+}\Delta^{+}$        &  $      8/20$ &   1/4      &  12/120   \\
$pn\to\Delta^{-}\Delta^{++}$       &  $    67/120$ &     1      &  67/120   \\
$pn\to\Delta^{+}\Delta^{0}$        &  $    43/120$ &     1      &  43/120   \\
$nn\to\Delta^{+}\Delta^{-}$        &  $      6/20$ &   1/2      &  18/120   \\
$nn\to\Delta^{0}\Delta^{0}$        &  $      8/20$ &   1/4      &  12/120   \\
\bottomrule
\end{tabular}
\caption{Expected isospin and symmetry factors for number of reactions within isospin groups at equilibrium. The first numeric column is a Clebsch-Gordan factor, the second column is symmetry factor, the third one is their product.}
\label{tab:clebsch}
\end{table}

\subsection{Thermodynamics} 
\label{sec:thermodynamics}

\begin{figure}
\includegraphics[width=0.48\textwidth]{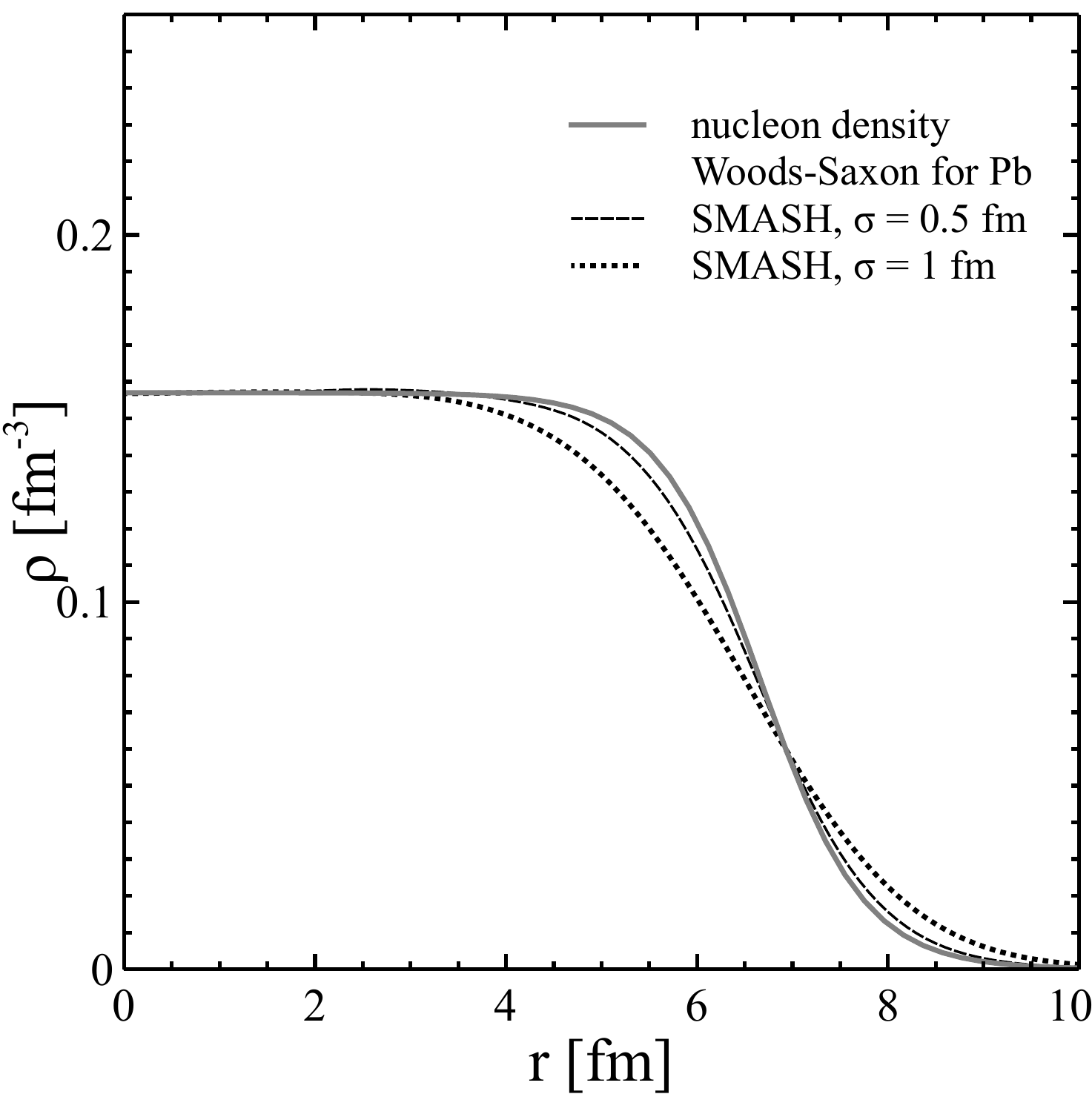}
\caption{Baryon density estimated in SMASH simulation with smearing $\sigma = 0.5$ fm (dashed line) and 1.0 fm (dotted line) is compared to the true density profile (solid line). Large $N_\text{test} = 1000$ for $\sigma = 1$ fm and $N_\text{test} = 10000$ for $\sigma = 0.5$ fm is taken to diminish fluctuations.}
\label{fig:density:smear}
\end{figure}

To investigate the thermodynamic properties of the hadron gas, the energy-momentum tensor $T^{\mu\nu}(\vec{r})$ and four-currents $j^{\mu}(\vec{r})$ can be calculated from the particle distribution functions. These two quantities provide access to the energy density and particle number density in the corresponding rest frames. Assuming that the potential energies of particles are small compared to their kinetic energies and taking into account that collisions happen instantaneously, the corresponding equations for non-interacting particles are applied:
\begin{align}
T^{\mu\nu}(\vec{r}) &= \int \frac{p^{\mu} p^{\nu}}{p^0} \mathit{f}(\vec{r}, \vec{p}) d^3p \\
j^{\mu}(\vec{r})    &= \int \frac{p^{\mu}}{p^0} \mathit{f}(\vec{r}, \vec{p}) d^3p \,,
\end{align}
where $\mathit{f}(\vec{r}, \vec{p})$ is the single-particle distribution function. For a discrete set of particles it reads
\begin{align}
\mathit{f}(\vec{r}, \vec{p}) = \sum_\text{part}  \delta^3(\vec{p} - \vec{p_\text{part}}) \delta^3 (\vec{r} - \vec{r_\text{part}})
\end{align}
For numerical calculations we substitute the delta-function by the smearing kernel
\begin{equation}
K(\Delta \vec{r}) = \frac{\gamma}{(2\pi \sigma^2)^{3/2}} \exp\left(-\frac{\Delta \vec{r}^2 + \gamma^2(\Delta \vec{r} \cdot \vec{\beta})^2}{2 \sigma^2} \right) \,,
\label{eqn:smearing_kernel}
\end{equation}
where $\Delta \vec{r} = \vec{r} - \vec{r}_\text{part}$, $\vec{\beta} = \vec{p}_\text{part}/E_\text{part}$ is the 3-velocity of the particle and $\gamma = (1-\vec{\beta}^2)^{-1/2}$. It is shown in \cite{Oliinychenko:2015lva} that this kernel has proper Lorentz-transformation properties, is normalized to 1 and represents a simple 3D-Gaussian in the rest frame of the particle. The equations for the numerical evaluation of thermodynamic quantities are then
\begin{align}
T^{\mu\nu}(\vec{r}) &= \frac{1}{N_\text{ev} N_\text{test}} \sum_\text{events} \sum_i \frac{p^{\mu}_i p^{\nu}_i}{p^0_i} K(\vec{r} - \vec{r_i}, p_i)  \\
j^{\mu}(\vec{r})    &= \frac{1}{N_\text{ev} N_\text{test}} \sum_\text{events} \sum_i \frac{p^{\mu}_i}{p^0_i} K(\vec{r} - \vec{r_i}, p_i) \,,
\label{eq:tmn_jmu}
\end{align}
where $N_\text{ev}$ is the number of events and $N_\text{test}$ is the test particle number. In the limit of the smearing width $\sigma \to 0$ and $N_\text{ev}N_\text{test} \to \infty$ the full smooth quantities are obtained. This limit is numerically challenging, because when reducing the smearing width $\sigma$, one has to increase statistics, keeping $\sigma^3 N_\text{ev}N_\text{test}=\text{const}$. Therefore, we take reasonably small $\sigma = 1$ fm and keep in mind the smearing effect, which is demonstrated in \cref{fig:density:smear} for the density calculation of a Pb nucleus comparing $\sigma = 0.5$ fm and 1 fm.

\begin{figure}
\includegraphics[width=0.48\textwidth]{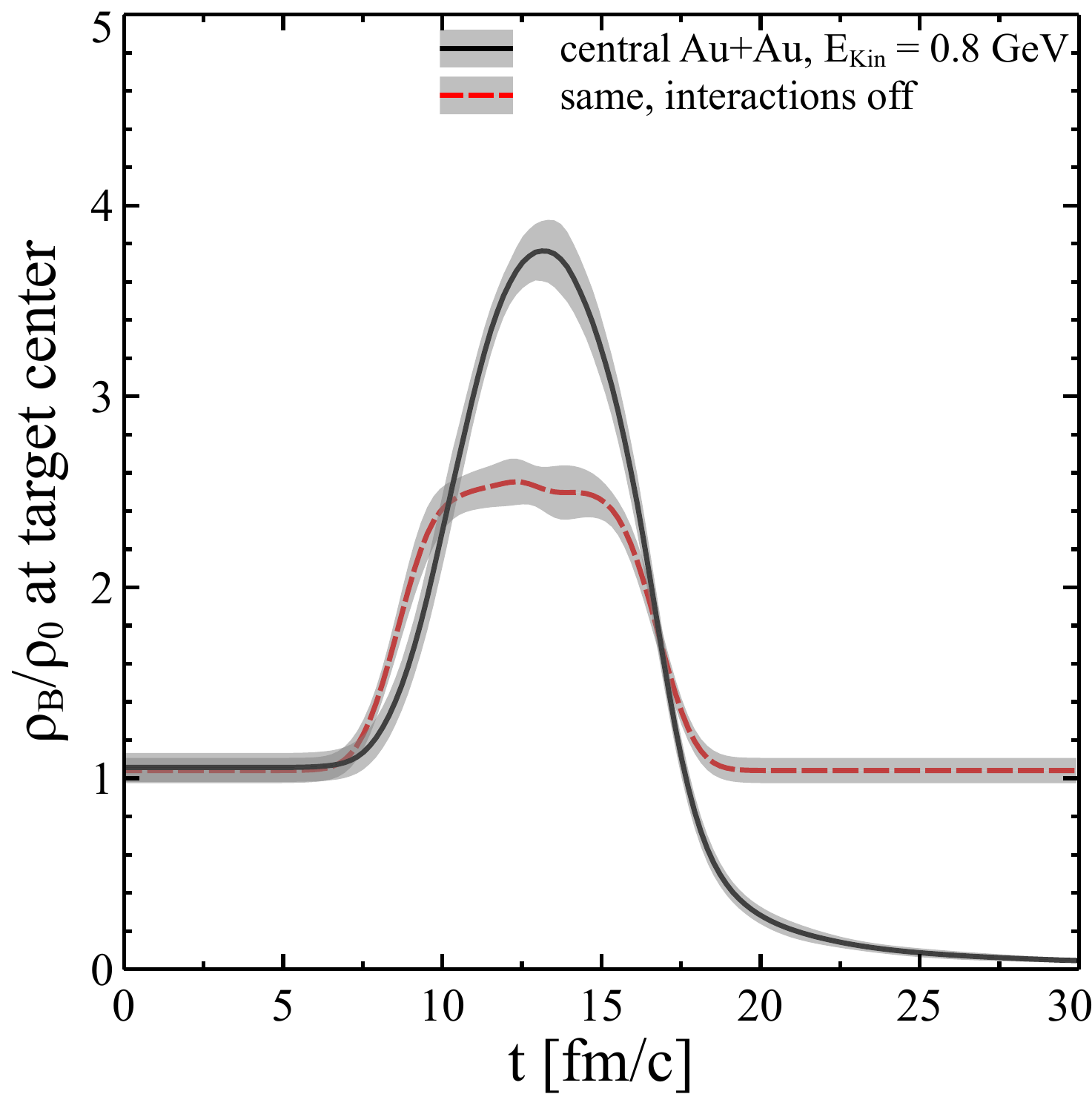}
\caption{Eckart rest frame net baryon density $\rho_B$ at the target center in central Au+Au collision at $E_\text{kin} = 0.8A\,\text{GeV}$ in units of the ground state nuclear density $\rho_0$. Time dependence $\rho_B(t)$ of the full SMASH simulation (full line) is compared to $\rho_B(t)$ of the SMASH simulation with all interactions off (dashed line).}
\label{fig:density:AuAu}
\end{figure}

\begin{figure}
\includegraphics[width=0.48\textwidth]{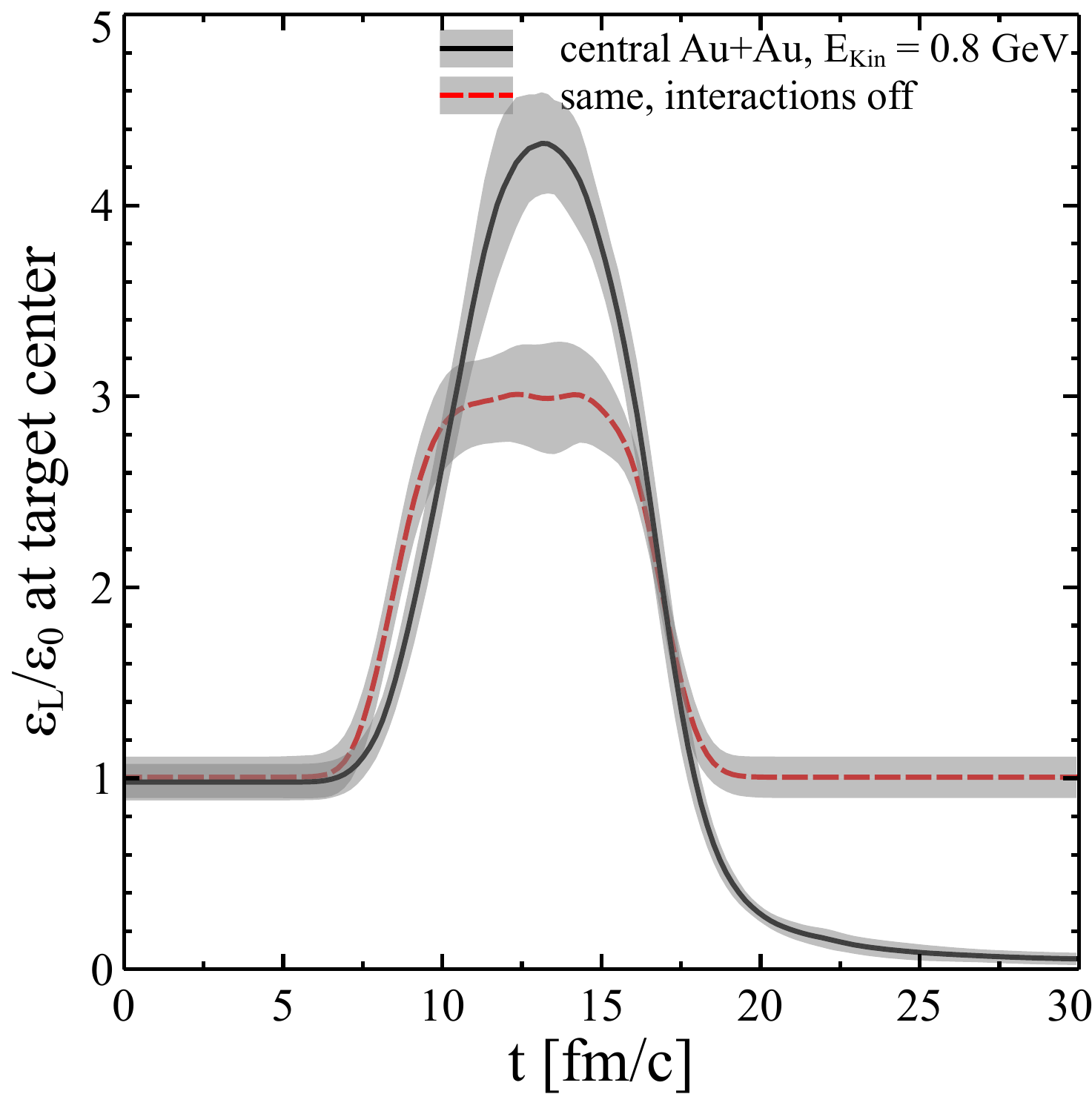}
\caption{Landau rest frame hadron density $\epsilon$ at the target center in central Au+Au collision at $E_\text{kin} = 0.8A\,\text{GeV}$ in units of the ground state nuclear energy density $\epsilon_0 = 0.150$ GeV/fm$^3$. Time dependence $\epsilon(t)$ of the full SMASH simulation (solid line) is compared to $\epsilon(t)$ of the SMASH simulation with all interactions off (dashed line).}
\label{fig:e:AuAu}
\end{figure}

The Eckart rest frame density is obtained as $\rho_\text{Eck} =\sqrt{j^{\mu}j_{\mu}}$. For net baryon (charge, isospin projection) density a naive weighting of particles in \cref{eq:tmn_jmu} with their baryon numbers can give rise to $j^{\mu}j_{\mu} <0$. Therefore, we compute $\rho = \rho^+ - \rho^-$, where~$+$ corresponds to positive baryon number (charge, isospin projection) and~$-$ corresponds to negative ones. In \cref{fig:density:AuAu} the dependence of the net baryon density versus time in the middle of the target in the central Au+Au collision at $E_\text{kin} = 0.8A$ GeV in the fixed-target frame is shown. The energy density in the Landau frame is depicted in \cref{fig:e:AuAu}. Both figures show that the ground state baryon/energy density values are reproduced, when the collision term is disabled. Including interactions the baryon/energy density rises to about 4 times the nuclear ground state densities.

\begin{figure*}
\centering
\includegraphics[width=0.35\textwidth]{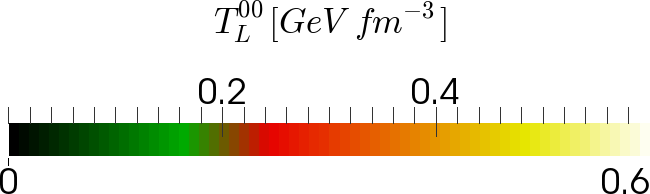}  \\
\includegraphics[width=0.19\textwidth]{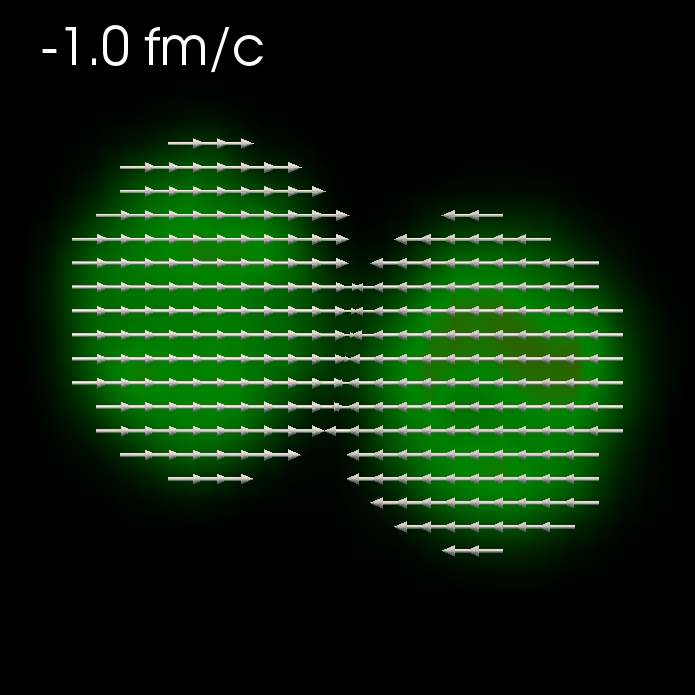}
\includegraphics[width=0.19\textwidth]{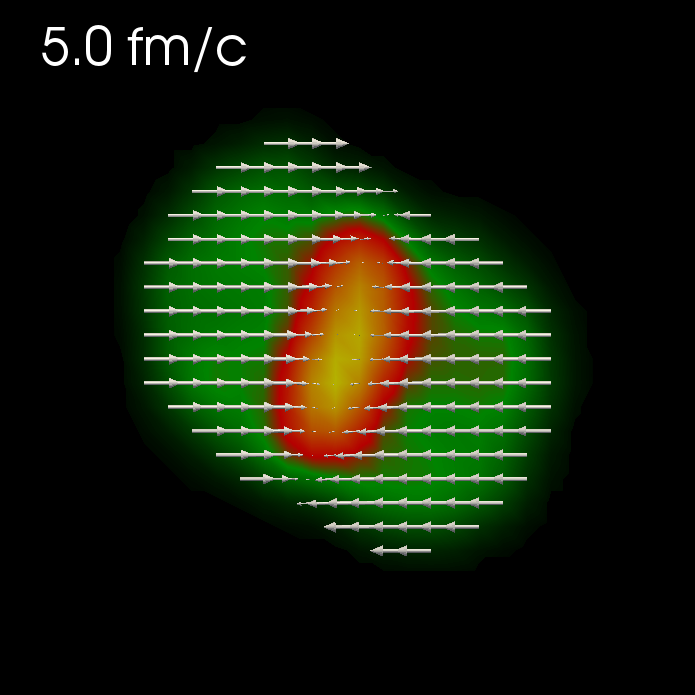}
\includegraphics[width=0.19\textwidth]{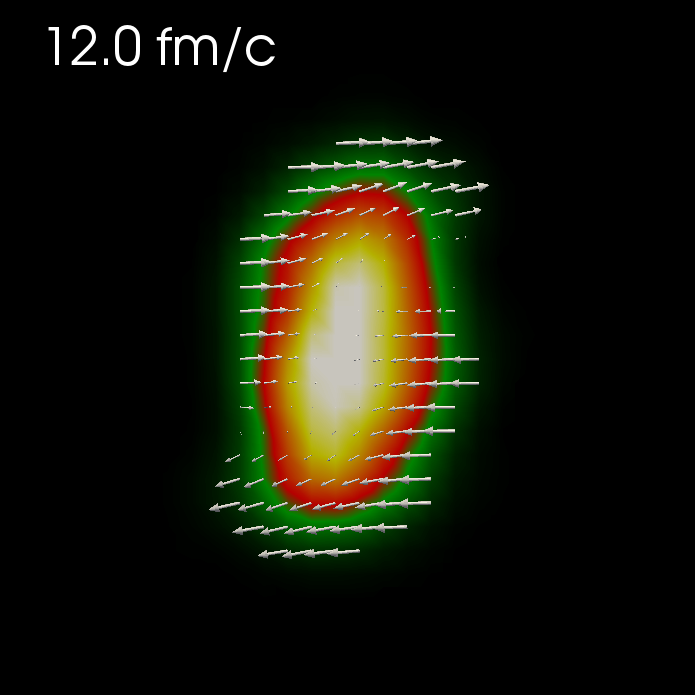}
\includegraphics[width=0.19\textwidth]{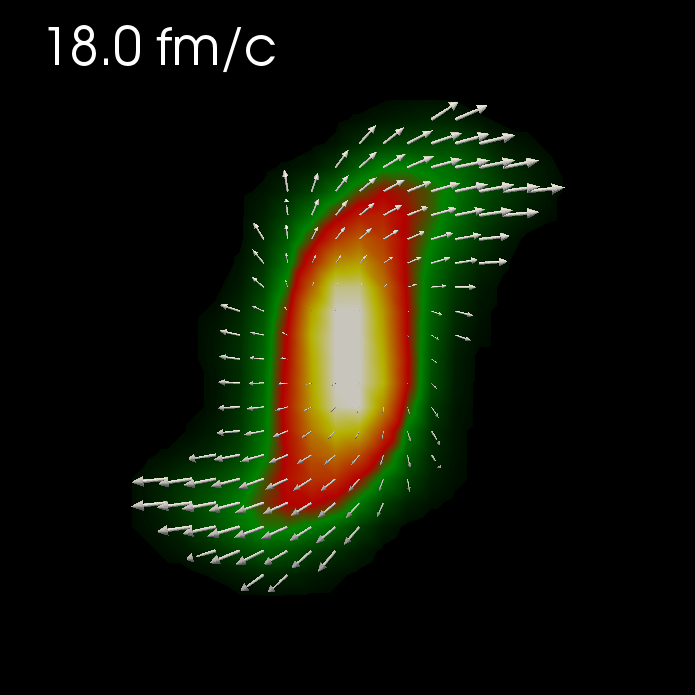}
\includegraphics[width=0.19\textwidth]{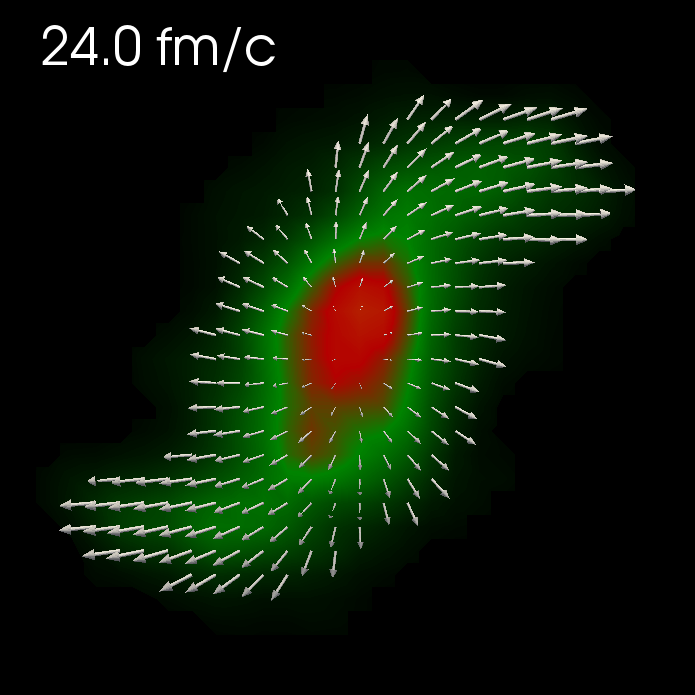}
\caption{
    Landau rest frame energy density $T^{00}_L$ (background color) and velocity of Landau frame (arrows), both for baryons. Au+Au collision at $E_\text{kin} = 0.8A\,\text{GeV}$ with impact parameter $b = 3\,\text{fm}$, $N_\text{test} = 20$. Color legend is given above. Velocity is proportional to the arrow length, maximal arrow length corresponds to velocity of 0.55 $c$.
}
\label{fig:landau_e_v}
\end{figure*}

In many applications (e.g., connecting non-equilibrium initial states to relativistic hydrodynamics) the Landau rest frame (LRF) quantities are needed. By definition, $T^{0i}_\text{LRF} = 0$, the energy flow in the LRF is zero. To find the LRF we solve the generalized eigenvalue problem $(T^{\mu \nu} - \lambda g^{\mu \nu})h_{\nu} = 0$, where $g^{\mu \nu}$ is the metric tensor. The eigenvector corresponding to the largest eigenvalue is proportional to
the 4-velocity of the LRF and the proportionality constant is fixed by the constraint that $\sqrt{u_\mu u^\mu}=1$. To demonstrate the result of this transformation the LRF energy density and collective velocities $u^\mu$ are plotted in the $x$-$z$-plane in \cref{fig:landau_e_v} for a Au+Au collision. One can observe the onset of radial flow after the initial collision of the two nuclei. We note that the LRF energy density before collision reproduces again the nuclear ground state energy density.

\section{Results for heavy-ion collisions}
\label{sec:results}

In this section we compare particle yields and spectra in heavy-ion collisions calculated with SMASH to experimental data from the HADES and FOPI collaborations. The focus for the current analysis lies on pions, because they contribute the majority of the newly produced particles; and on protons, because they are part of the initial system before the collision.

Some time after the collision, the particles don't interact anymore and thus their momenta are frozen. Therefore, the basic bulk observables to quantify the dynamics of the collision are rapidity and transverse momentum spectra. To obtain Lorentz-invariant spectra, the longitudinal rapidity~$y$ and the transverse mass~$m_T$ are used as momentum coordinates:
\begin{align}
y := \operatorname{atanh}\Big(\frac{p_z}{E}\Big) \quad
m_T := \sqrt{m^2 + p_x^2 + p_y^2}
\end{align}
Usually the rapidity~$y$ is rescaled to~$y_0$ such that the nuclei are located at~$y_0 = \pm 1$ before the collision:
\begin{equation}
y_0 := \frac{y - y_{\rm cm}}{y_{\rm cm}}
\end{equation}
where $y_{\rm cm}$ is the rapidity in the center-of-mass frame.

To obtain sensible comparisons between our calculation and experimental data the procedure to select centrality classes needs to be the same. See \cref{sec:erat} for how this is done for the FOPI data.

\begin{figure}
\centering
\includegraphics[width=\linewidth]{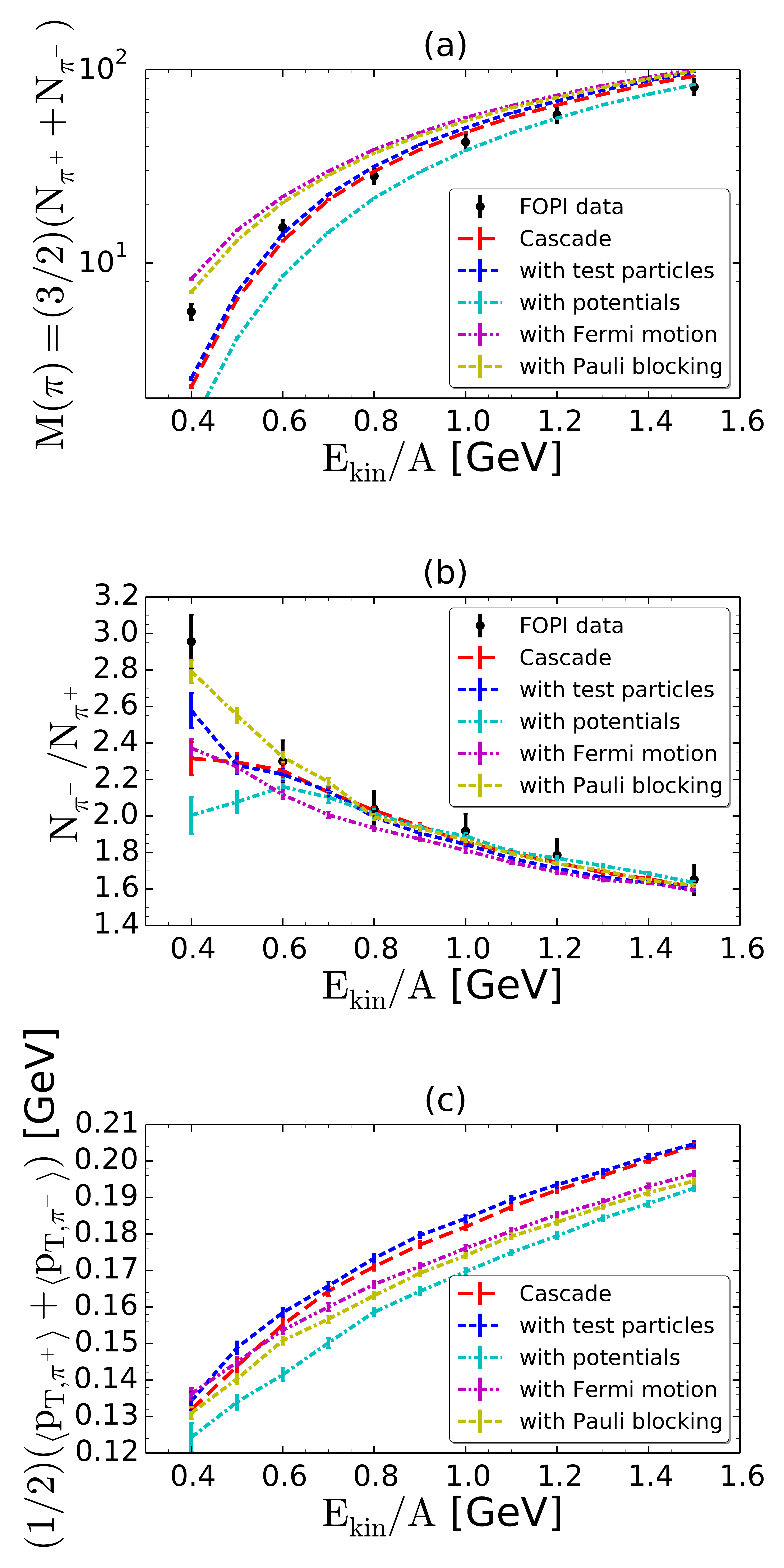}
\caption{
Pion production in gold-gold collisions at kinetic energies ranging from $0.4A\,\text{GeV}$ to $1.6A\,\text{GeV}$, as measured by FOPI~\cite{Reisdorf:2006ie} (markers), in comparison to SMASH (lines).
The upper plot~(a) shows the excitation function of~$\pi^+$ and $\pi^-$ multiplicities, the plot in the middle~(b) shows the ratio.
The lower plot~(c) shows the average transverse momentum of the pions.
The impact parameter was set to $b = 1.33\,\text{fm}$.
The results of the SMASH simulation are shown for the cascade with the following features successively switched on: 20 test particles per real particle, Skyrme and symmetry potentials, Fermi motion, Pauli blocking.
}
\label{fig:fopi_mult_vs_energy}
\end{figure}

First, let us have a look at the total pion multiplicities and their averaged transverse momentum over a broad range of energies. In \cref{fig:fopi_mult_vs_energy} the total multiplicities of charged pions in central Au+Au collisions at kinetic energies from $0.4A\,\text{GeV}$ to $1.6A\,\text{GeV}$ are compared to FOPI measurements~\cite{Reisdorf:2006ie}. The upper plot shows the total pion multiplicity, the one in the middle shows the ratio of negative pions to positive pions to indicate the isospin asymmetry. The lower plot shows the average transverse momentum of the pions.
The impact parameter~$b$ for the SMASH events was sampled from a minimum bias distribution with $b < 2\,\text{fm}$ onto which the corresponding $ERAT$ cuts have been applied. The simulations were run successively with and without potentials (see \cref{eq:potential}), Fermi motion (see \cref{sec:nuclear_collisions}) and Pauli blocking (see \cref{sec:pauli_blocking}).
Potentials and Pauli blocking require a sufficient number of test particles to function properly. When any of these features was enabled, 20~test particles were used instead of one.

Without potentials (and Fermi motion and Pauli blocking) the SMASH results agree well with the data, except for the lowest energy at $0.4A\,\text{GeV}$. A deviation at low energies is expected, because potentials should have a strong effect there.
Running the cascade with 20 instead of one test particle per real particle, there is a slight increase in multiplicity. This effect should be considered a systematic error of the model, since changing the test particle number is not supposed to affect the physics.
Additionally enabling the potentials (which are soft, see \cref{sec:nuclear_collisions}) decreases the pion multiplicities by a large amount.
Adding Fermi motion to the simulation yields the strongest effect and increases the multiplicities.
Pauli blocking causes a small decrease in multiplicity.
For the more physical scenario with all features enabled there is an overestimation of the number of pions at all energies.
Such an overestimation and a decrease of the multiplicities due to soft potentials and Pauli blocking has been observed with one of the first transport models as well~\cite{Kruse:1985hy}.

The pion ratios look similar with and without potentials. Only for the lowest energy the results with potentials are a bit closer to the experimental values. Please note that no Coulomb potentials are included in this calculation. In an earlier comparison with the FOPI data for gold-gold collisions at $1.5A\,\text{GeV}$, it has been suggested that Coulomb potentials "almost exclusively" account for the difference in the momentum spectra of the charged pion species~\cite{Reisdorf:2006ie}. The results here do not support this claim, because the total relative multiplicities of the pions are reproduced without any Coulomb potentials.

The transverse momenta do not vary significantly among the different pion species or with and without potentials.
It is difficult to pin down the reason for the overestimation of the pion multiplicity. At this energy a lot of implementation details can influence the multiplicity significantly: Fermi momenta, potentials and the $N\Delta$ cross sections (which haven't been measured) introduce some uncertainties. The cross sections can be reduced by in-medium effects~\cite{TerHaar:1987ce,Li:1993ef}, which is unaccounted for in SMASH. These in-medium effects would reduce the number of produced pions. More work is needed to understand the exact reasons for the discrepancy.
On the other hand, SMASH is primarily designed for FAIR energies, where potentials will be less important, and the results are similar to those of other approaches.

\begin{figure}
\centering
\includegraphics[width=\linewidth]{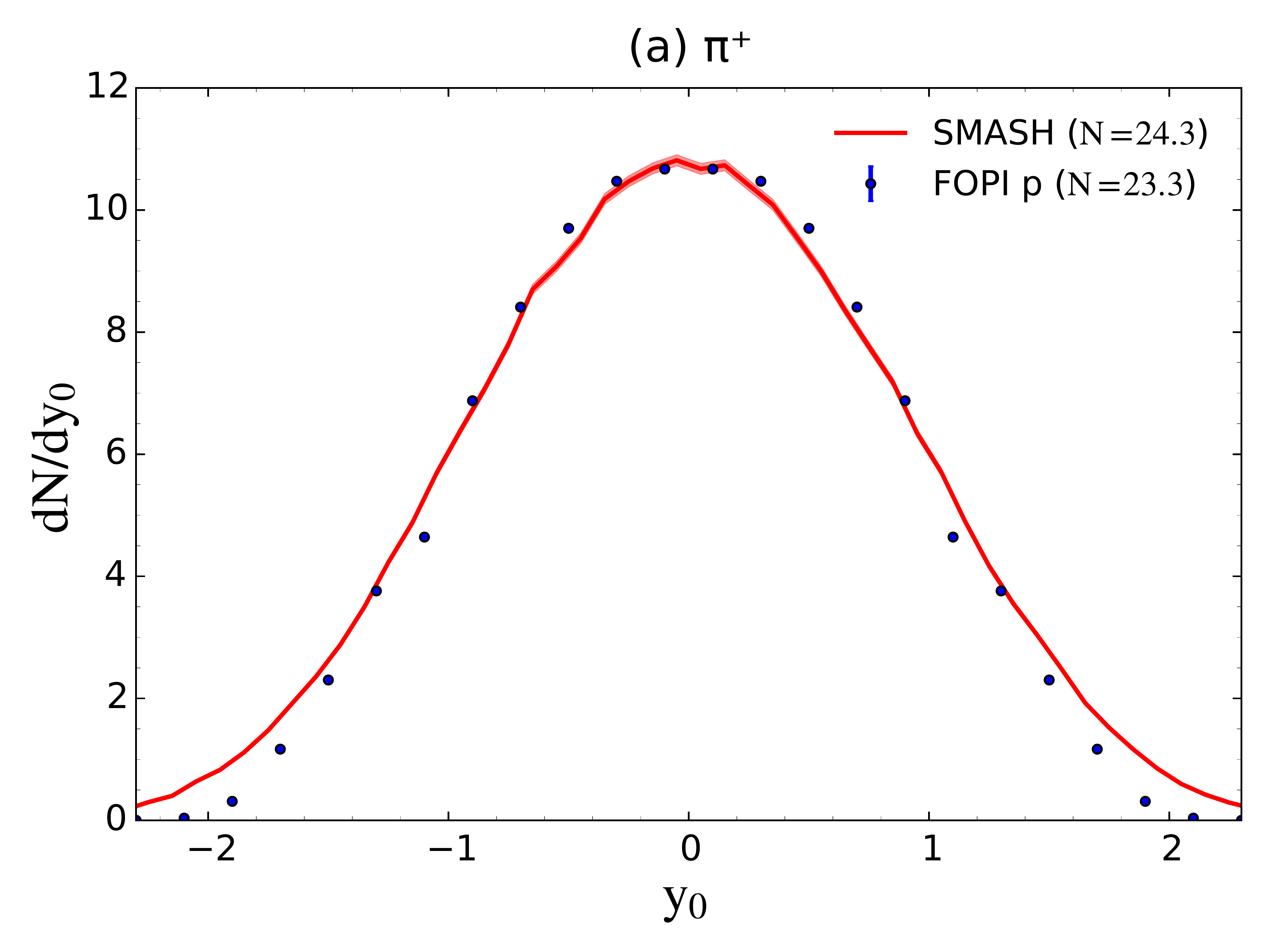}
\includegraphics[width=\linewidth]{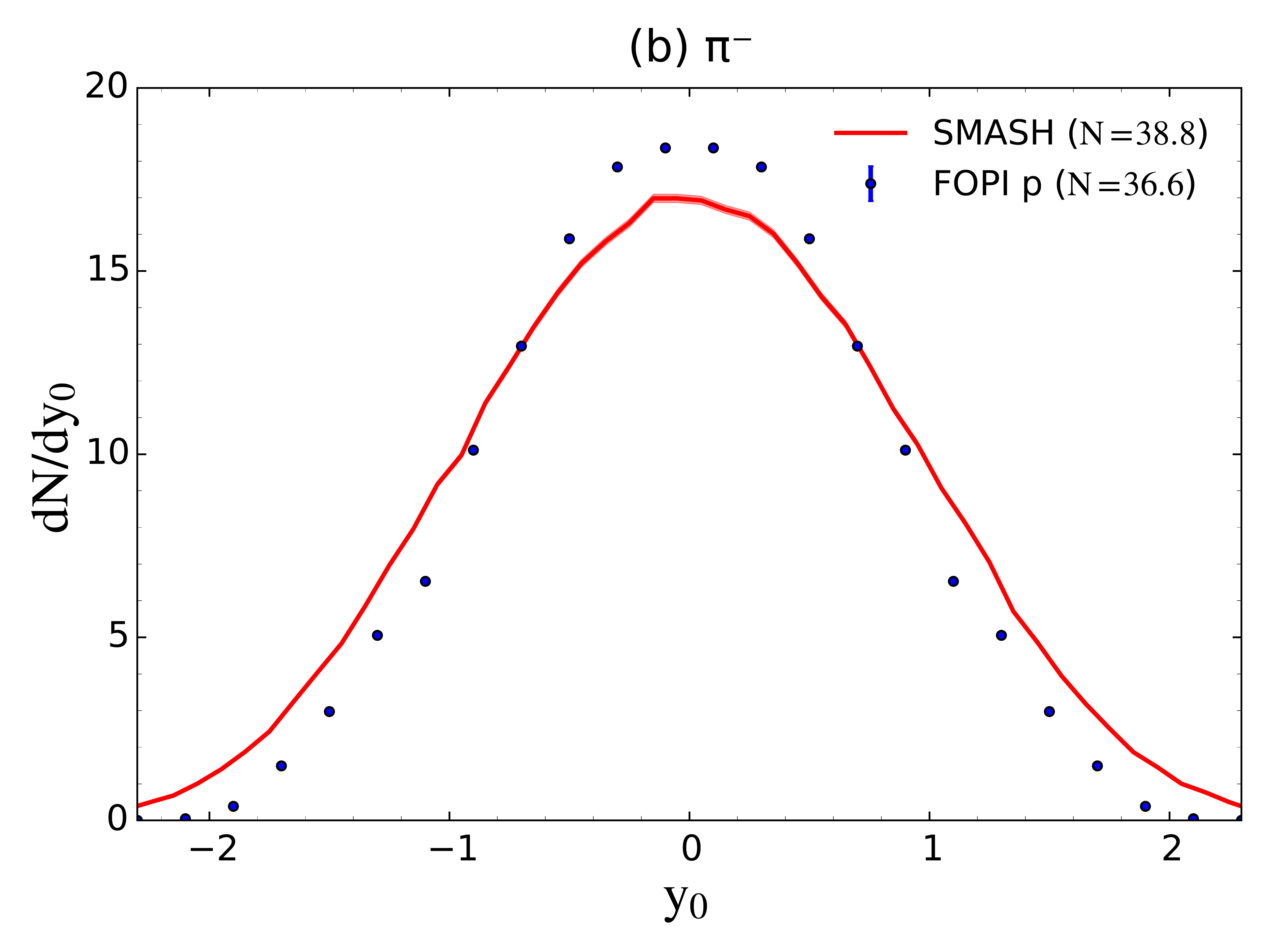}
\caption{
Rapidity spectra for pions measured by FOPI in gold-gold collisions at~1.5~AGeV~\cite{Reisdorf:2006ie}.
The experimental data (markers) is compared to the corresponding SMASH results (lines).
$N$ is the total number of pions obtained by integrating the spectrum.
The normalized rapidity~$y_0 = (y - y_\text{cm}) / y_\text{cm}$ was used.
The impact parameter for the simulated events was sampled from the distribution given by the $\mathit{ERAT}$ cuts corresponding to the experimental data (see \cref{sec:erat}).
The SMASH simulations were performed with potentials, Pauli blocking and Fermi motion.
}
\label{fig:fopi_y}
\end{figure}

\begin{figure}
\centering
\includegraphics[width=\linewidth]{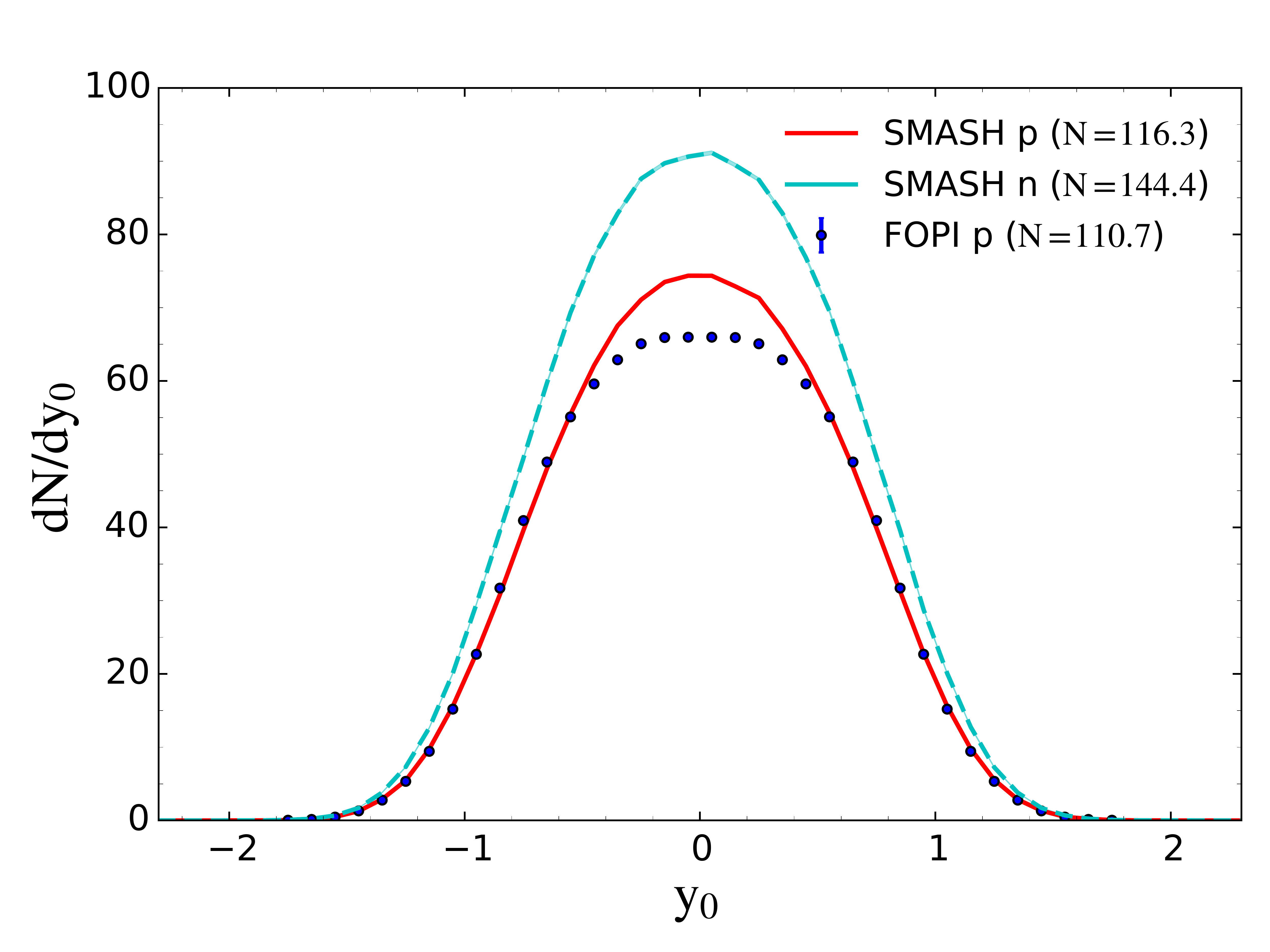}
\caption{
Rapidity spectra for protons measured by FOPI in gold-gold collisions at~1.5~AGeV~\cite{Reisdorf:2006ie}.
The experimental data (markers) is compared to the corresponding SMASH results (lines).
See \cref{fig:fopi_y} for more details.
$N$ is the total number of particles obtained by integrating the spectrum.
Spectators (particles only interacting elastically) have been ignored.
To distinguish between unbound protons and deuteron or other nuclei, a coalescence afterburner with parameters $p_0 = 0.3\,\text{GeV}, r_0 = 0.9\,\text{fm}$ was used to model the clustering (see \cref{sec:clustering}).
}
\label{fig:fopi_protons_y}
\end{figure}

Since the multiplicities agree rather well, let us move on to more differential observables. \cref{fig:fopi_y} shows charged pion multiplicities as a function of the scaled rapidity~$y_0$, comparing the spectra obtained from SMASH to the experimental results of the FOPI collaboration, for Au+Au collisions at a kinetic energy of~$1.5A \,\text{GeV}$. SMASH reproduces the shape of the rapidity spectra fairly well, overestimating the total multiplicities by a few percent as seen before.

In \cref{fig:fopi_protons_y} the proton rapidity spectrum yielded by SMASH is compared to FOPI measuremnts. The parameters are the same as just discussed. To get rid of spectators, all nucleons that interact only elastically have been ignored. SMASH does not model the production of nuclei formed by clustered nucleons. To be able to compare to the experimental data, a simple coalescence afterburner described in \cref{sec:clustering} has been employed. Any pairs of nucleons with momentum distance $\Delta p < 0.3\,\text{GeV}$ and spatial distance $\Delta x < 0.9\,\text{fm}$ have been ignored.
These parameters were chosen to fit the data.
The shape is very well reproduced at the tails, but the number of protons is overestimated at mid-rapidity.

\begin{figure*}
\centering
\includegraphics[width=0.49\textwidth]{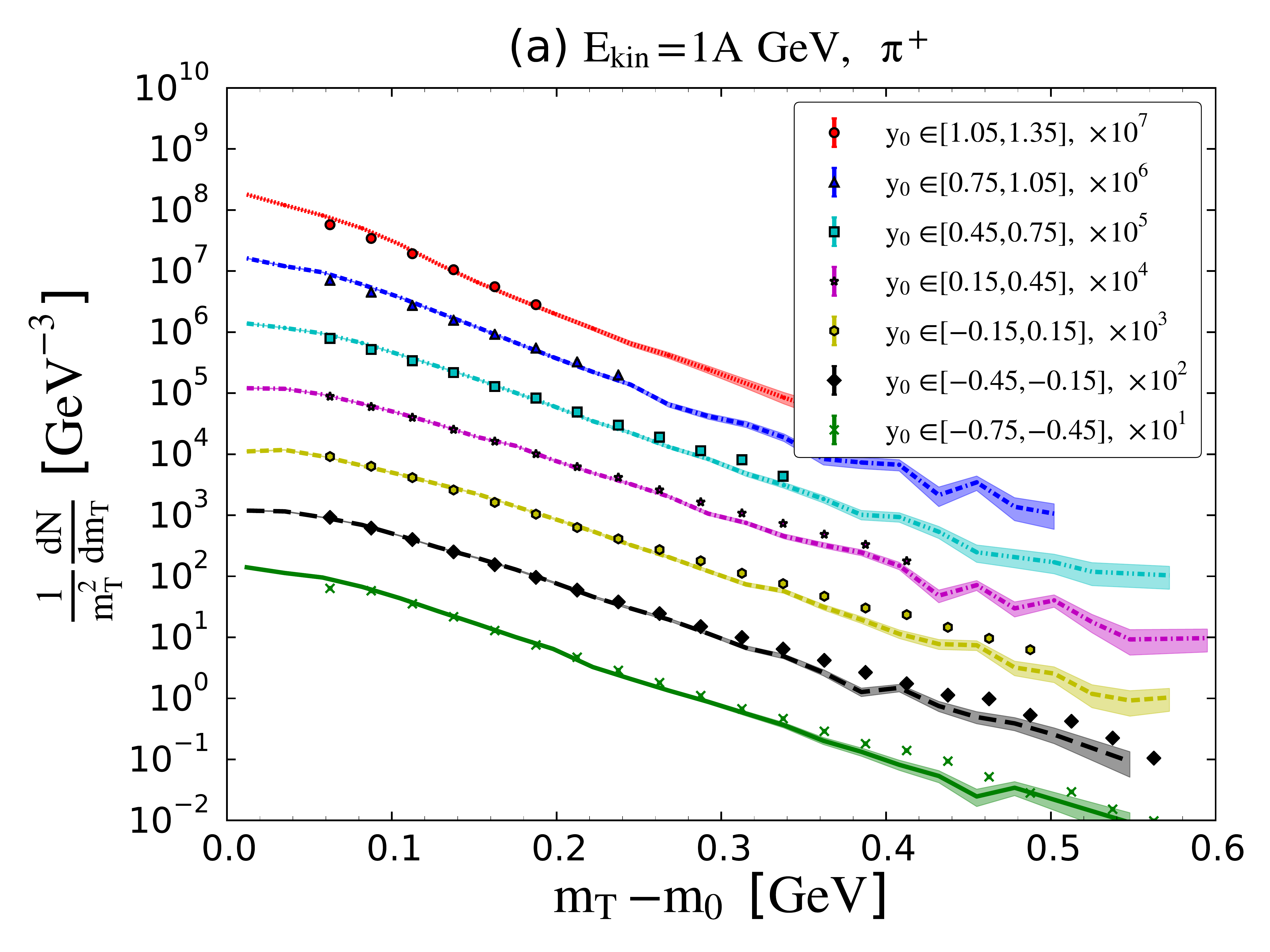}
\includegraphics[width=0.49\textwidth]{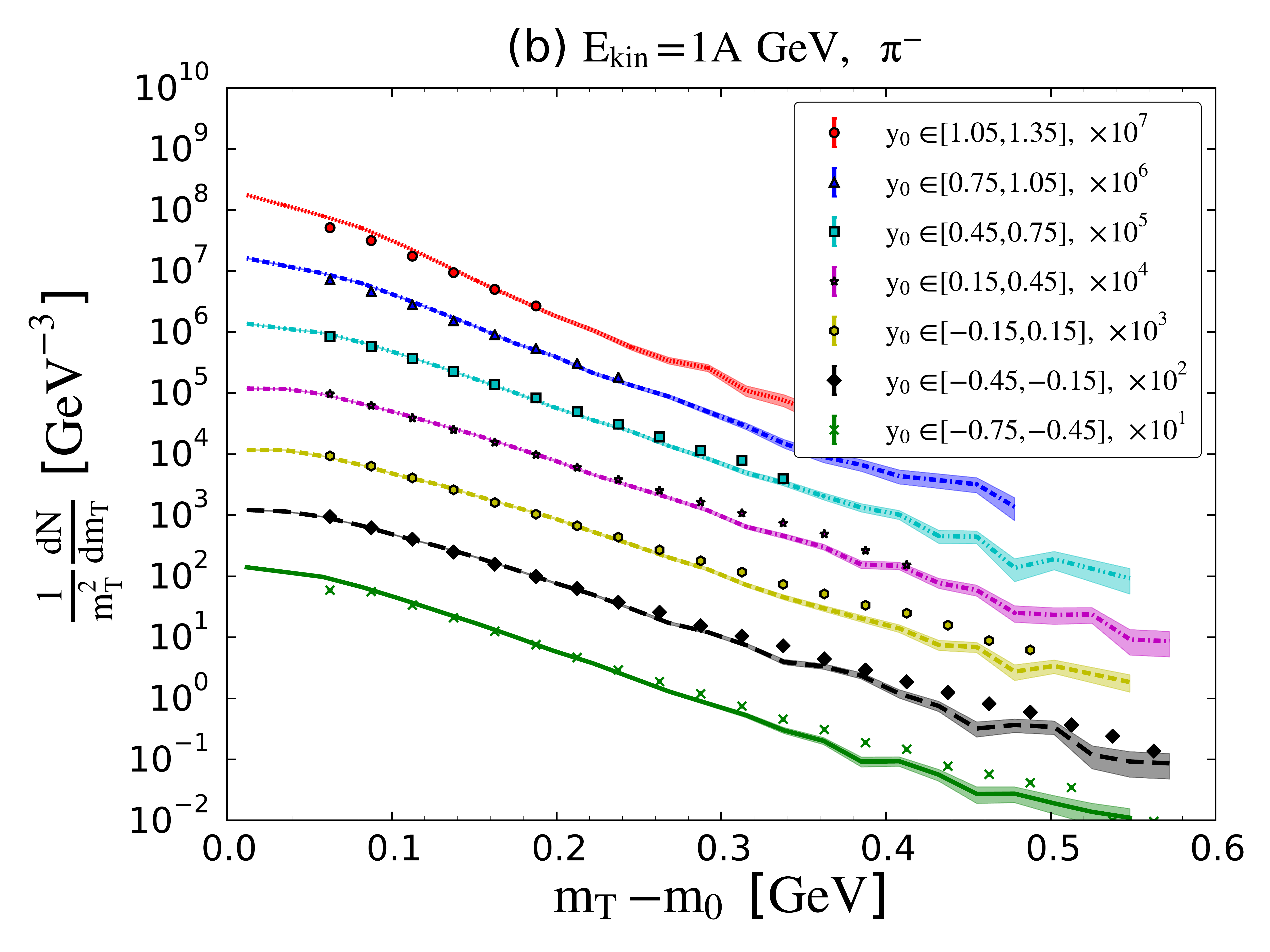}
\includegraphics[width=0.49\textwidth]{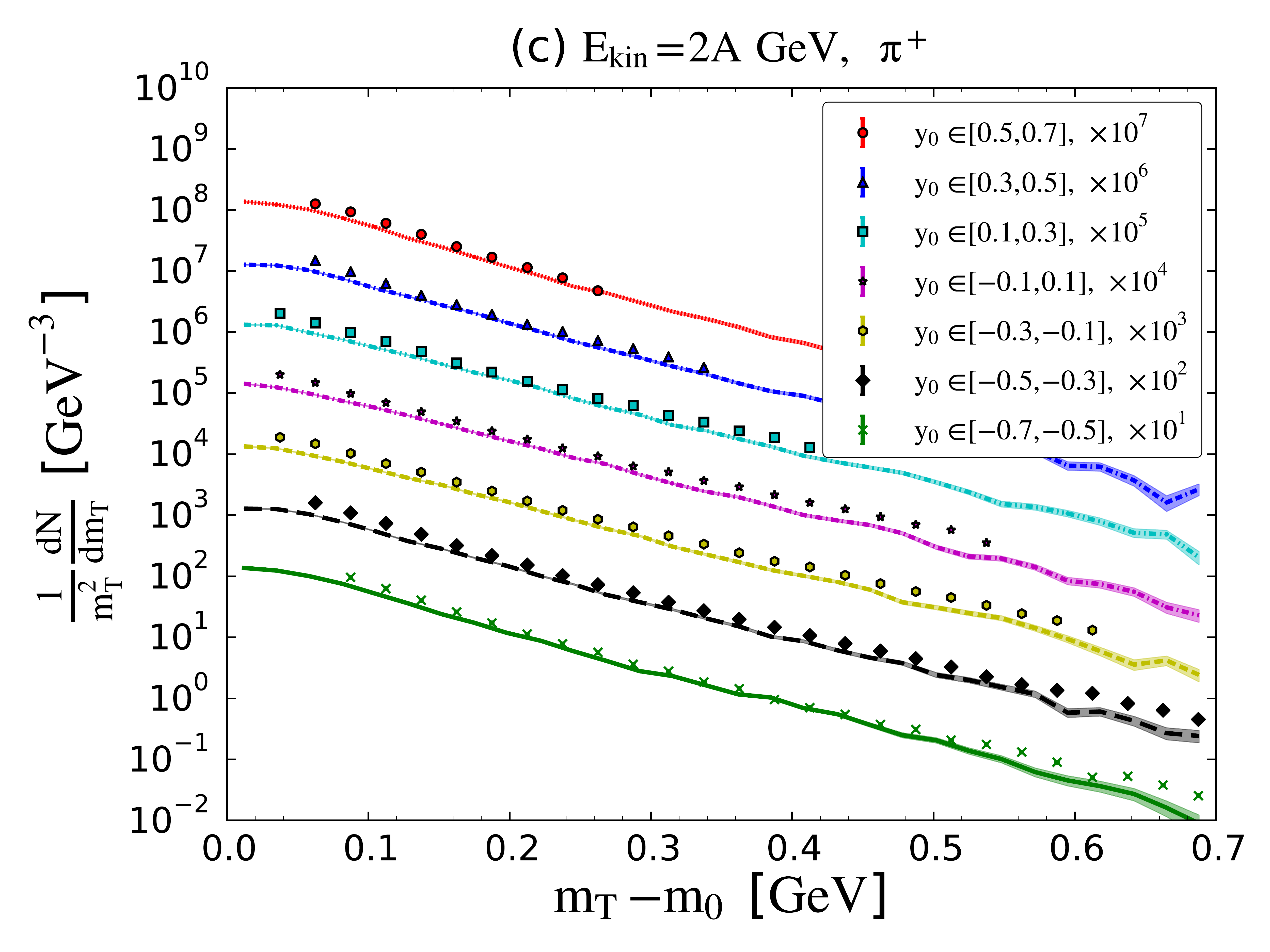}
\includegraphics[width=0.49\textwidth]{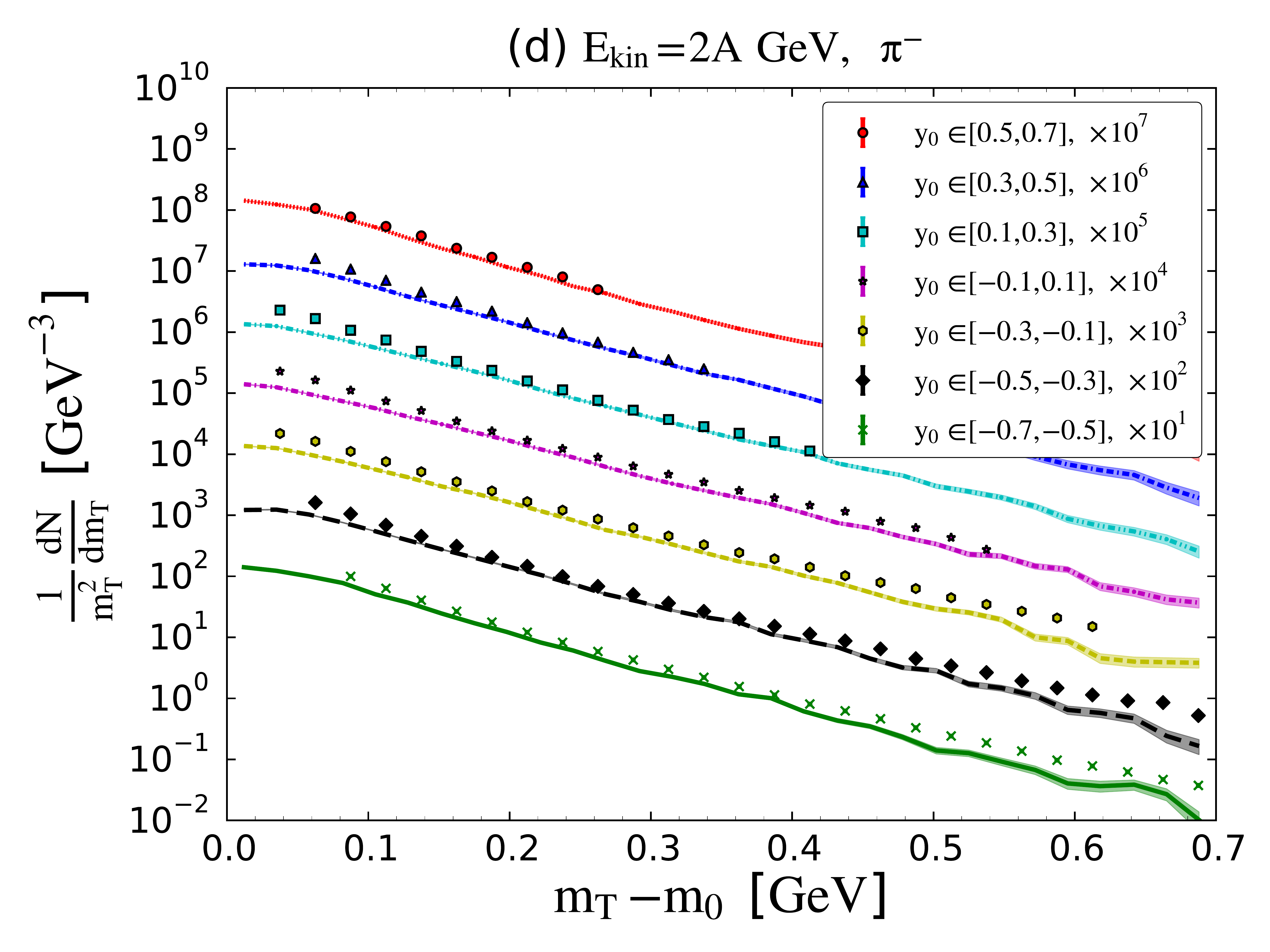}
\caption{
Transverse mass spectra for pions measured by HADES in carbon-carbon collisions at~1 and $2A\,\text{GeV}$~\cite{Agakishiev:2009zv}.
The experimental data (markers) is shown for different longitudinal rapidity bins and compared to the corresponding SMASH results (lines).
For readability, the data corresponding to each bin was multiplied with a different power of 10.
The impact parameter distribution provided by HADES was used for sampling the events with SMASH.
}
\label{fig:hades_CC_spectra}
\end{figure*}

In \cref{fig:hades_CC_spectra}, the multiplicity of charged pions is shown as a function of the transverse mass~$m_T$ for different windows of normalized rapidity~$y_0$, for C+C collisions at energies~$E_{\rm kin} \in \{1, 2\}A \,\mathrm{GeV}$ as measured by the HADES collaboration~\cite{Agakishiev:2009zv}. For a purely thermal spectrum one would expect a straight line in the logarithmic plot, with the slope corresponding to the effective temperature. The events were generated with SMASH by sampling the impact parameter distribution provided by HADES (which was reconstructed using another transport model~\cite{Agakishiev:2009zv}).
The calculations were performed with Skyrme and symmetry potentials, Fermi motion and Pauli blocking. It can be seen that SMASH describes the experimental data reasonably well. There are some deviations for large rapidities at $1A \,\mathrm{GeV}$ and for small transverse mass at $2A \,\text{GeV}$. In comparison to the UrQMD transport model~\cite{Agakishiev:2009zv}, SMASH gives a similarly good agreement with the HADES data.

\begin{figure*}
\centering
\includegraphics[width=0.49\textwidth]{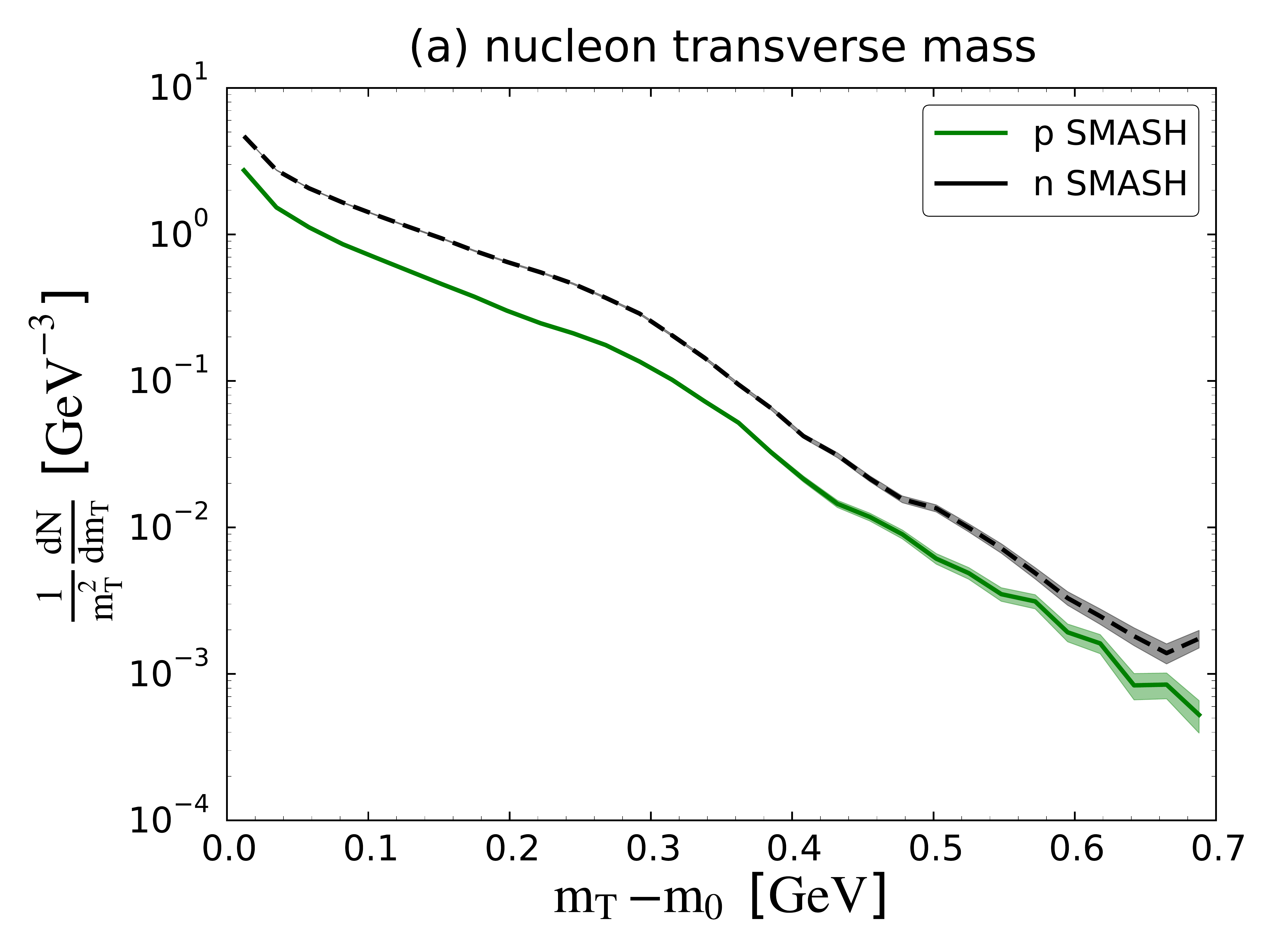}
\includegraphics[width=0.49\textwidth]{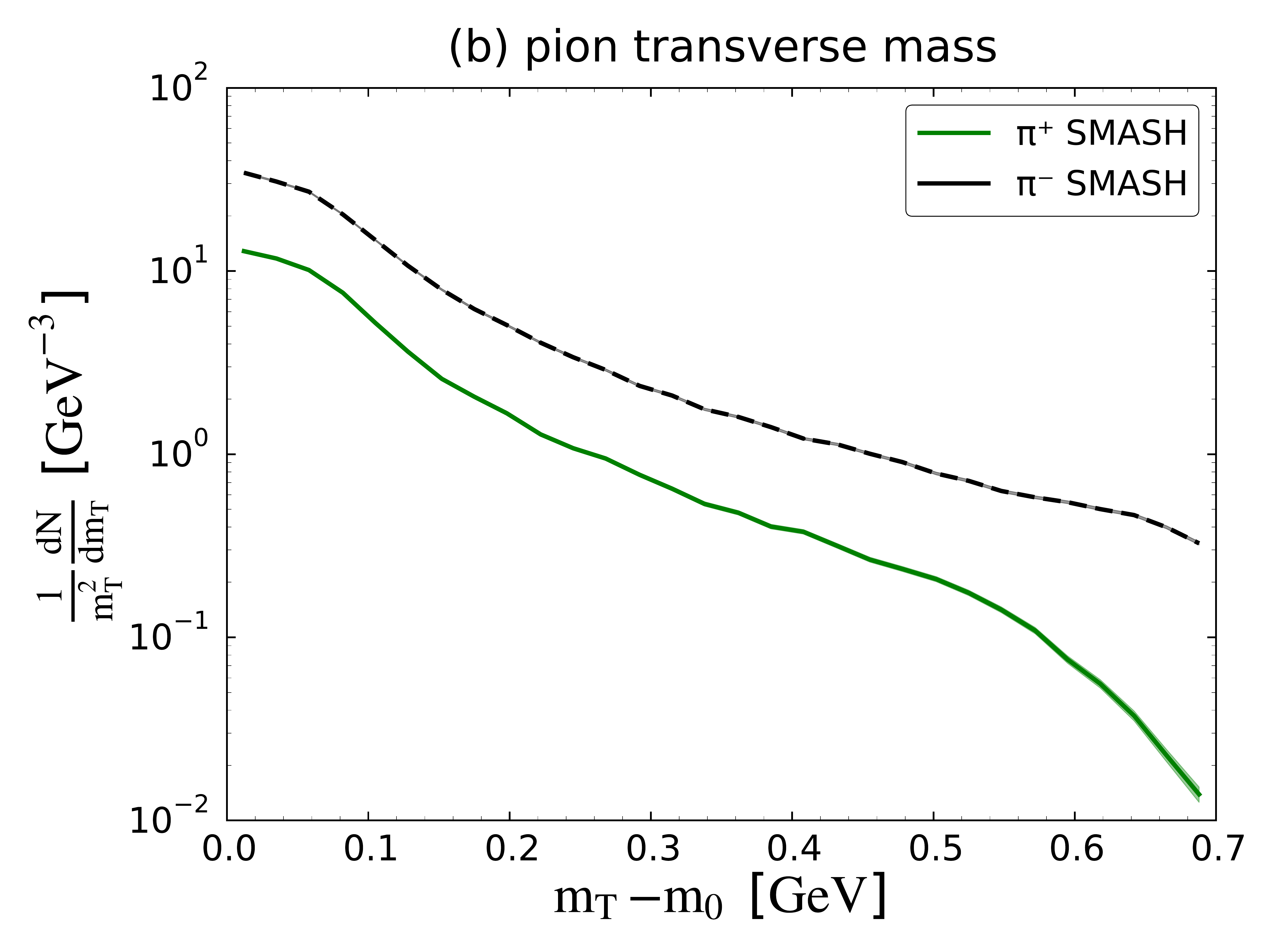}
\includegraphics[width=0.49\textwidth]{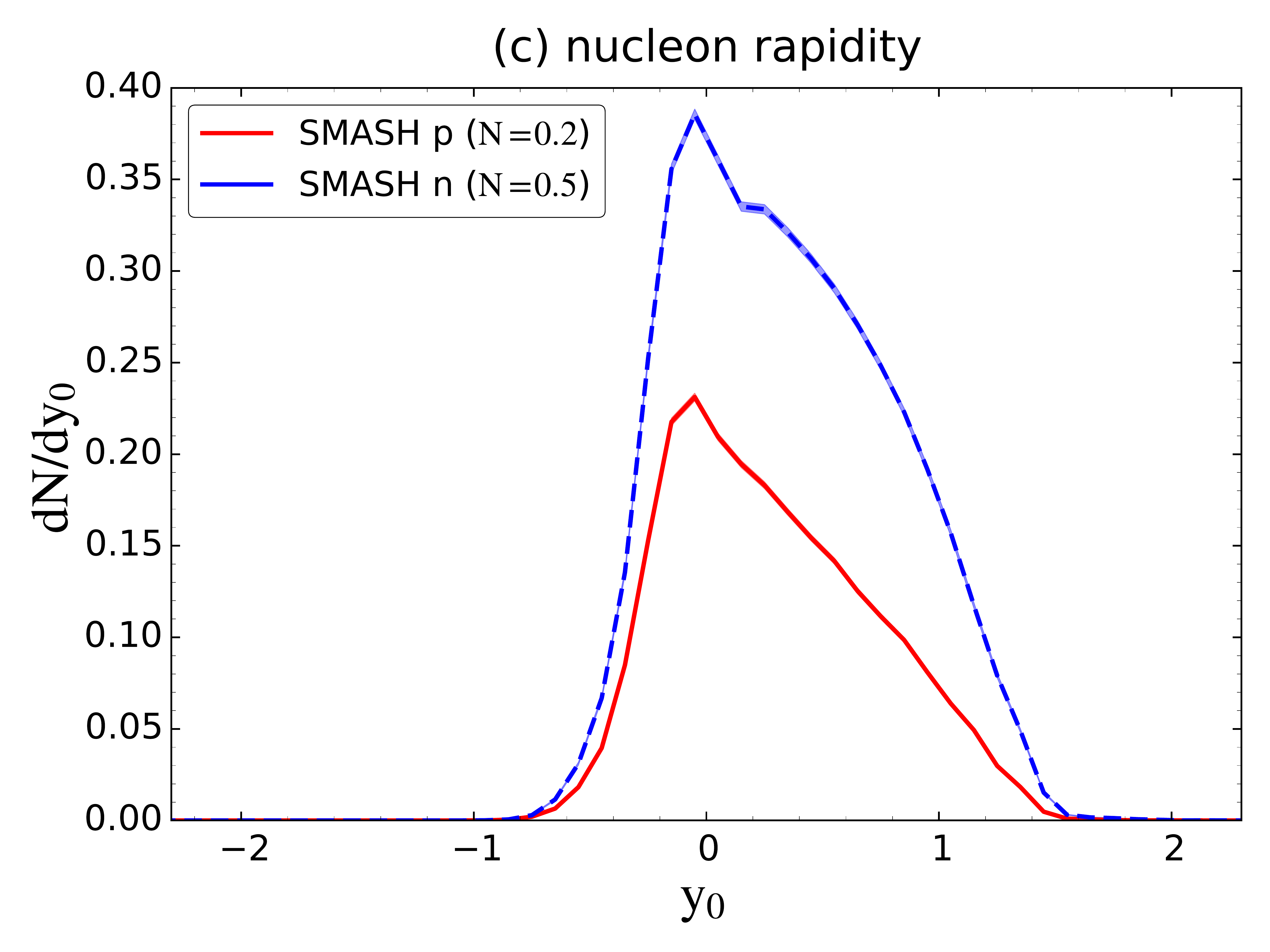}
\includegraphics[width=0.49\textwidth]{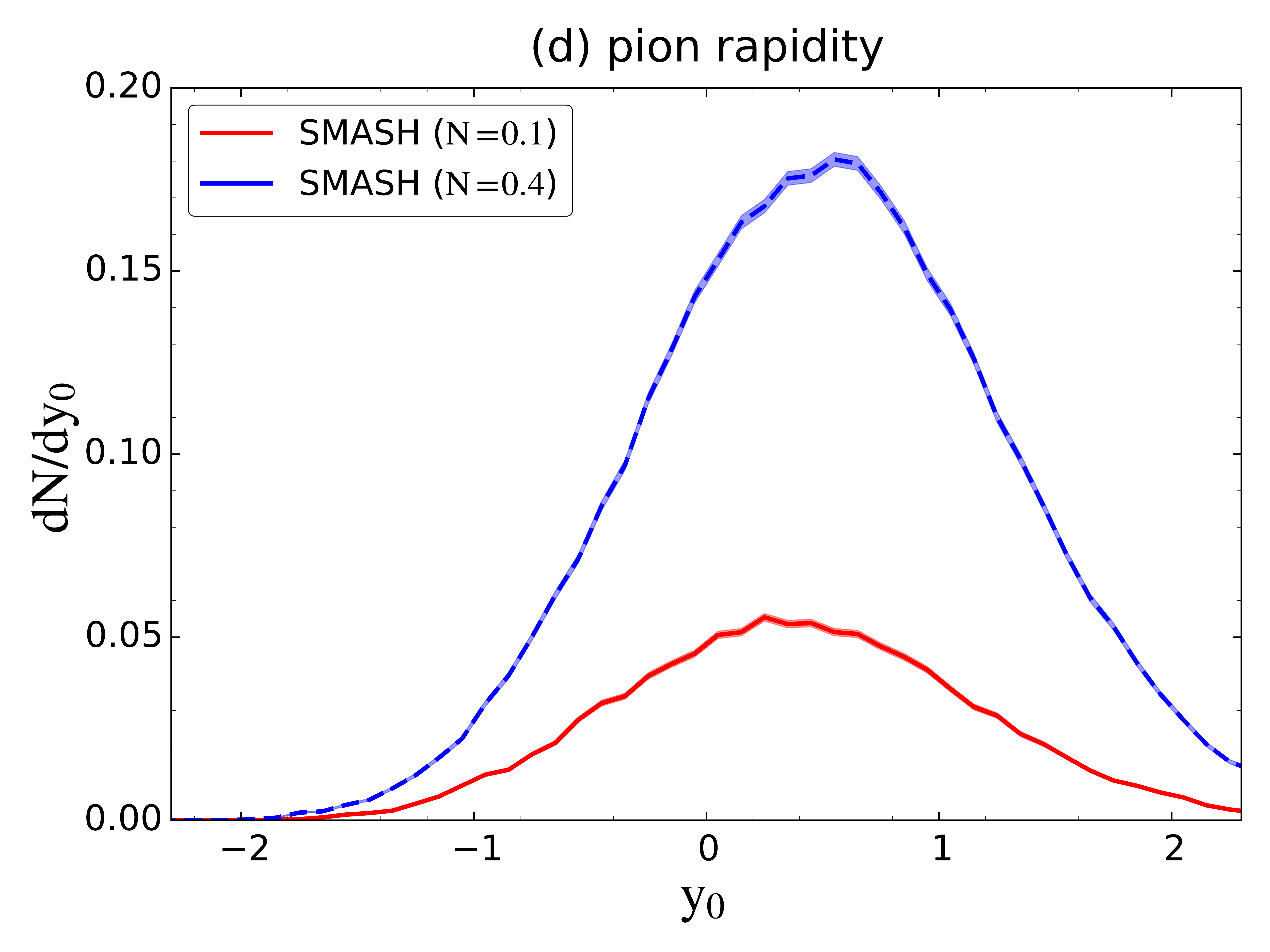}
\caption{
Transverse mass and rapidity spectra for charged pions and nucleons in $\pi^-$-carbon collisions at~$E_\text{kin} = 1.7\,\text{GeV}$.
They were obtained from a SMASH simulation with 20~test particles per real particle, including potentials, Fermi motion and Pauli blocking.
Spectators (particles only interacting elastically) were ignored.
Data for this scenario has been measured by HADES, but is not yet published.
The legend shows the total multiplicity~$N$ obtained from integrating the rapidity spectrum.
}
\label{fig:hades_pi-C_spectra}
\end{figure*}

\begin{figure}
\centering
\includegraphics[width=0.48\textwidth]{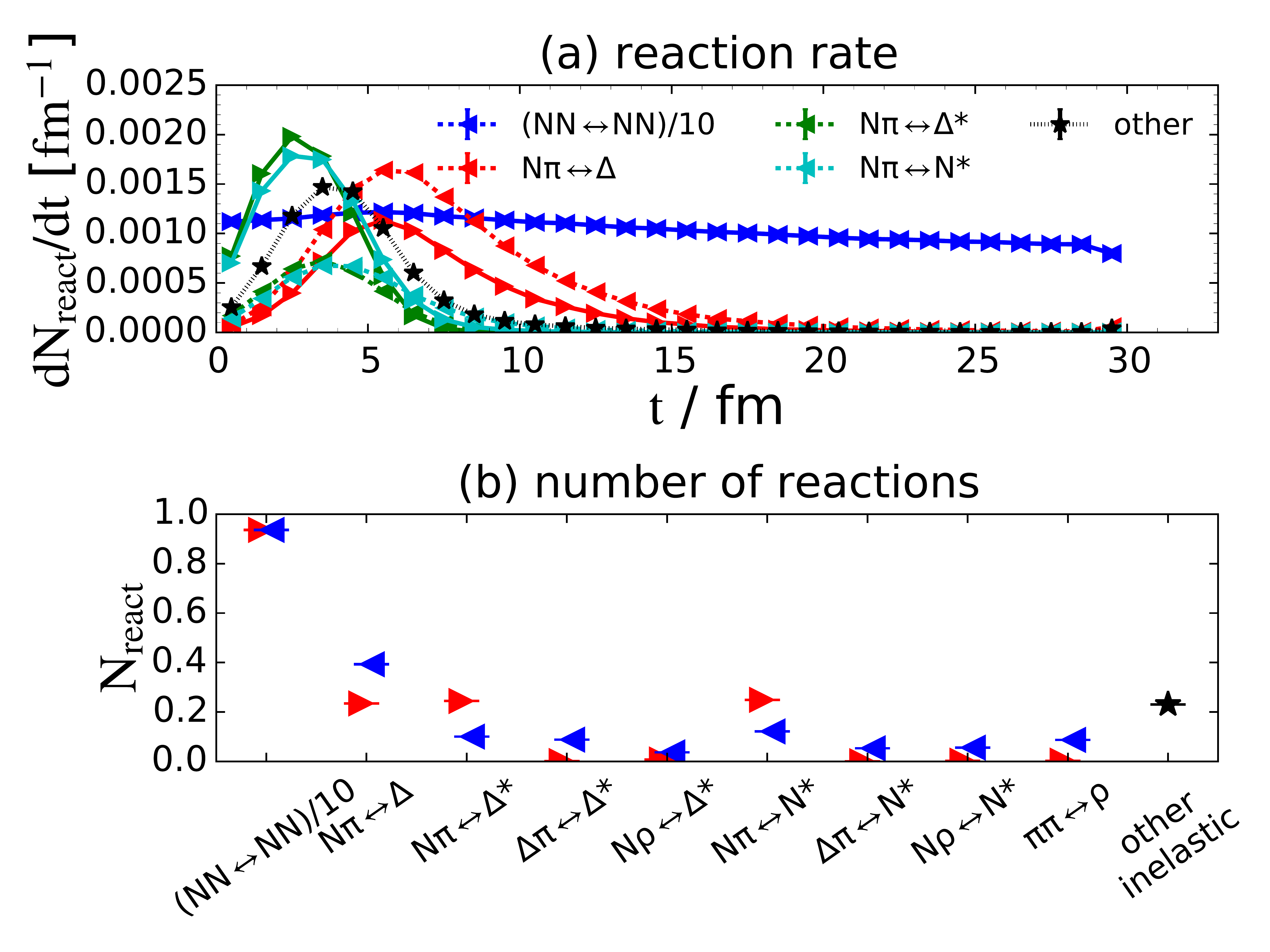}
\caption{
Number of reactions in a $\pi^-$-carbon collision at $E_\text{kin} = 1.7\,\text{GeV}$, averaged over 25000 events.
The upper plot (a) shows the forward (right arrows, solid lines) and backward rates (left arrows, dashed lines) per event for the most important reactions as a function of time.
The lower plot (b) shows the total number of forward and backward reactions per event for various reactions.
The same SMASH simulation results as in \cref{fig:hades_pi-C_spectra} were used.
}
\label{fig:hades_pi-C_reactions}
\end{figure}

In \cref{fig:hades_pi-C_spectra}, the multiplicity of nucleons and charged pions is shown as a function of transverse mass~$m_T$ and rapidity~$y$, for $\pi^-$+C collisions at $1.7\,\text{GeV}$.
The impact parameter was sampled from a minimum-bias distribution over the full range $b \in [0,3]\,\text{fm}$.
Like before, the SMASH simulation included potentials, Fermi motion and Pauli blocking.
20~test particles were used per real particle.
Spectators (particles that only interact elastically) have been ignored.
This scenario has been experimentally studied by the HADES collaboration, the results are however not yet public.
The spike in the nucleonic rapidity spectrum corresponds to slow participants in the nucleus.

Unlike in experiment or in a simple thermal model, we can look at whole time evolution in a transport code. This enables us to study the different reaction rates. In \cref{fig:hades_pi-C_reactions} various forward (right) and backward (left) reaction rates are shown, as a function of time and as a total per event.
Elastic nucleon-nucleon interactions are dominating and are divided by 10 in the plot. These are mostly due to the combination of Fermi motion and potentials, causing the nucleons in the target to interact with each other. 

The first inelastic reactions are excitations of $N^*$ and $\Delta^*$ resonances. Production of $\Delta$ resonances or elastic nucleon-nucleon collisions happen at later times.
Producing $N^*$ and $\Delta^*$ consumes more pions than it directly yields, but these excitations decay mostly into $\Delta$ and $\rho$, which finally produce pions again.
It is remarkable that this system is far from chemical equilibrium, unlike symmetric collisions of heavy nuclei like gold or lead.

All in all, the transport approach presented here matches the experimental data on pion and proton production reasonably well, passes equilibrium/detailed balance tests and compares well to elementary particle production cross sections.

\section{Summary and Outlook}

To summarize, a new hadronic transport approach (SMASH) has been introduced. It is aimed at providing a dynamical description of heavy-ion reactions in the low and intermediate beam energy range. The relativistic Boltzmann equation with hadronic degrees of freedom is solved including a basic version of nuclear mean-field potentials. Interactions proceed via resonance excitation and decay, where all resonances have vacuum properties only. The initial conditions are demonstrated explicitly and it is shown that the approach maintains detailed balance. The elementary cross sections and angular distributions are in agreement with experimental data. The comparison of proton and pion spectra to experimental data from $E_{\rm Kin} = 1-2A$ GeV hints at missing medium modifications of the cross sections, but there is still reasonable agreement in the current approach. Predictions for particle production in $\pi$-A collisions are made.
In this case the meson-baryon interactions play a more dominant role than in heavy-ion reactions.

In the future, the approach will be enhanced to include the full strangeness production and the cross sections are going to be extended to higher energies by including string excitation and fragmentation. In addition, photon and dilepton production \cite{Weil:2016fxr} is going to be studied in detail. Here it is of special interest to compare the non-equilibrium hadronic production with the one from thermal rates as currently employed in hydrodynamic approaches. In general, this approach will be very useful to study the effects of hadronic rescattering on flow and correlation observables at RHIC and LHC energies. Infinite-matter calculations are going to be employed to study transport coefficients of hadronic matter as a function of temperature and baryo-chemical potential. Also the effects of kinematic cuts and baryon diffusion on higher moments will be investigated \cite{Petersen:2015pcy}.
Overall, this approach constitutes a very flexible hadronic transport approach that is going to shed light on the properties of hot and dense strongly interacting matter as created in heavy-ion reactions in a large range of beam energies.

\section{Acknowledgements}
\label{sec:ack} The authors thank M.~Bleicher, K.~Gallmeister, S.~A.~Bass, Y.~Nara, J.~Gerhard, J.~M.~Torres-Rincon and W.~Bauer for fruitful discussions and Y.~Leifels for providing the FOPI data tables. Discussions with S.~Pratt, E.~Bratkovskaya, W.~Cassing and J.~Aichelin during the early stages of development are acknowledged. Computational resources have been provided by the Center
for Scientific Computing (CSC) at the Goethe-University of Frankfurt. The authors acknowledge funding of a Helmholtz Young Investigator Group VH-NG-822 from the Helmholtz Association and GSI. This work was supported by the Helmholtz International
Center for the Facility for Antiproton and Ion Research (HIC for FAIR) within the
framework of the Landes-Offensive zur Entwicklung Wissenschaftlich-Ökonomischer Exzellenz (LOEWE) program launched by the State of Hesse. D.~O. acknowledges support by the Deutsche Telekom Stiftung. M.~A. is supported by the Marie Skodowska-Curie Individual Fellowship
658574 FastTh.
D.~O. and V.~S. acknowledge support by the Helmholtz Graduate School for Hadron and Ion Research (HGS-HIRe).

\appendix
\section{Integrals used in Pauli blocking}
\label{sec:paulibl_append}

Determining if the reaction is Pauli-blocked requires calculation of the phase-space density at a given point $(\vec{r}, \vec{p})$. Whilst in the momentum space we just count momenta in the sphere around $\vec{p}$, in the coordinate space we take advantage of the function that was suggested in the GiBUU model, see section D.4.3 in \cite{Buss:2011mx}. In GiBUU, however, the integrals in the smearing function are computed numerically. We have found the following analytical expressions for them:

\begin{align*}
& \frac{1}{2\pi} \int_{\Delta V_r, |\vec{r}-\vec{r}_j|<r_c} d^3r \, \exp \left( - \frac{(\vec{r}-\vec{r}_j)^2}{2 \sigma^2} \right) \\
= &\begin{cases}
    \alpha & r_c > r_r,\, r_j = 0 \\
    \beta & r_c > r_r + r_j \\
    \gamma & r_c < r_r + r_j
\end{cases} \\
\alpha &= - 2 r_r \sigma^2 e^{-\frac{r_r^2}{2\sigma^2}} + \sqrt{2\pi} \sigma^3 \erf\big(\frac{r_r}{\sqrt{2} \sigma}\big) \\
\beta &= \frac{\sigma^4}{r_j} \left(e^{-\frac{(r_j+r_r)^2}{2\sigma^2}} - e^{-\frac{(r_j-r_r)^2}{2\sigma^2}} \right) \\
&\phantom{{}={}}+ \sqrt{\frac{\pi}{2}} \sigma^3 \left(\erf\big(\frac{r_j+r_r}{\sqrt{2} \sigma}\big) - \erf\big(\frac{r_j-r_r}{\sqrt{2} \sigma}\big) \right)\\
\gamma &= \frac{\sigma^2}{r_j} \left(\frac{1}{2}e^{-\frac{r_c^2}{2\sigma^2}} ((r_c-r_j)^2-r_r^2+2\sigma^2) - \sigma^2 e^{-\frac{(r_j-r_r)^2}{2\sigma^2}}\right) \\
&\phantom{{}={}} + \sqrt{\frac{\pi}{2}} \sigma^3 \left(\erf\big(\frac{r_c}{\sqrt{2} \sigma}\big) - \erf\big(\frac{r_j-r_r}{\sqrt{2} \sigma}\big) \right) \\
\kappa &= \frac{2 \Delta V_r \Delta V_p N}{(2 \pi \hbar c)^3} \left(\erf\big(\frac{r_c}{\sqrt{2} \sigma}\big) - \frac{r_c}{\sigma} \sqrt{\frac{2}{\pi}} e^{-\frac{r_c^2}{2\sigma^2}} \right)
\end{align*}

\section{Infrastructure and Technology}
A hadronic transport code needs to be maintainable and well-documented.
SMASH is written in object-oriented modular C++11 and under Git version control~\cite{git}. The code repository is linked to the project management platform Redmine~\cite{redmine} which allows for easy collaborative work on the project and issue tracking. The whole documentation (internal and external) is generated with Doxygen~\cite{doxygen}.
As output formats, the well established OSCAR 1997 \cite{OSCAR1997} and 2013 \cite{OSCAR2013} formats are supported for particle lists and collision history output in ASCII text and binary format. In addition, ROOT trees \cite{ROOT} can be generated and VTK output \cite{VTK} can be used to visualize the simulation.

\section{Centrality selection for FOPI data}
\label{sec:erat}

The FOPI collaboration introduces an $\mathit{ERAT}$ cut determined by the $b_0 < 0.15$ bin, where
\begin{equation}
b_0 := \frac{b}{b_{\rm max}}
\quad
b_{\rm max} := 1.15\,\mathrm{fm}\, \big(A_P^\frac{1}{3} + A_T^\frac{1}{3}\big)
\end{equation}
for an impact parameter~$b$ and given number of nucleons in the projectile~($A_P$) and in the target~($A_T$).
$\mathit{ERAT}$ is defined as a ratio of the transverse kinetic energy and the longitudinal kinetic energy~\cite{Reisdorf:1996qj}, which can be directly calculated from the momenta:
\begin{equation}
\mathit{ERAT}
:= \frac{E_T}{E_L}
:= \frac{\sum_i p_{Ti}^2 / (m_i + E_i)}{\sum_i p_{Li}^2 / (m_i + E_i)}
\end{equation}
It has been shown that this quantity is monotonic in the impact parameter~$b$ and can thus be used for constraining the centrality, while being much easier to access experimentally.
The $\mathit{ERAT}$~cut corresponding to the desired $b$~cut can be obtained in the following way:
\begin{enumerate}
    \item Sample events using a minimum bias distribution with~$b \in [0, b_{\rm max}]$, for a sufficiently large~$b_{\rm max}$.
    \item Calculate the $\mathit{ERAT}$ histogram from the events.
    \item Renormalize the histogram to the maximal cross section~$\pi b_{\rm max}^2$.
    \item Calculate the cross section corresponding to the cut: $\sigma := \pi b_{\rm cut}^2$.
    \item Find the largest $\mathit{ERAT}$ corresponding to~$\sigma$.
    \item Ignore all events beyond that $\mathit{ERAT}$~value.
\end{enumerate}
After this procedure, the remaining events should belong to the same centrality class as the experimental events.

Note that $\mathit{ERAT}$ is frame-dependent. For the purpose of this paper, it has been calculated in the fixed-target frame.

\section{Nucleon clustering}
\label{sec:clustering}

A hadronic transport code does not have a concept of nuclei, because it considers only hadronic degrees of freedom. However when comparing to experiment, it is important to know which nucleons are bound in a cluster, because only unbound protons are considered as protons by the detector.

To model clustering we use a simple coalescence afterburner inspired by the work of \citeauthor{Li:2015pta}~\cite{Li:2015pta} that considers the pairwise distance in position and momentum space. Any pair of nucleons with a relative distance $\Delta r < r_0$ and a relative momentum $\Delta p < p_0$ is considered to be part of a cluster and will be ignored when calculating the nucleon spectra. To make this procedure Lorentz-invariant, before calculating the distances the particles are boosted to the center-of-momentum frame and their position is extrapolated so the boosted four-vectors correspond to the same time.

It is usually experimentally known how many protons are bound in a cluster, so the parameters~$(r_0, p_0)$ can be chosen such that the correct multiplicities are obtained. Care has to be taken that the simulation runs long enough, otherwise $r_0$ strongly depends on the time at which the simulation is stopped.

\bibliography{inspire,non_inspire}

\end{document}